\documentclass[12pt]{article}
\usepackage{amsmath,bm,amsfonts,color,amsthm, amssymb}
\usepackage[margin=2.5cm]{geometry}
\usepackage{graphicx}
\usepackage{natbib}
\usepackage{xcolor}
\usepackage{enumitem}
\usepackage{subcaption,siunitx,booktabs}
\usepackage[citecolor=blue,urlcolor=blue,allcolors=blue]{hyperref}
\usepackage{arydshln}

\usepackage{titlesec}
\titlelabel{\thetitle.\quad}

\usepackage{tikz}
\usepackage{transparent}

\newenvironment{psmallmatrix}
  {\left(\begin{smallmatrix}}
  {\end{smallmatrix}\right)} 

\usetikzlibrary{positioning}
\usetikzlibrary{shadows}
\usepgflibrary{shapes}

\usepackage{mathtools}

\DeclarePairedDelimiter\floor{\lfloor}{\rfloor}

\usepackage{apptools}
\AtAppendix{\counterwithin{lemma}{section}}
\newtheorem{lemma}{Lemma}
\AtAppendix{\counterwithin{proposition}{section}}
\newtheorem{proposition}{Proposition}

\usepackage{titlesec}

\titleformat*{\section}{\large\bfseries}
\titleformat*{\subsection}{\large\bfseries}
\titleformat*{\subsubsection}{\large\bfseries}
\titleformat*{\paragraph}{\large\bfseries}
\titleformat*{\subparagraph}{\large\bfseries}

\usepackage{titlesec}
\titlelabel{\thetitle\quad}

\theoremstyle{definition}

\newtheorem{mytheorem}{Theorem}
\numberwithin{mytheorem}{section}
\usepackage{setspace}
\newcommand*{\QEDA}{\hfill\ensuremath{\square}}

\newtheorem{myremark}{Remark}
\numberwithin{myremark}{section}

\newtheorem{myassumption}{Assumption}
\numberwithin{myassumption}{section}

\newtheorem{definition}{Definition}
\numberwithin{definition}{section}

\title{Nonparametric Cointegrating Regression Functions with Endogeneity and Semi-Long Memory}

\author{Sepideh Mosaferi \footnote{Corresponding author; E-mail address: \href{mailto:smosaferi@umass.edu}{smosaferi@umass.edu} \newline Sepideh Mosaferi is a Visiting Assistant Professor with the Department of Mathematics and Statistics, University of Massachusetts Amherst. Mark S. Kaiser is a Professor with the Department of Statistics and Statistical Laboratory, Iowa State University.} \\University of Massachusetts Amherst 
\\ and \\ Mark S. Kaiser \\ Iowa State University}

\begin{document}

\maketitle
\begin{abstract}
This article develops nonparametric cointegrating regression models with endogeneity and semi-long memory. 
We assume that semi-long memory is produced in the regressor process by tempering of random shock coefficients.  The fundamental properties of long memory processes are thus retained in the regressor process.   
Nonparametric nonlinear cointegrating regressions with serially dependent errors and endogenous regressors driven by long memory innovations have been considered in \cite{WP2016}. That work also implemented a statistical specification test for testing whether the regression function follows a parametric form. The limit theory of test statistic involves the local time of fractional Brownian motion. 
The present paper modifies the test statistic to be suitable for the semi-long memory case.  With this modification, the limit theory for the test involves the local time of the standard Brownian motion and is free of the unknown parameter $d$.
Through simulation studies, we investigate the properties of nonparametric regression function estimation as well as test statistic. We also demonstrate the use of test statistic through actual data sets. 

\vspace{0.25cm}


\end{abstract}

\section{Introduction}  \label{sec:Intro}
Regression models that relate two time series have long been of interest in economics, and any number of quantities of interest in such models are nonstationary.  If two variables in a linear regression model are integrated (in the time series sense) of the same order but some linear combinations of them are stationary, then the model is said to be a cointegrated regression model. That label is also used to refer to models in which the covariate or regressor process is nonstationary and the response process is related to it through a nonlinear function (see \citealp{tjostheim2020some} for a recent review).  Nonlinear cointegrated regression models have been applied, among others, to problems in environmental Kuznets curve, wind turbine data sets, and stock prices (e.g. \citealp{fasen2013time} and references therein). 
In particular, let
\begin{equation}\label{genmodel}
y_k=f(x_k)+u_k, \qquad k=1, \ldots, N, 
\end{equation}
be a nonlinear cointegrating regression model, where $x_k$ is a nonstationary regressor, $u_k$ is an error process, and $f(\cdot)$ is an unknown real function.

Two problems that have been the focus of much research are the estimation of the unknown regression function $f(\cdot)$ nonparametrically and the testing that $f(\cdot)$ has a particular parametric form.  Many of the theoretical results related to these problems
have assumed strict exogeneity in which the regressor $x_k$ is assumed to be uncorrelated with the regression error $u_k$ (e.g. \citealp{karlsen2007nonparametric,cai2009functional, WP2009a, wang2014martingale}).  Often, the regressor $x_k$ is assumed to be a short memory process.
Extending this framework by allowing the regressor $x_k$ to be driven by long memory innovations and permitting correlation with $u_k$ (so-called endogeneity) has received less attention in the literature and leads to some technical problems that have not been completely resolved. 

Recently, \cite{WP2016} established a limit theory for nonparametric estimation of regression models that allow for both endogeneity and long memory in the regressor process. They also developed a test statistic for parametric forms of regression functions of type $f(x)=g(x,\theta_0)$ where $g(\cdot,\theta)$ represents a parametric family of functions with an unknown true parameter value $\theta_0 \in \Theta$. The authors assumed that $\Theta$ is a compact subspace of $\mathbb{R}^m$ for some finite $m$.        
The test of \cite{WP2016} is a modification of a test statistic given by \cite{hardle1993comparing} for the random sample case. The test was also used in \cite{gao2012model} for a nonlinear cointegrating model with a martingale error structure and no endogeneity. 

A property of the test under long memory and endogeneity is that the limit distribution depends on the value of the fractional differencing parameter $d$ in the nonstationary regressor $x_k$. 
The motivation for the work reported here was that if one employs the idea of tempering and assumes that there are semi-long memory input shocks to the regressor process $x_k$, the limit distribution of the test statistic given by \cite{WP2016} no longer depends on the unknown parameter $d$. 
We describe this idea in detail and list the consequences which are related to the asymptotic theory for nonparametric cointegrating regression functions.  

\subsection{Our contributions} \label{sec:contribution}

Here, we summarize the main contributions of our paper as follows. First, we demonstrate the asymptotic properties of the kernel estimator of the unknown regression function $f(x)$ in (\ref{genmodel}) through semi-long memory and endogeneity and propose its confidence intervals.
Second, we develop test statistic for the parametric forms of the regression function under the null hypothesis, where we show that the limit distribution of the test statistic under semi-long memory involves the local time of standard Brownian motion and is free from the fractional differencing parameter $d$.
Third, we investigate the properties of the regression function estimator as well as the performance of the test statistic through simulation studies. Finally, we explain how the test statistic can be used in real applications. To the best of our knowledge, these objectives have not been formally studied in the existing literature.

\subsection{Organization of the paper} \label{sec:layout}

The remainder of the article is organized as follows. The regressor process $x_k$ is described in Section \ref{sec:context}. In Section \ref{sec:InitRes}, we provide some initial results that are suitable for developing the limit theory presented in the following sections. In Section \ref{sec:Reg}, we introduce nonlinear cointegrating regression models under the assumption that the regressor process involves strongly tempered shocks.
We establish the limit distribution of the estimator of the regression function.  

In Section \ref{sec:Test}, we present a test for parametric forms of the regression function, show that its asymptotic distribution is free of the fractional differencing parameter, and consider its power under local alternatives. Section \ref{sec:Simulation} contains the results of simulation studies for the estimation of the regression function and the properties of the test statistic. Section \ref{sec:application} describes the application of the test statistic utilizing an actual data set. 
Section \ref{sec:Discussion} contains concluding remarks and some directions for future work.
The proofs of all technical results and additional simulations are contained in the Appendix. All the \texttt{R} code implementing the proposed methodology is available at Github repository \url{https://github.com/SepidehMosaferi/Nonparametric-Regression-Functions}.

Throughout the paper, we denote $C, C_1, C_2, ...$ as generic constants which may differ at each appearance. We use $\rightarrow_{P}$ for convergence in probability, $\Rightarrow$ for weak convergence of the associated probability measures, $\rightarrow_{D}$ for convergence in distribution, and $=_{D}$ for equivalence in distribution. For any two functions $f$ and $g$, $f \asymp g $ means $C_1\leq f/g \leq C_2$.
We use $||x||=\max_i|x_i|$ for the vector $x=(x_i)$, $\wedge$ for the minimum between two numbers, and a.s. for almost surely. Finally,
i.i.d. and f.d.d. mean independent and identically distributed and finite-dimensional distribution, respectively.

\section{The regressor process} \label{sec:context}

The models we consider throughout this article are of the general form (\ref{genmodel}) in which the regressor process is the sum of input shocks that result in either a long memory process or a semi-long memory process.  In the long memory process, we take for $k=1, \ldots, N$
\begin{align}\label{LMx}
x_k^\circ & = \sum_{s=1}^{k} X_d(s), \nonumber \\
X_d(s) & = \sum_{j=0}^{\infty} b_d(j) \zeta(s-j), 
\end{align}
where $\zeta(s)$ is an i.i.d. noise with $\mathbb{E}(\zeta(0))=0$ and $\mathbb{E}(\zeta^2(0))=1$. The coefficient $b_d(j)$ regularly varies at infinity as $j^{d-1}$. In this article, we use
\begin{equation*} 
b_d (j)\ \sim \ \frac{c_d}{\Gamma(d)} \, j^{d-1}, \quad j \to \infty,  \quad c_d \ne 0,  \quad d \ne 0, 
\end{equation*}
where $\Gamma(d)$ is the gamma function defined as $\int_{0}^{\infty} e^{-x} x^{d-1} dx$, and $ 0<d<1/2$ is the fractional differencing parameter. 

In contrast to the long memory process, a semi-long memory process contains strongly tempered shocks as
\begin{align}\label{SLMx}
x_k & = \sum_{s=1}^{k} X_{d,\lambda}(s), \nonumber \\
X_{d,\lambda}(s) & = \sum_{j=0}^{\infty} e^{-\lambda j} b_d(j)\zeta(s-j), 
\end{align}
where $\zeta(s)$ and $b_d(j)$ are the same as for the long memory case.  In (\ref{SLMx}), 
$\lambda \equiv \lambda_N >0$ is called the tempering parameter and is sample size dependent, which satisfies the following main assumption:

\begin{itemize}[leftmargin=*]
\item {\bf Semi-Long Memory Assumption}: The tempering parameter $\lambda\to 0$ and $N\lambda\to\infty$ as $N\to\infty$.
\end{itemize}

\noindent The tempering parameter $\lambda$ in expression (\ref{SLMx}) allows the range of the fractional differencing parameter $d$ to be extended from $(0,1/2)$ to $(0,\infty)$. 

The strongly tempered process $\eqref{SLMx}$ belongs to the general class of stochastic processes called tempered linear processes; see \citet{sabzikar2018invariance}. These processes have a semi-long memory property in the sense that their autocovariance functions initially resemble that of a long memory process but eventually decay fast at an exponential rate. One special case of such processes is the autoregressive tempered fractionally integrated moving average ARTFIMA$(p,d,\lambda,q)$ process with $p$ as the order of the  autoregressive polynomial and $q$ as the order of the moving average polynomial, which has been studied by \cite{meerschaert2014tempered} and \cite{sabzikar2019parameter}.

The class of ARTFIMA$(0,d,\lambda,0)$ that does not have autoregressive and moving average components is highly applicable. Some empirical examples include modeling logarithmic returns for AMZN stock prices, modeling geophysical turbulence in water velocity data, and modeling climate data sets given in \cite{sabzikar2019parameter}. In all cases, the authors have found that by using an ARTFIMA model, we are able to capture aspects of the low-frequency activity of time series better than an ARFIMA model without considering $\lambda$ in the model.

\section{Initial results} \label{sec:InitRes}

The local time process of a stochastic process $G(x)$ is defined as
\begin{equation*}
L_G(t,s)=\lim\limits_{\epsilon \rightarrow 0} \frac{1}{2 \epsilon} \int_{0}^{t}I\{|G(r)-s| \leq \epsilon \} \, dr, 
\end{equation*}
where $(t,s) \in \mathbb{R}_{+} \times \mathbb{R}$. Let $d_N:= [\mathbb{E}(x_N^2)]^{1/2}$ and consider $x_{k,N}:=x_k/d_N$ for $1 \leq k \leq N$ to be a triangular array. A function of $x_{k,N}$ that will be used in the sequel takes the form of a sample average of functions of $x_{k,N}$ as
\begin{equation*}
S_N := \frac{c_N}{N} \sum_{k=1}^{N} g(c_N x_{k,N}),
\end{equation*}
where $g(x)$ is a bounded function such that $\int_{\mathbb{R}} |g(x)| dx< \infty$ and $c_N:= d_N/h$ where $c_N \rightarrow \infty$ and $c_N/N \rightarrow 0$. 
The bandwidth parameter is denoted by $h$ and satisfies $h \equiv h_N \rightarrow 0$ as $N \rightarrow \infty$. 
Functions in the form of $S_N$ commonly arise in nonlinear cointegrating regressions, which could be the kernel function $K(.)$ or its squared $K^2(.)$; see \cite{karlsen2001nonparametric}, \cite{karlsen2007nonparametric}, and \cite{WP2009a}.

The limit behavior of $S_N$ will be important in establishing the limit behavior of the kernel estimator of $f(x)$ in the context of nonparametric regression. The limit distribution of $S_N$ is as follows
\begin{equation} \label{local}
S_N \rightarrow_{D} \int_{\mathbb{R}} g(x) dx \, L_{B}(1,0),
\end{equation}
where $L_{B}(t,s)$ is the local time of standard Brownian motion $B(t)$ at the spatial point $s$; see Proposition \ref{Prop9} part (i) to follow. When the function $g(.)$ is a kernel density, the integral in expression (\ref{local}) is unity, and the limit is then the local time of $B(.)$ at the origin; see \cite{jeganathan2004convergence} and \cite{WP2009a} for some of the related results.   
In the case of long memory, the local time given in expression (\ref{local}) is in the form of $L_{B_{d+1/2}}(t,s)$ and therefore depends on the unknown parameter $d$ (\citealp{WP2009a}). This complicates the limit theory of the kernel estimator of regression function as well as the associated test statistic for long memory processes.
 
\section{Nonparametric regression function estimation} \label{sec:Reg}

\noindent Assume the model (\ref{genmodel}) where $f(\cdot)$ is an unknown real regression function. To induce endogeneity, let $\eta_k=(\zeta(k),\epsilon(k))'$ be a sequence of random vectors with $\mathbb{E}(\eta_0)=0$ and $\mathbb{E}(\eta_0 \eta_0')=\Sigma$ where

\begin{equation} \label{corr-matrix}
\Sigma = \begin{pmatrix}
1 & \mathbb{E}(\zeta(0) \epsilon(0))\\
\mathbb{E}(\epsilon(0) \zeta(0)) & \mathbb{E}(\epsilon(0)^2)
\end{pmatrix},
\end{equation}
and take $u_k$ in (\ref{genmodel}) to be $u_k=\sum_{j=0}^{\infty} \psi_j \eta_{k-j}$ for  the coefficient vector $\psi_j=(\psi_{j1},\psi_{j2})$. Therefore, $\mathbb{E}(u_0^2)=\sum_{j=0}^{\infty} \psi_j \Sigma \psi_j'$ and $cov(u_k,x_k) \neq 0$ by recalling expression (\ref{SLMx}).
Now, assume $\eta_0$ and $\psi_j$ satisfy the following assumption.

\begin{myassumption} \label{ass:coef}
Let $\sum_{j=0}^{\infty}\psi_j \neq 0$ and $\sum_{j=0}^{\infty}j^{1/4}(|\psi_{j1}|+|\psi_{j2}|)< \infty$. Also, let $\mathbb{E}||\eta_0||^{\alpha} < \infty$ for $\alpha>2$.
\end{myassumption}

\noindent Assumption \ref{ass:coef} allows the error term $u_k$ to be cross-correlated with the regressor $x_k$. Furthermore, assume that the characteristic function $\varphi(t)$ of $\zeta(0)$ satisfies $\int_\mathbb{R} (1+|t|) |\varphi(t)| dt < \infty,$ that ensures smoothness in the corresponding kernel density (see \citealp{WP2016}).

The Nadaraya-Watson regression estimator of $f(x)$ is
\begin{equation} \label{fhat}
\hat{f}(x)=\frac{\sum_{k=1}^{N}y_k K_h(x_k-x)}{\sum_{k=1}^{N}K_h(x_k-x)}, 
\end{equation}
where $K_h(s)=h^{-1}K(s/h)$, and $K(\cdot)$ is a non-negative bounded continuous function. 
In order to establish the asymptotic behavior for the kernel estimate $f(x)$, $\{ x_{k,N} \}_{k \geq 1, N \geq 1}$ needs to be a strong smooth array (see Definition \ref{def.smootharray} as well as Proposition \ref{Prop1} in the Appendix). 
Now, we impose the following two assumptions on the kernel $K(.)$:

\begin{myassumption} \label{ass:kernel}
$K(x)$ is a non-negative bounded continuous function that satisfies
$\int_{\mathbb{R}} K(x) dx=1$ and $\int_{\mathbb{R}} |\hat{K}(x)|dx < \infty$, where $\hat{K}(x)=\int_{\mathbb{R}} e^{ixt} K(t) dt$.
\end{myassumption}

\begin{myassumption} \label{ass:function}
For a given $x$, there exists a real positive function $f_1(s,x)$ and $\gamma \in (0,1]$ such that when $h$ is sufficiently small,
$|f(hy+x)-f(x)| \leq h^\gamma f_1(y,x) \, \forall y \in \mathbb{R}$
and $\int_{\mathbb{R}}K(s) [f_1(s,x)+f_1^2(s,x)]ds < \infty$ hold.
\end{myassumption}

Assumption \ref{ass:kernel} is needed for some of the technical proofs and is satisfied for many commonly used kernels that the Fourier transformation of $K(x)$ is integrable. The assumption \ref{ass:function} can be used in developing analytic forms for the asymptotic bias function in kernel estimation and can be verified for various kernels $K(x)$ and regression functions $f(x)$. 
The following theorem is the main result on the Nadaraya-Watson kernel estimator of (\ref{fhat}) and recall that $\lambda$ is a function of $N$ such that $\lambda \rightarrow 0$ and $N \lambda \rightarrow \infty$ as $N \rightarrow \infty$.

\begin{mytheorem} \label{Theo.f}
Under Assumptions \ref{ass:coef}--\ref{ass:function} and for any $h$ satisfying $\sqrt{N}\lambda^d h \rightarrow \infty$ and $ \sqrt{N} \lambda^d h^{1+2\gamma} \rightarrow 0$ where $\gamma \in (0,1]$, we have
\begin{equation} \label{maintheo}
\Big\{\sqrt{N}\lambda^d h\Big\}^{1/2} \Big(\hat{f}(x)-f(x)\Big) \rightarrow_D d_0 \, N(0,1) \, L_{B}^{-1/2}(1,0), 
\end{equation}
where $d_0^2=\mathbb{E}(u^2_{0}) \int_{\mathbb{R}} K^2(s) ds$ and $N(0,1)$ denotes a standard normal which is independent from the local time of Brownian motion $L_{B}(1,0)$.
If we normalize the limit form (\ref{maintheo}), we have:
\begin{equation} \label{normalized}
\Big\{h \sum_{k=1}^{N} K_h(x_k-x) \Big\}^{1/2} \Big(\hat{f}(x)-f(x)\Big) \rightarrow_D N(0,\sigma^2), 
\end{equation}
where $\sigma^2= d_0^2$; see also expression (\ref{th2.1proof}) given in the Appendix.
\end{mytheorem}

For a fixed value of $x$, $f(x)$ has a continuous $(p+1)$ derivative in a small neighborhood of $x$. For any $h$ satisfying $\sqrt{N} \lambda^d h \rightarrow \infty$ as $h \rightarrow 0$ and $\sqrt{N} \lambda^d h^{2(p+1)+1} \rightarrow 0$, we have $\hat{f}(x) \rightarrow_{P} f(x)$. Therefore, $\hat{f}(x)$ is a consistent estimator of $f(x)$. The consistency of $\hat{f}(x)$ could be followed from Theorem 3.1 of \cite{WP2009b}.

\begin{myremark}
The local time given in Theorem \ref{Theo.f} is the local time of standard Brownian motion, which is independent of the unknown parameter $d$. This is a direct consequence of tempering the time series in the regressor $x_{k}$. In this regard, the result of Theorem \ref{Theo.f} is different from Theorem 2.1 in \cite{WP2016}, which involves the local time process $L_{B_{d+1/2}}(1,0)$ and is related to a fractional Brownian motion with parameter $d+1/2$.
\end{myremark}

\begin{myremark}
Based on Theorem \ref{Theo.f}, the bandwidth $h$ has to satisfy certain rate conditions to ensure that the asymptotic distribution on the right side of expression (\ref{normalized}) holds. Under semi-long memory and for $d > 0$, we require $\sqrt{N} \lambda^d h \rightarrow \infty$ and $\sqrt{N} \lambda^d h^{1+2 \gamma} \rightarrow 0$. 
\end{myremark}

The term $\mathbb{E}(u^2_0)$ in $d_0^2$ can be estimated by 
\begin{equation} \label{sigma2}
\hat{\sigma}^2_N=\frac{\sum_{k=1}^{N}[y_k-\hat{f}(x_k)]^2 K_h(x_k-x)}{\sum_{k=1}^{N}K_h(x_k-x)}.
\end{equation}
Expression (\ref{sigma2}) is appropriate for use in the interval estimation and other inferential procedures.
When $\mathbb{E}(u_0^8)< \infty$ and $\int_{\mathbb{R}} K(s) f_1^2(s,x) ds < \infty$ for a given $x$, for any $h$ satisfying $\sqrt{N} \lambda^d h \rightarrow \infty$ and $h \rightarrow 0$, we have $\hat{\sigma}^2_N \rightarrow_{P} \mathbb{E}(u_0^2)$. The consistency of $\hat{\sigma}^2_N$ is stated in Theorem 3.2 in \cite{WP2009b}, and we refer an interested reader to their manuscript.
 
\section{Specification test for the regression function} \label{sec:Test}
\noindent When there is no reason to believe that $f(x)$ in (\ref{genmodel}) follows a particular parametric form, the use of a nonparametric estimator is attractive, but nonparametric estimators generally have a slow convergence rate compared to parametric estimators. It is often possible to determine a plausible parametric regression function and then conduct a test of a hypothesis formulated as,
\begin{equation} \label{null}
H_0: f(x)=g(x,\theta_0), 
\end{equation}
where $\theta_0$ in (\ref{null}) is a vector of unknown parameters that belongs to a compact and convex space $\Theta$.

The tests of (\ref{null}) have been previously considered by \cite{hardle1993comparing}, \cite{horowitz2001adaptive}, \cite{gao2009specification}, \cite{wang2012specification}, and \cite{WP2016} under different assumptions on the data generating mechanism.    
In fact, \cite{gao2009specification} and \cite{wang2012specification} considered a kernel-smoothed U statistic of the form $\sum_{j,k=1, j \neq k}^{N} \hat{u}_k \hat{u}_j K[(x_k-x_j)/h]$ with $\hat{u}_k=y_k-g(x_k,\hat{\theta}_N)$, where $\hat{\theta}_N$ is an estimator of $\theta$. 
The estimator $\hat{\theta}_N$, can be based on a non-linear least squares method by setting $Q_N(\theta)=\sum_{k=1}^{N}(y_k-g(x_k,\theta))^2$ and minimizing it over $\theta \in \Theta$ to obtain
$\hat{\theta}_N = \text{argmin}_{\theta \in \Theta} Q_N(\theta).$

The asymptotic for the U statistic is difficult to extend to the case of endogenous regressors. 
\cite{WP2016} modified a test statistic previously suggested by  \cite{hardle1993comparing} for the consideration of endogenous regressors with long memory.  In that case, both the limit theory and the convergence rate of the test statistic depend on the fractional differencing parameter $d$ through the local time of fractional Brownian motion $L_{B_{d+1/2}}(t,s)$, which complicates its use in actual problems. In this section, we consider the statistic under the assumption of semi-long memory input shocks to the regressors $x_k$.  Both the limit distribution of  \cite{WP2016} under long memory, and our limit distribution for use with semi-long memory regressors are based on the following statistic,
\begin{equation} \label{test}
T_{N} := \int_{\mathbb{R}} \Big\{\sum_{k=1}^{N} K\Big[\frac{x_k-x}{h} \Big] [y_k-g(x_k,\hat{\theta}_N)] \Big\}^2 \pi(x) dx.
\end{equation}

The term $\pi(x)$ in (\ref{test}) is a positive integrable weight function with a compact support. To develop the asymptotic theory of $T_{N}$ in the context of semi-long memory regressors and endogeneity, we provide three assumptions as follows.

\begin{myassumption} \label{ass:k_compact}
$K(x) \pi(x)$ has a compact support such that $\int_{\mathbb{R}} K(x)dx=1$ and $|K(x)-K(y)| \leq C |x-y|$ whenever $|x-y|$ is sufficiently small.
\end{myassumption}

\begin{myassumption} \label{ass:gs}
There exist $g_1(x)$ and $g_2(x)$ such that for each $\theta, \theta_0 \in \Theta$, $|g(x,\theta)-g(x,\theta_0)| \leq C ||\theta-\theta_0|| g_1(x)$ holds, and for some $0< \beta \leq 1$, $|g_1(x+y)-g_1(x)| \leq C |y|^{\beta} g_2(x)$ holds, whenever $y$ is sufficiently small. Furthermore, $\int_{\mathbb{R}} [1+g_1^2(x)+g^2_2(x)] \pi(x) dx < \infty$ holds.
\end{myassumption}

\begin{myassumption} \label{ass:H0}
Under $H_0$, if $\theta_0$ is the true value of $\theta$,
$||\hat{\theta}_N-\theta_0||=o_P\Big(\{\sqrt{N} \lambda^d h \}^{-1/2}\Big)$.
\end{myassumption}

As noted in \cite{WP2016}, Assumption \ref{ass:gs} covers a wide range of functions $g(x,\theta)$ and $\pi(x)$. Typical examples of $g(x,\theta)$ include $\theta e^x/(1+e^x)$, $e^{- \theta |x|}$, $e^{- \theta x^2}$, $e^{\theta |x|}/(1+e^{\theta |x|})$, $\theta \log|x|$, $(x+\theta)^2$, etc.
Note that Assumptions \ref{ass:k_compact}--\ref{ass:H0} can be compared with those imposed by \cite{WP2016}. In particular, the convergence rate imposed on $\hat{\theta}_N$ by Assumption \ref{ass:H0} has been verified in Section 4 of \cite{WP2016} and will be required in our proof of the theorems to follow.
We now consider the asymptotic behavior of $T_{N}$ in (\ref{test}) and its convergence rate in the semi-long memory setting.

\begin{mytheorem} \label{Theo.test}
Let Assumptions \ref{ass:coef} and \ref{ass:k_compact}--\ref{ass:H0} hold. Then, under $H_0$, we have
\begin{equation} \label{normalizedtest}
T_{\lambda,d}:= \frac{1}{\sqrt{N} \lambda^d h} T_{N} \rightarrow_D d_{(0)}^2 L_{B}(1,0),
\end{equation}
where $d_{(0)}^2=\mathbb{E}(u^2_0) \int_{\mathbb{R}}K^2(s)ds \int_{\mathbb{R}} \pi(x) dx$,  $N^{1/2-\delta_0} \lambda^d h \rightarrow \infty$, and $\delta_0$ is as small as required.
\end{mytheorem}

\begin{myremark} \label{localtimes}
(i) Under semi-long memory settings for $d \in \mathbb{R}_{+}$, the limit distribution of (\ref{normalizedtest})  is free of the unknown fractional differencing parameter $d$; however, the convergence rate depends on the unknown parameter $d$ even when $\lambda \rightarrow 0$ and $N \lambda \rightarrow \infty$.
(ii) Under the long memory setting for $d \in (0,1/2)$, the test statistic and limit distribution given by \cite{WP2016} are
$T_{N,d}:= (d_N/Nh) T_N \rightarrow_{D} d^2_{(0)} L_{B_{d+1/2}}(1,0)$,
where $d_N \sim N^{d+1/2}$ is $N \rightarrow \infty$, and $T_N$ is as given in (\ref{test}).  The limit distribution of this version of the test statistic relies on $d$ through the fractional Brownian motion process $B_{d+1/2}$.
(iii) Under the short memory setting for $d=0$ and $d_N \sim N^{1/2}$, the test statistic and limit distribution are 
$T_{N,0}=(1/\sqrt{N}h) T_{N} \rightarrow_{D} d^2_{(0)} L_{B}(1,0)$.
(iv) Note that the limit distribution of test statistic (\ref{normalizedtest}) formulated under the semi-long memory case is similar to the short memory case.  In neither case does the limit distribution depend on $d$. 
\end{myremark}

To verify that the test has nontrivial power, one can assess the null hypothesis (\ref{null}) against a local alternative
\begin{equation} \label{alter}
H_A: f(x)=g(x,\theta_0)+ \rho_N m(x), \quad \text{for} \,\, x \in \mathbb{R}.
\end{equation} 
Here, $\rho_N$ is a sequence of numerical constants that measures the local deviation from the null hypothesis.  Also in (\ref{alter}), $m(x)$ is a real function free of $\theta$ without lying in the span of $g(x,\theta)$ and its derivative functions. 
To make $m(x)$ smooth enough under $H_A$ for the sake of asymptotic power development, we give the following two assumptions.

\begin{myassumption} \label{ass:m}
(i) There exist $m_1(x)$ and $\gamma \in (0,1]$ such that for any $y$ sufficiently small, $|m(x+y)-m(x)| \leq C |y|^{\gamma} m_1(x)$ holds. (ii) Let $\int_{\mathbb{R}} [1+m^2(x)+m^2_1(x)] \pi(x) dx < \infty$ and $\int_{\mathbb{R}} m^2(x) \pi(x) dx > 0$ hold. (iii) The function $m(x)$ is not an element of the space spanned by $g(x,\theta)$ and its derivative functions.
\end{myassumption}

\begin{myassumption} \label{ass:HA}
Under $H_A$, if $\theta_0$ is the true value of $\theta$, $||\hat{\theta}_N-\theta_0||=o_P\Big(\{\sqrt{N} \lambda^d h \}^{-1/2}\Big)$.
\end{myassumption}

The conditions on $m(x)$ in Assumption \ref{ass:m} are week and can be satisfied by a large class of real functions. They are required to ensure the divergence of the test statistic and its consistency under $H_A$.

\begin{mytheorem} \label{Theo.normalizedtest}
Let Assumptions \ref{ass:coef}, \ref{ass:k_compact}--\ref{ass:gs}, and \ref{ass:m}--\ref{ass:HA} hold. Then, under $H_A$, we have
\begin{equation*} 
\lim\limits_{N \rightarrow \infty} P \Big(T_{\lambda,d} \geq T_0 \Big)=1, 
\end{equation*}
where $T_0$ is positive, and $h \rightarrow 0$ satisfies $N^{1/2-\delta_0} \lambda^d h \rightarrow \infty$ for a small enough $\delta_0$ and any $\rho_N$ satisfying $N^{1/2} \lambda^d h \rho_N^2  \rightarrow \infty$.
\end{mytheorem}


\begin{myremark} \label{remark.power}
Based on Theorem \ref{Theo.normalizedtest}, the test statistic $T_{\lambda,d}$ has nontrivial power against the local alternatives in the form of (\ref{alter}) whenever $\rho_N \rightarrow 0$ at a rate that is slower than $\{\sqrt{N} \lambda^d h\}^{-1/2}$ as $\{\sqrt{N} \lambda^d h\}^{-1} \rightarrow 0$. The proof of Theorem \ref{Theo.normalizedtest} ensures the divergence of normalized test statistic $T_{\lambda,d}$ and test consistency under $H_A$; see Remarks 3.2 and 3.3 in \cite{WP2016} for a related discussion. 
\end{myremark}

\section{Simulation studies} \label{sec:Simulation}

In this section, we illustrate the properties of the Nadaraya-Watson regression estimator and its confidence intervals. In addition, we study the size and power of the specification test statistic using the same regression functions.

\subsection{Regression function properties} \label{sec:regression_sim}

\noindent To examine the behavior of regression function estimators, we generate data sets from a model with the form of (\ref{genmodel}), where we incorporate $\sigma$ as follows:

\begin{equation} \label{modelUS}
y_k=f(x_k)+\sigma u_k. 
\end{equation}
The regressor process $x_k$ is defined in (\ref{SLMx}) for the semi-long memory (SLM) setting and as $x_k=x_k^\circ$ where $x_k^\circ$ is given in (\ref{LMx}) for the long memory (LM) setting.
 Let $u_k=\psi u_{k-1}+\epsilon(k)$ with $\psi=0.25$ and $\mathbb{E}(\zeta(k) \epsilon(k)):= \rho_{\zeta,\epsilon}$ (c.f., (\ref{corr-matrix})) such that $(\zeta(k), \epsilon(k))$ are i.i.d. $N\Big(0,\begin{psmallmatrix}1 & \rho_{\zeta,\epsilon} \\ \rho_{\zeta,\epsilon} & 1 \end{psmallmatrix}\Big)$. 
Also, let $\rho_{\zeta,\epsilon}=0.5$ and $\sigma=0.2$ in (\ref{modelUS}).

We consider the following two regression functions:
\begin{itemize} 
\item straight line regression function: $f(x)= \theta_0 + \theta_1 x$, where $(\theta_0,\theta_1)=(0,1)$, and
\item quadratic regression function: $f(x)= \theta_0 + \theta_1 x + \theta_2 x^2$, where $(\theta_0,\theta_1,\theta_2)=(0,1,1)$.
\end{itemize}
We set the sample size at $N=\{ 50, 100, 500 \}$, and the number of Monte Carlo replications at $R=2000$.

We use the estimator (\ref{fhat}) with an Epanechnikov kernel in the form of $K(u)=0.75(1-u^2)1_{\{|u| \leq 1\}}$. For simplicity, we let $h = \{N^{-1/3}, N^{-1/7}\}$ and
$\lambda = N^{-1/5}$ under the assumption of $N \lambda \rightarrow \infty$. We assume $d=\{ 0.1, 0.2, 0.3, 0.4\}$. 
We study Monte Carlo approximations of  absolute bias ($\mathbb{E}\{|\hat{f}(x)-f(x)|\}$), standard deviation (std) ($[\mathbb{E}\{[\hat{f}(x)-E(\hat{f}(x))]^2\}]^{0.5}$) and root of mean squared error (rmse) ($[\mathbb{E}\{[\hat{f}(x)-f(x)]^2\}]^{0.5}$) of $\hat{f}(x)$ for the true regression functions over the interval $[-1,1]$ at $100$ points equally spaced between $-1$ and $1$.

\begin{figure}[ht]
\centering
\begin{tabular}{ cc }
 \includegraphics[width=0.45\textwidth]{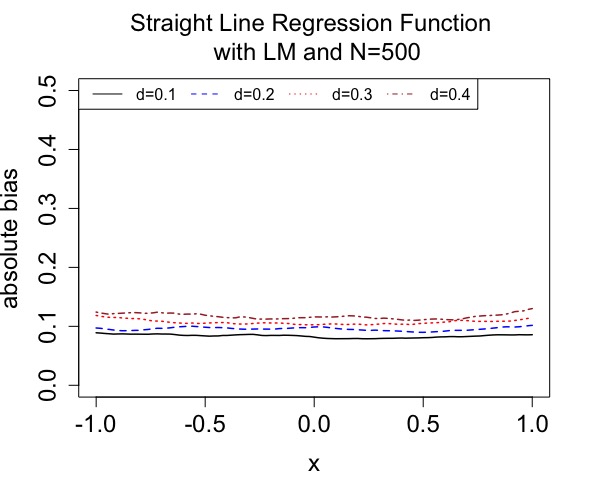} &
 \includegraphics[width=0.45\textwidth]{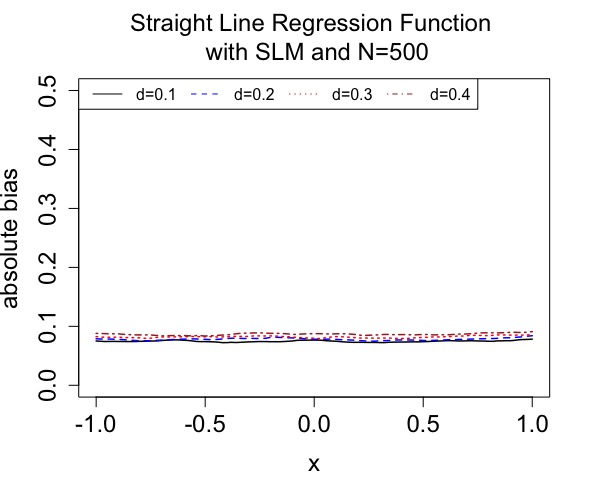}\\
  \includegraphics[width=0.45\textwidth]{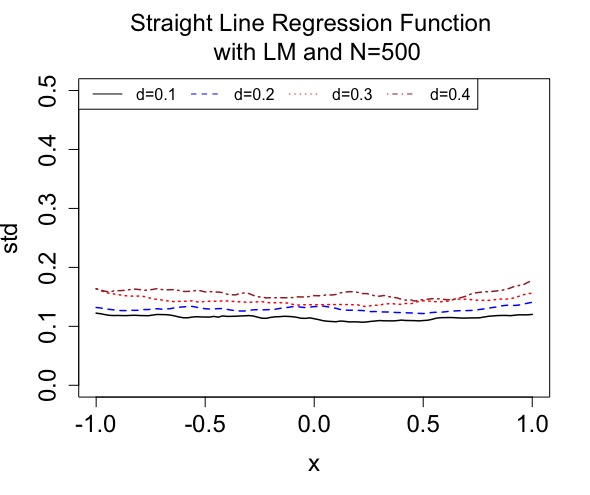} &
 \includegraphics[width=0.45\textwidth]{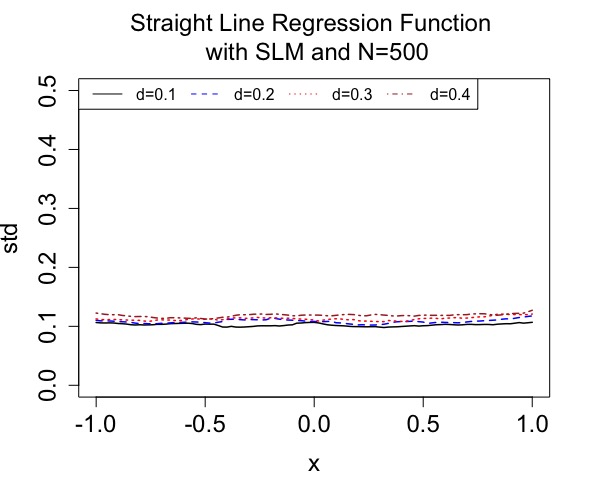} \\
  \includegraphics[width=0.45\textwidth]{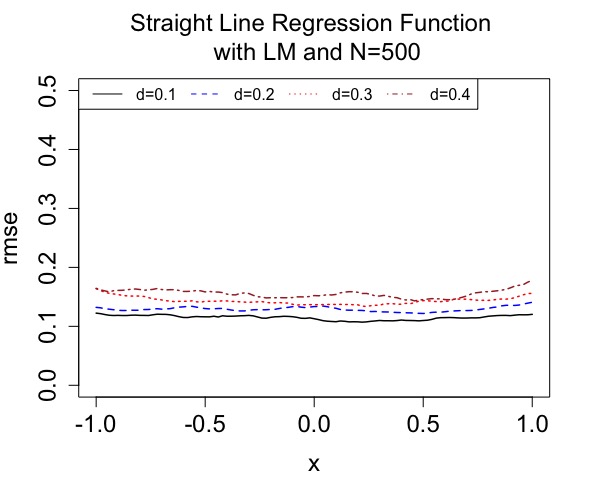} &
 \includegraphics[width=0.45\textwidth]{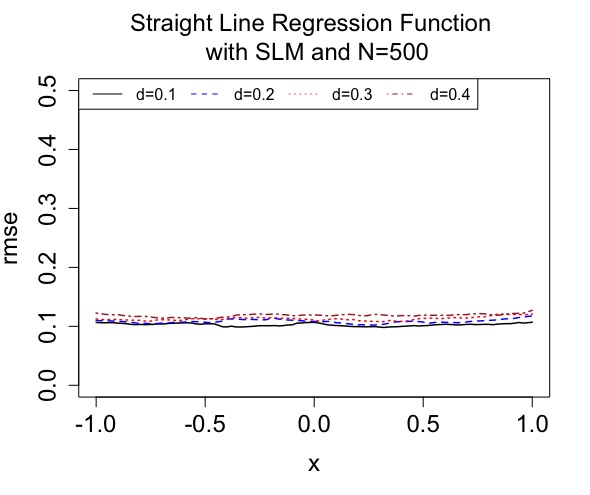}
 \end{tabular}
\caption{Comparison of empirical absolute bias, std, and rmse for the straight line regression function with $h=N^{-1/7}$ and $\lambda=N^{-1/5}$ under SLM.}
\label{Fig.line_N500_1}
\end{figure}

\begin{figure}[ht]
\centering
\begin{tabular}{ cc }
 \includegraphics[width=0.45\textwidth]{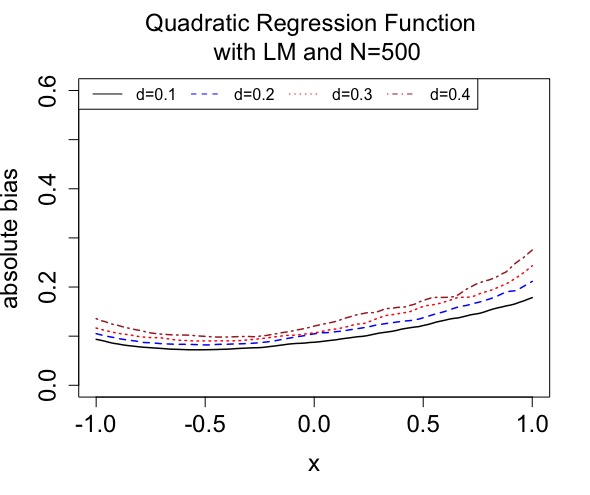} &
 \includegraphics[width=0.45\textwidth]{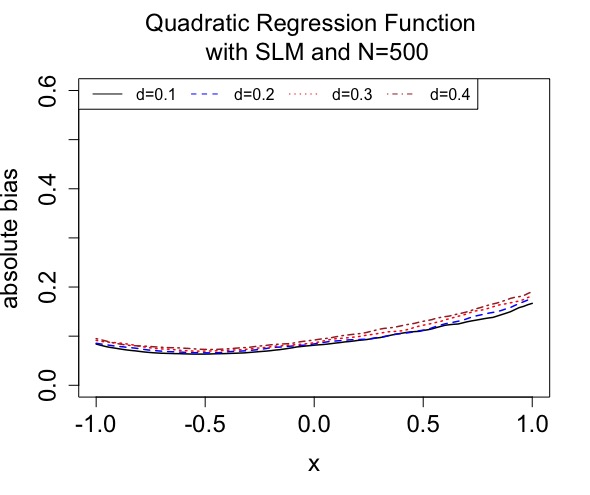}\\
  \includegraphics[width=0.45\textwidth]{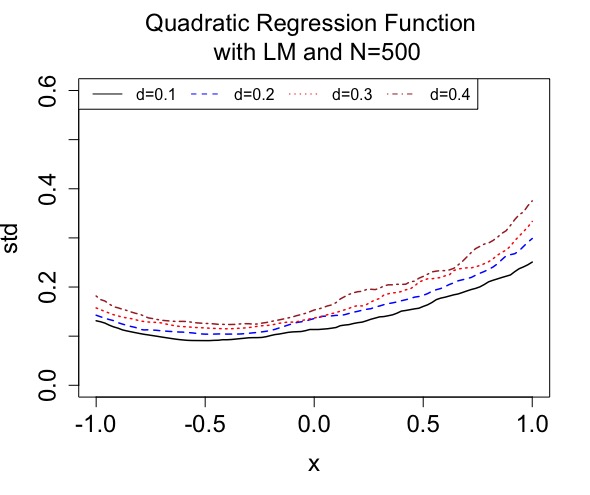} &
 \includegraphics[width=0.45\textwidth]{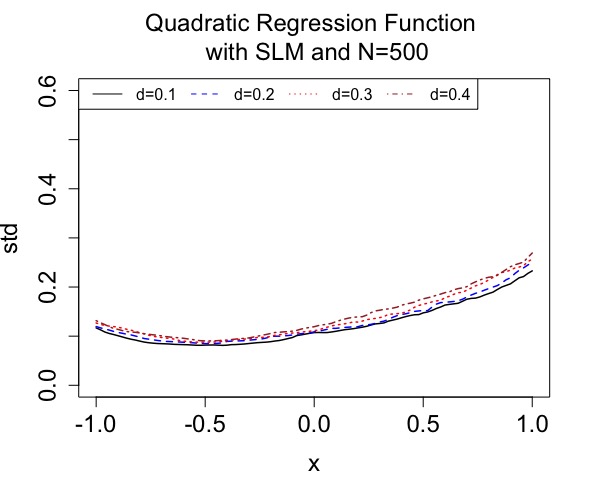} \\
  \includegraphics[width=0.45\textwidth]{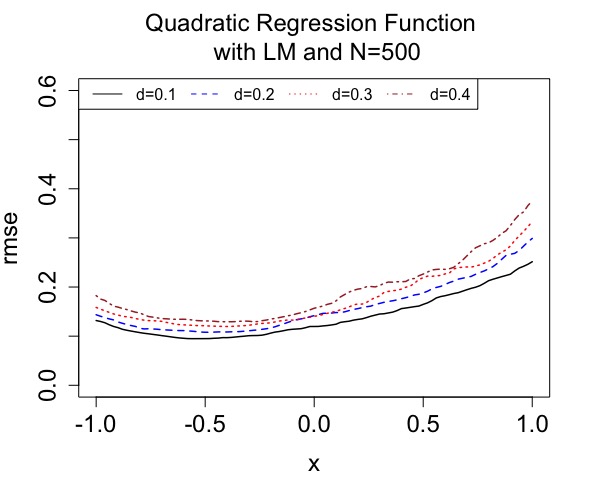} &
 \includegraphics[width=0.45\textwidth]{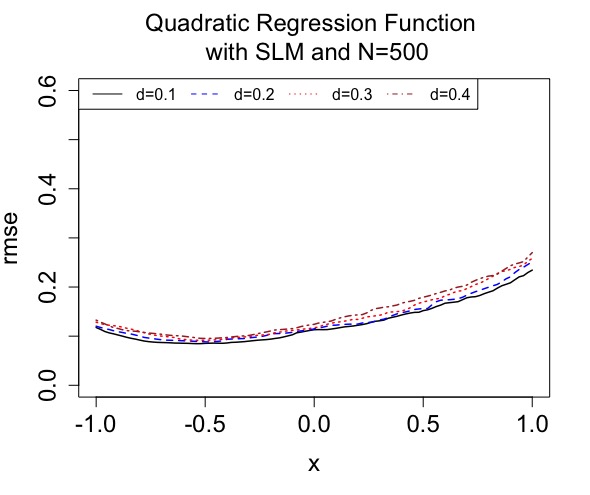}
 \end{tabular}
\caption{Comparison of empirical absolute bias, std, and rmse for the quadratic regression function with $h=N^{-1/7}$ and $\lambda=N^{-1/5}$ under SLM.}
\label{Fig.quad_N500_1}
\end{figure}

To construct point-wise confidence intervals for the regression function $f(x)$, we use the limit distribution given in (\ref{normalized}). An asymptotic $100(1-\alpha)\%$ level confidence interval for $f(x)$ is then given by
\begin{equation*}
\hat{f}(x) \pm z_{\alpha/2} \Big\{ \hat{\sigma}^2_N \int_{\mathbb{R}} K^2(s)ds \Big/ h \sum_{k=1}^{N} K_h(x_k-x)  \Big\}^{1/2}, 
\end{equation*}
where $\hat{\sigma}^2_N$ is given in expression (\ref{sigma2}). 
We present the values of absolute bias, std, and rmse graphically for all the values of $x$ and for the two regression functions with $N=500$ and $h=N^{-1/7}$ in Figures \ref{Fig.line_N500_1} and \ref{Fig.quad_N500_1}. For the SLM case, these criteria are stable across values of $d$, while they vary as $d$ changes in the LM case and this is true for both straight-line and quadratic regression functions.

\begin{figure}[ht]
\centering
\begin{tabular}{ cc }
 \includegraphics[width=0.45\textwidth]{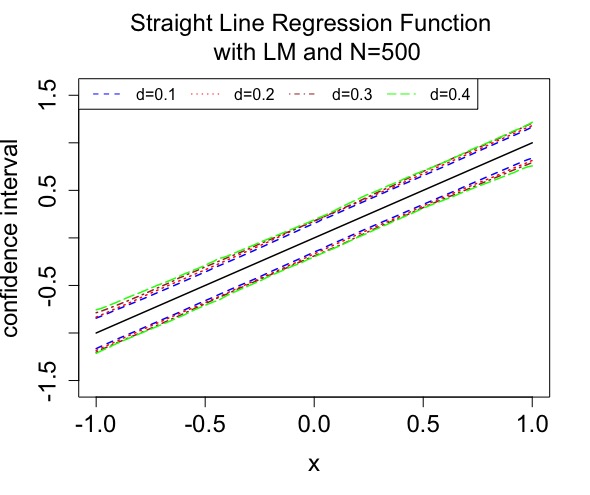} &
 \includegraphics[width=0.45\textwidth]{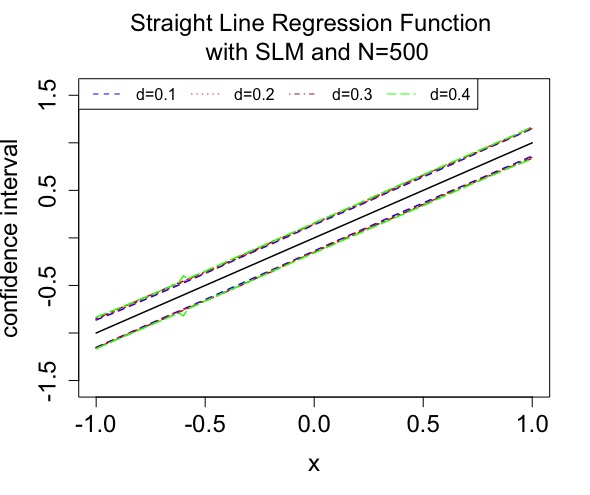}\\
  \includegraphics[width=0.45\textwidth]{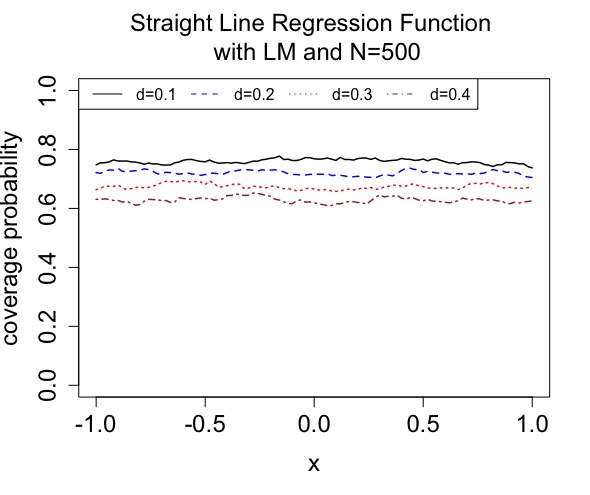} &
 \includegraphics[width=0.45\textwidth]{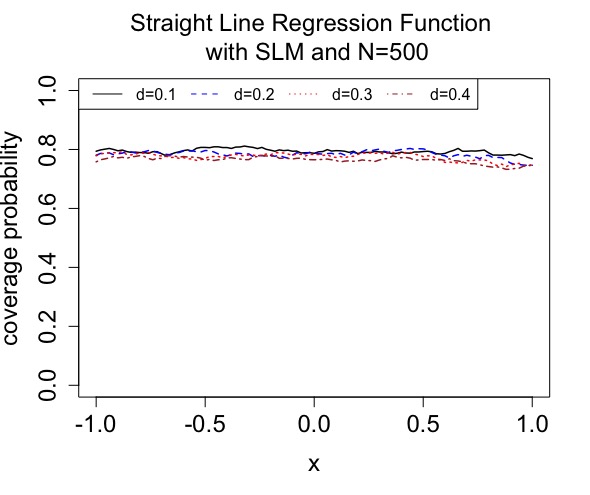} \\
  \includegraphics[width=0.45\textwidth]{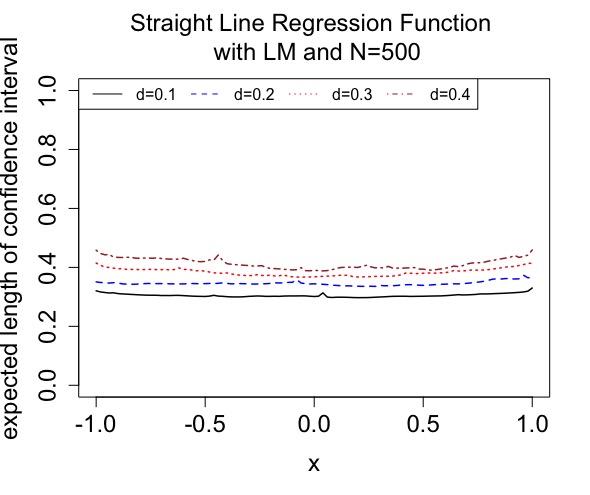} &
 \includegraphics[width=0.45\textwidth]{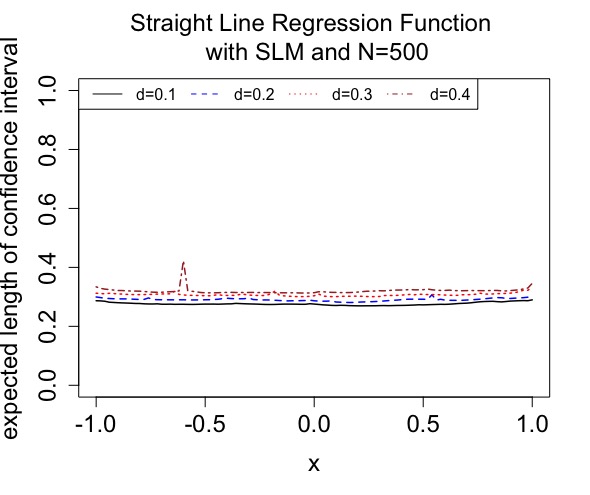}
 \end{tabular}
\caption{Comparison of empirical confidence intervals, coverage probabilities and expected lengths of confidence intervals for the straight line regression function with $h=N^{-1/7}$ and $\lambda=N^{-1/5}$ under SLM.}
\label{Fig.line_N500_2}
\end{figure}

\begin{figure}[ht]
\centering
\begin{tabular}{ cc }
 \includegraphics[width=0.45\textwidth]{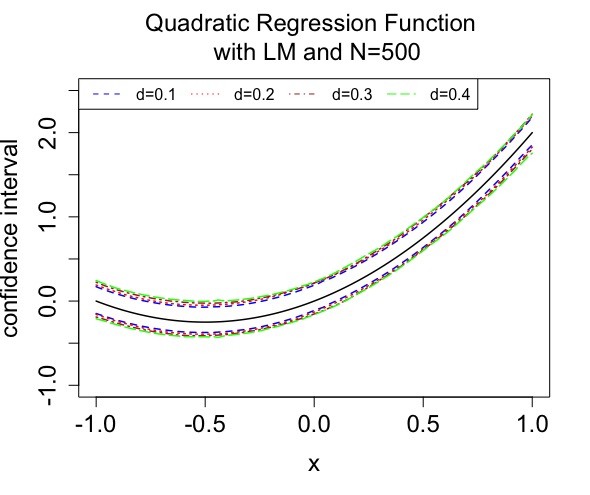} &
 \includegraphics[width=0.45\textwidth]{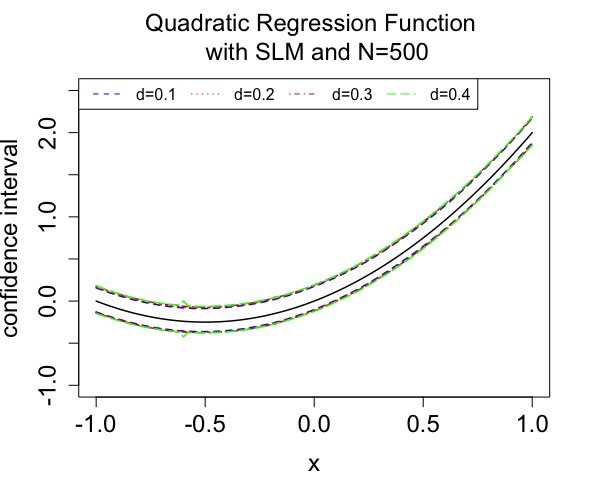}\\
  \includegraphics[width=0.45\textwidth]{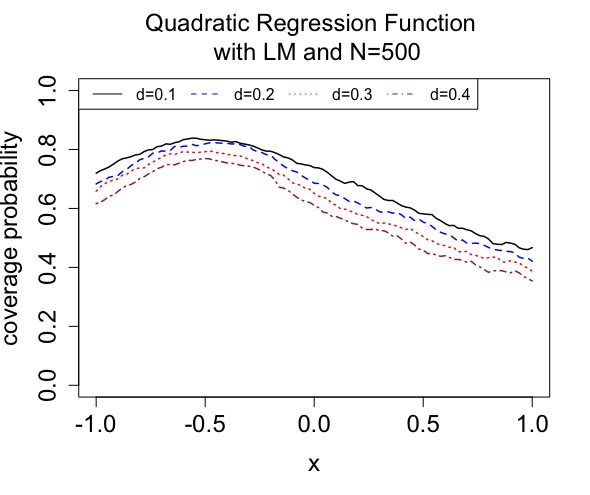} &
 \includegraphics[width=0.45\textwidth]{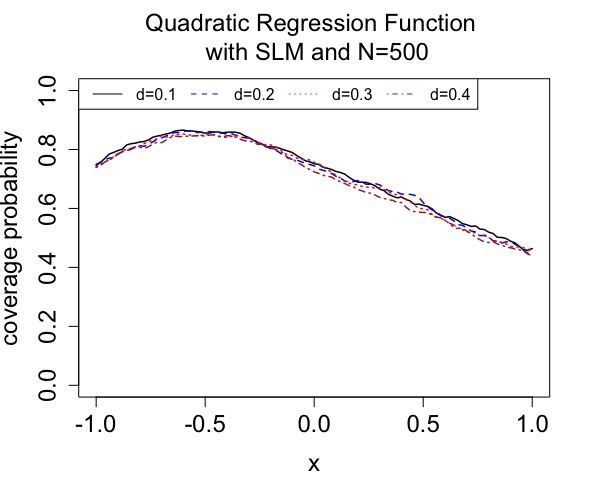} \\
  \includegraphics[width=0.45\textwidth]{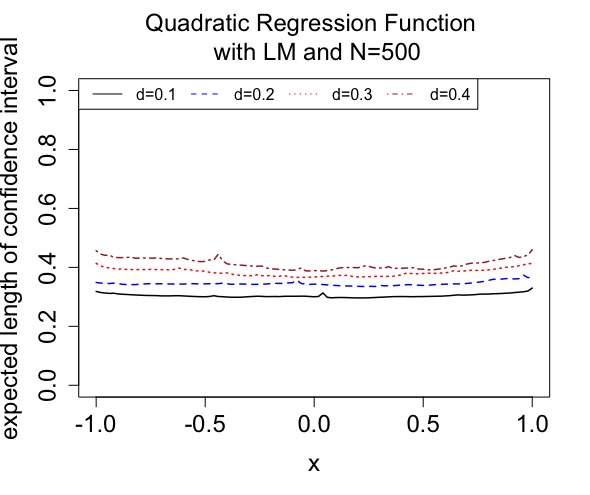} &
 \includegraphics[width=0.45\textwidth]{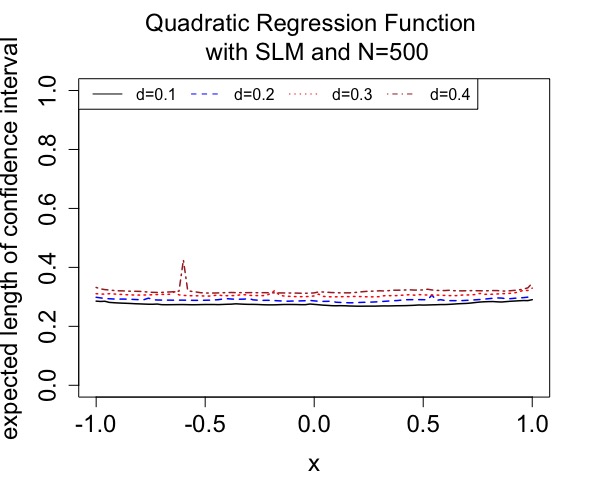}
 \end{tabular}
\caption{Comparison of empirical confidence intervals, coverage probabilities and expected lengths of confidence intervals for the quadratic regression function with $h=N^{-1/7}$ and $\lambda=N^{-1/5}$ under SLM.}
\label{Fig.quad_N500_2}
\end{figure}

In Figures \ref{Fig.line_N500_2} and \ref{Fig.quad_N500_2}, we illustrate Monte Carlo coverage probabilities of confidence intervals and $(1-\alpha)=0.95$ point-wise intervals with their expected lengths. We observe that the expected lengths and coverage probabilities remain roughly similar for different values of $d$ under the SLM scenario, which is not the case for the LM scenario. 
Note that coverage was always reduced from the nominal level of $0.95$, but much less so for SLM processes than for LM processes.  Coverage also decreased as $d$ increased.
Additional results for smaller sample sizes and $h=N^{-1/3}$ are given in the Appendix.

\subsection{Test statistic properties} \label{sec:test_sim}

In order to investigate the finite sample performance of test statistic, we first construct the Monte Carlo distributions of test statistic under both long and semi-long memory processes with endogeneity. Then, we use resampling methods to study the size and power of test statistic since their limiting distributions have complex forms and are not practical. In particular, we use the subsampling technique developed by \cite{mosaferi2024properties}. 

Drawing on the application to be presented in Section \ref{sec:application} we
consider two regression functions, a straight line ($y_k=\theta_0+\theta_1 x_k + \sigma u_k$) and a  quadratic ($y_k=\theta_0+\theta_1 x_k + \theta_2 x_k^2+ \sigma u_k$), 
where $(\theta_0,\theta_1,\theta_2)=(0,1,1)$. For simplicity, we let $h=N^{-1/3}$ and $N=\{50, 100, 500 \}$. We assume that the amount of endogeneity is $\rho_{\zeta,\epsilon}=0.5$ and $d=\{0.1, 0.2, 0.3, 0.4, 0.5, 1, 1.5\}$, where $d \geq 0.5$ is only for the case of SLM, again based on experience with the application of Section \ref{sec:application}. Additionally, we let $\lambda=N^{-1/5}$.
Other details of the simulation studies conducted to examine the behavior of the tests are the same as those described in Section \ref{sec:regression_sim}. 

\begin{figure}[ht]
\centering
\begin{tabular}{ ccc }
\includegraphics[width=0.35\textwidth]{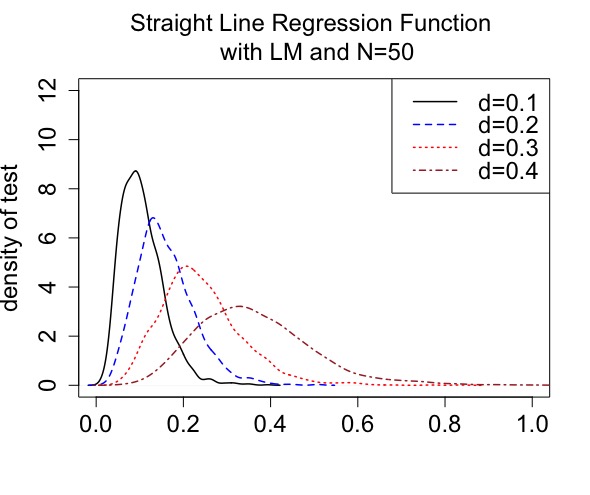} &
\includegraphics[width=0.35\textwidth]{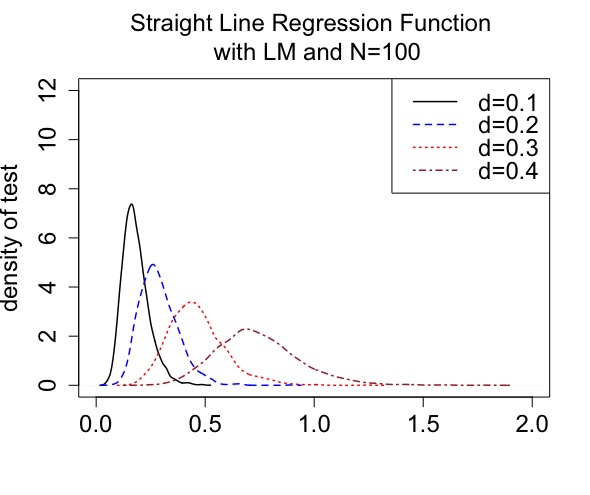} &
\includegraphics[width=0.35\textwidth]{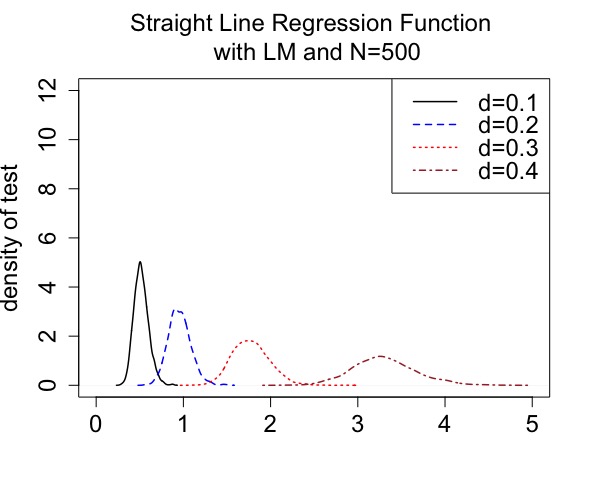} \\
\includegraphics[width=0.35\textwidth]{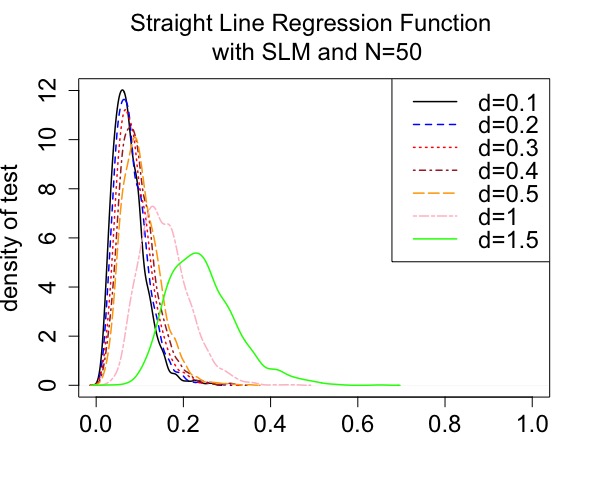} &
\includegraphics[width=0.35\textwidth]{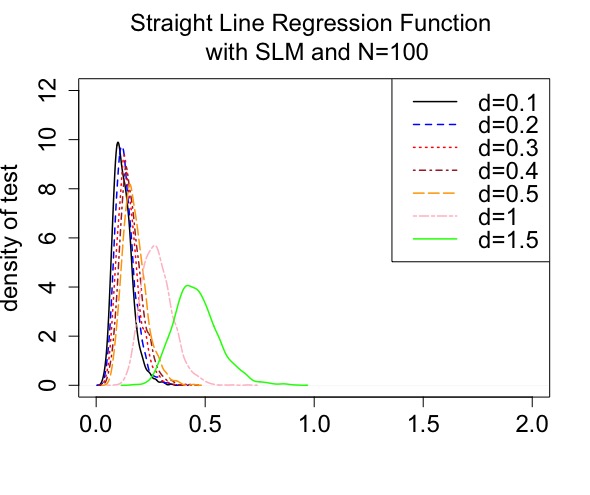} &
\includegraphics[width=0.35\textwidth]{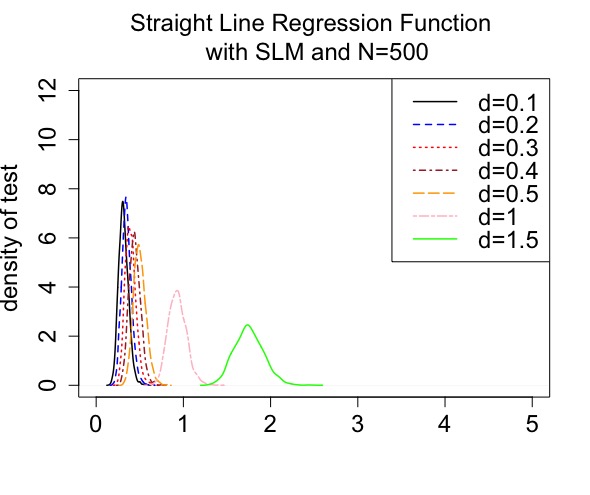}
\end{tabular}
\caption{Monte Carlo densities of test statistic for the straight line regression function under LM and SLM cases.}
\label{Fig.densities_line}
\end{figure}

\begin{figure}[ht]
\centering
\begin{tabular}{ ccc }
\includegraphics[width=0.35\textwidth]
{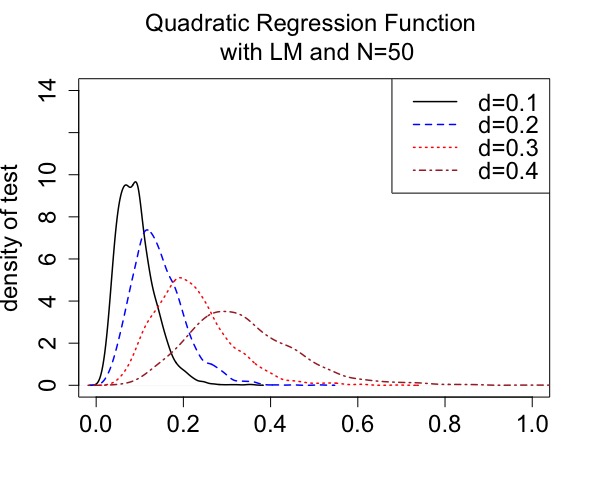} &
\includegraphics[width=0.35\textwidth]{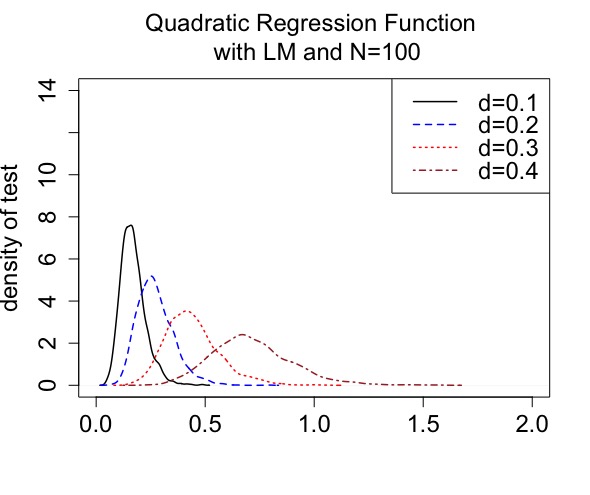} &
\includegraphics[width=0.35\textwidth]{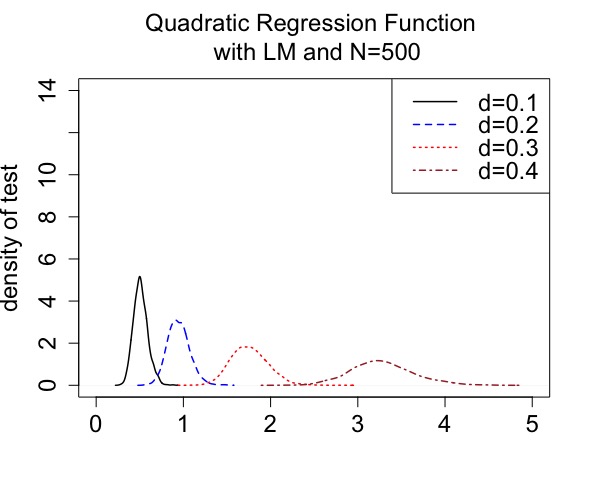} \\
\includegraphics[width=0.35\textwidth]{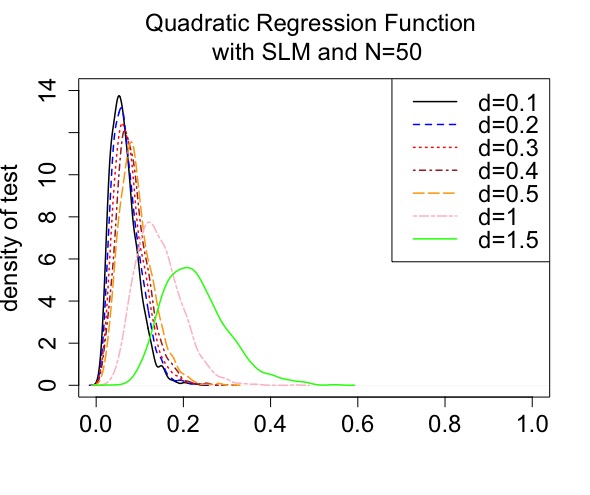} &
\includegraphics[width=0.35\textwidth]{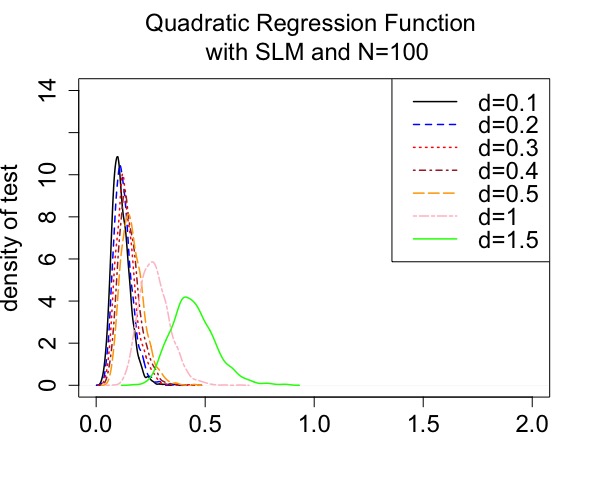} &
\includegraphics[width=0.35\textwidth]{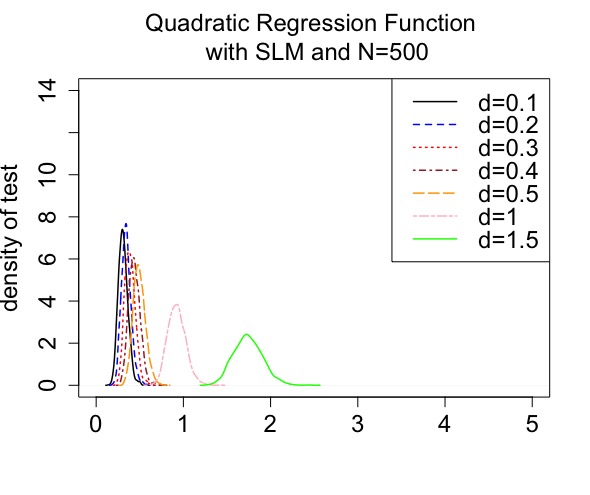}
\end{tabular}
\caption{Monte Carlo densities of test statistic for the quadratic regression function under LM and SLM cases.}
\label{Fig.densities_quad}
\end{figure}

Monte Carlo densities of the test statistic are given in Figures \ref{Fig.densities_line} and \ref{Fig.densities_quad}. We observe that the densities become further from zero as the value of $d$ increases under the LM case. On the other hand, the densities overlap quite a lot for the SLM case and in particular for $0<d<1/2$. 

To study the size of the test, we use the straight line and quadratic regression function assumptions to formulate $H_0$. To study the power of test, we use the $H_A$ given in equation (\ref{alter}). We assume $\rho_N=1/N$ and $m(x_k)=|x_k|$. Therefore, we have $y_k=\theta_0+\theta_1 x_k + \frac{1}{N} |x_k| + \sigma u_k$ and $y_k=\theta_0+\theta_1 x_k + \theta_2 x_k^2+ \frac{1}{N} |x_k| + \sigma u_k$. 
For the block sizes, we choose the values of $\{[0.5 \sqrt{N}], [\sqrt{N}], [2 \sqrt{N}], [4 \sqrt{N}] \}$. The size and power values were close to $1$ for the LM and SLM cases. These results are not given in the manuscript, but allowed us to identify a bias in the form of the test statistic. We rectify this bias issue and construct a de-biased test statistic using the subsampling procedure and the algorithm described in Section C of \cite{mosaferi2024properties}. 

\clearpage

The values of size and power for the de-biased test statistic are given in Tables \ref{Tab:sizepower_line} and \ref{Tab:sizepower_quad}. Values of size for the straight line model are quite reasonable in both LM and SLM cases for small block sizes but decrease as block size increases. The size for the quadratic model is depressed from nominal, also decreases as block size increases, but is considerably more stable for SLM than LM situations.  Power is low across the board, also decreases with block size, and is less stable for LM cases than for SLM ones. Additional results for $d \geq 0.5$ under SLM and smaller sample sizes are given in the Appendix. 

\begin{table}[!ht]
\footnotesize
\caption{The values of size and power for de-biased test statistic at $5\%$ level for the straight line regression function $f(x)$ with $N=500$. Values in brackets denote different block sizes -- see text.} 
\label{Tab:sizepower_line}
\centering
\setlength{\tabcolsep}{10pt} 
\renewcommand{\arraystretch}{1.25} 
\begin{tabular}{cccc}
\toprule
size/power & d & LM & SLM \\
\midrule
size & 0.1 & $\{0.052, 0.028, 0.021, 0.020\}$ & $\{0.054, 0.027,      0.017, 0.020 \}$ \\
 & 0.2 & $\{0.050, 0.034, 0.024, 0.027\}$ & $\{0.054, 0.029, 0.022, 0.025\}$ \\
 & 0.3 & $\{0.058, 0.041, 0.029, 0.023\}$ & $\{0.052, 0.030, 0.017,   0.023\}$ \\
 & 0.4 & $\{0.060, 0.041, 0.031, 0.025\}$ & $\{0.050, 0.031, 0.021, 0.021\}$ \\ \midrule
 power & 0.1 & $\{0.055, 0.032, 0.022, 0.021\}$ & $\{ 0.057, 0.030,    0.019, 0.021 \}$ \\
  & 0.2 & $\{0.057, 0.040, 0.032, 0.032 \}$ & $\{0.054, 0.031,      0.024, 0.026 \}$ \\
 & 0.3 & $\{ 0.083, 0.059, 0.045, 0.038 \}$ & $\{0.058, 0.034,      0.023, 0.026\}$ \\
 & 0.4 & $\{0.116, 0.096, 0.073, 0.061 \}$ & $\{0.058, 0.035,      0.025, 0.024\}$ \\
\bottomrule
\end{tabular}
\end{table} 

\clearpage

\begin{table}[!ht]
\footnotesize
\caption{The values of size and power for de-biased test statistic at $5\%$ level for the quadratic regression function $f(x)$ with $N=500$. Values in brackets denote different block sizes -- see text.} 
\label{Tab:sizepower_quad}
\centering
\setlength{\tabcolsep}{10pt} 
\renewcommand{\arraystretch}{1.25} 
\begin{tabular}{cccc}
\toprule
size/power & d & LM & SLM \\
\midrule
size & 0.1 & $\{0.031, 0.017, 0.009, 0.009\}$ & $\{0.028, 0.014,     0.009, 0.009\}$ \\
 & 0.2 & $\{0.025, 0.014, 0.007, 0.007\}$ & $\{0.028, 0.014, 0.007,   0.011\}$ \\
 & 0.3 & $\{0.058, 0.014, 0.008, 0.009\}$ & $\{0.028, 0.017, 0.008,    0.010 \}$ \\
 & 0.4 & $\{0.200, 0.049, 0.019, 0.016\}$ & $\{0.027, 0.014, 0.009,    0.009\}$ \\ \midrule
power & 0.1 & $\{0.031, 0.018, 0.010, 0.010 \}$ & $\{0.030, 0.013,     0.009, 0.009 \}$ \\
  & 0.2 & $\{ 0.027, 0.016, 0.008, 0.007 \}$ & $\{0.029, 0.015,    0.009, 0.012 \}$ \\
 & 0.3 & $\{ 0.061, 0.018, 0.009, 0.011 \}$ & $\{ 0.028, 0.016,     0.009, 0.010 \}$ \\
 & 0.4 & $\{0.208, 0.054, 0.023, 0.021 \}$ & $\{0.028, 0.015,     0.009, 0.010 \}$ \\
\bottomrule
\end{tabular}
\end{table}

\section{Application} \label{sec:application}

In this section, we illustrate the performance of the test statistic given in equation (\ref{normalizedtest}) in the analysis of environmental Kuznets curves.  In particular, we focus on the relationship of carbon dioxide emissions (CO$_2$) v.s. gross domestic product (GDP) per capita in Spain and sulfur dioxide emission (SO$_2$) v.s. GDP per capita in the United Kingdom. We consider the natural logarithmic transformation of these quantities ($x_k$ and $y_k$) before proceeding. The data set for Spain is from 1950 to 2008 and contains 59 observations. The data set for the United Kingdom is from 1870 to 1999 and contains 130 observations. 
The data sets are provided on the Github repository of the first author.
As mentioned in the previous section, the limit theory of this test does not have a simple form and is not practical, so we again rely on the use of subsampling to compute benchmarks. 

\begin{figure}[ht]
\centering
\begin{tabular}{ cc }
\includegraphics[width=0.5\textwidth]{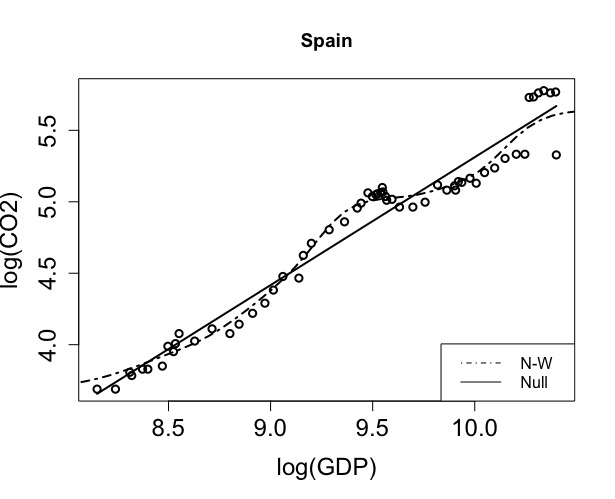} &
\includegraphics[width=0.5\textwidth]{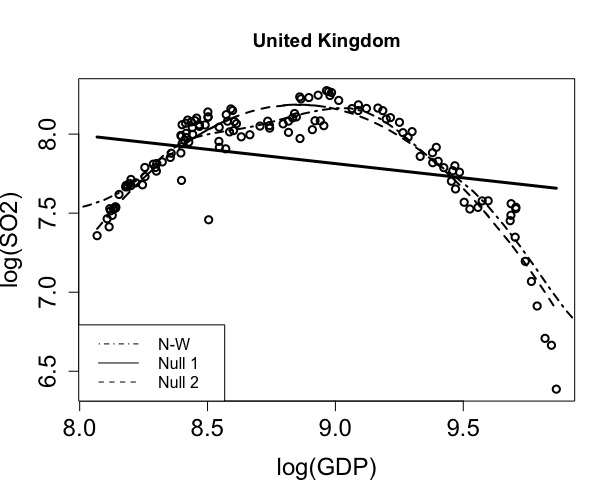} 
\end{tabular}
\caption{Scatter-pots of environmental Kuznets curves for Spain and United Kingdom with the null hypotheses and overlaid N-W regression functions.}
\label{Fig.countries}
\end{figure}

\begin{figure}[ht]
\centering
\begin{tabular}{ cc }
\includegraphics[width=0.5\textwidth]{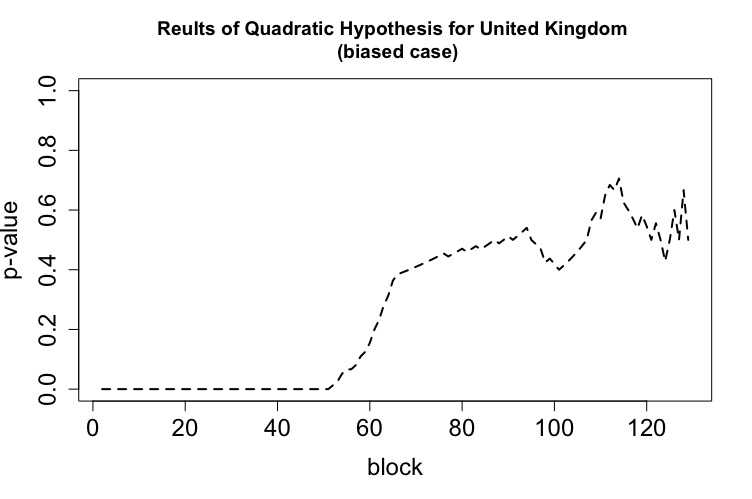} &
\includegraphics[width=0.5\textwidth]{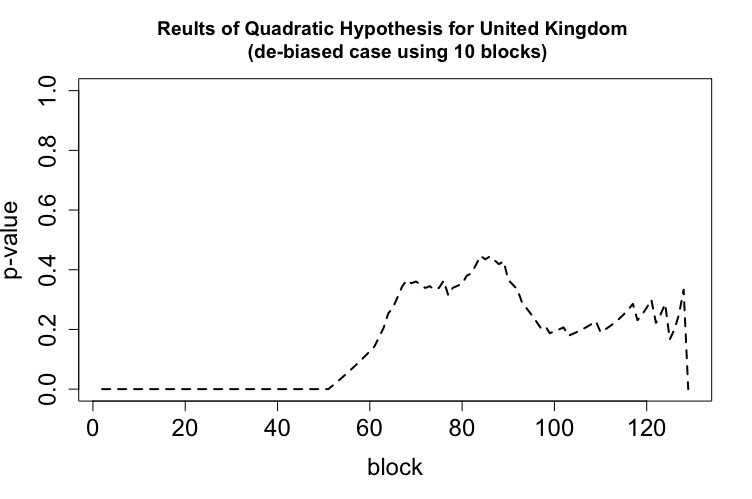} \\
\includegraphics[width=0.5\textwidth]{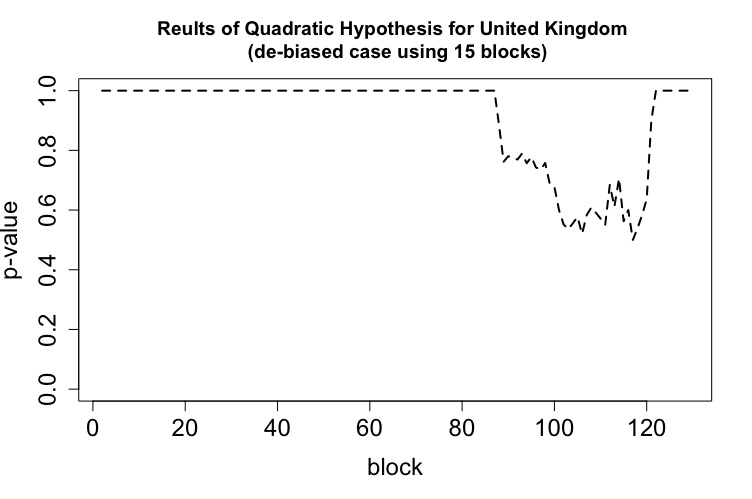} &
\includegraphics[width=0.5\textwidth]{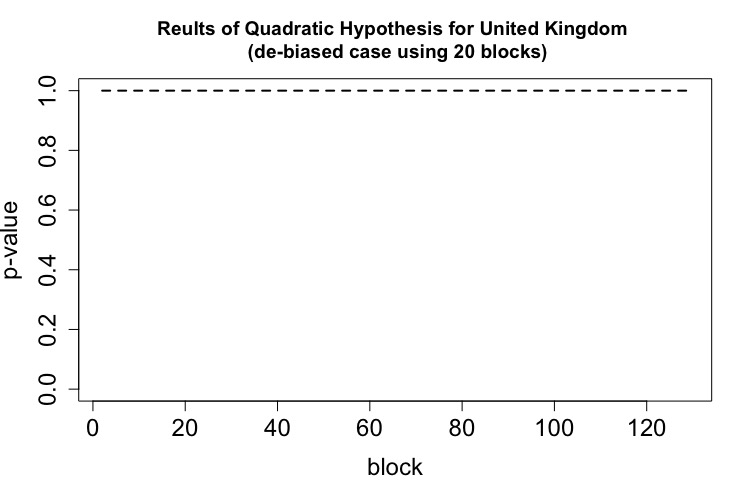}
\end{tabular}
\caption{P-values v.s. block sizes for the biased and de-biased forms of test statistic for the United Kingdom under the quadratic regression function null hypothesis.}
\label{Fig.pvalues_UK}
\end{figure}

Scatter-plots of the relevant variables are shown in Figure \ref{Fig.countries}. For Spain, the pattern of points suggests a straight-line regression function, while for the United Kingdom the pattern appears to be more quadratic in form. Thus, we test the null hypothesis of $f(x)=\theta_0+\theta_1 x$ for Spain and two null hypotheses of (1) $f(x)=\theta_0+\theta_1 x$ and (2) $f(x)=\theta_0+\theta_1 x + \theta_2 x^2$ for the United Kingdom. 
We also overlay the Nadaraya-Watson (N-W) regression function estimator $\hat{f}(x)$ from (\ref{fhat}) on the plots in Figure \ref{Fig.countries}. Optimal values for the bandwidth $h$ were selected using cross-validation (e.g. \citealp{park1990comparison,hardlebook}) with the help of N-W regression functions. The optimal bandwidth was $h=0.151$ for Spain and $h=0.107$ for the United Kingdom. The associated least squares values based on $\text{LCV}(h)=\sum_{k=1}^{N}(y_k - \hat{f}_{-k}(x_k))^2$ are given in Table \ref{Tab:application_pars}, where $\hat{h}_{\text{optimal}}=\text{argmin}_{h>0}\text{LCV}(h)$. 
To estimate the values of the fractional differencing parameter $d$ and the tempering parameter $\lambda$, we use their Whittle estimators from \texttt{artfima} package in \texttt{R} (e.g., \citealp{sabzikar2019parameter}). The estimated values of these parameters are also given in Table \ref{Tab:application_pars}. Based on the previous section and due to the biased form of test statistic $T_{\lambda,d}$, we consider both the biased and de-biased form of tests for use with subsampling. The resultant p-values are summarized in Table \ref{Tab:applicationpvals}. Furthermore, the p-values for the quadratic null hypothesis for the United Kingdom based on all possible block sizes values are displayed in Figure \ref{Fig.pvalues_UK}.
\begin{table}[!ht]
\footnotesize
\caption{Estimated values of $h$, $d$, and $\lambda$.} 
\label{Tab:application_pars}
\centering
\setlength{\tabcolsep}{10pt} 
\renewcommand{\arraystretch}{1.25} 
\begin{tabular}{ccccc}
\toprule
country & h & LCV & d & $\lambda$ \\
\midrule
Spain & 0.151 & 0.005 & 1.079 & 0.138  \\ 
United Kingdom & 0.107 & 0.008 & 1.020 & 0.056  \\
\bottomrule
\end{tabular}
\end{table}

For Spain, we observe that the null hypothesis of a straight line would be rejected when we use the biased form of test statistic, but with the de-biased test statistic this hypothesis would not be rejected. For the United Kingdom, the null hypothesis of a straight line is rejected for both the biased and de-biased forms of the test statistic. A hypothesized quadratic form for the regression function would not be rejected for either the biased or de-biased test statistics. Based on Figure \ref{Fig.pvalues_UK}, including a greater number of blocks in the de-biasing process increases the p-values associated with the hypothesis of a quadratic regression function. In particular, for the last plot in Figure \ref{Fig.pvalues_UK}, we can see that when 20 blocks are included in de-biasing, the p-value is one for all block sizes.  

\begin{table}[!ht]
\footnotesize
\caption{P-values for testing the functional forms of environmental Kuznets curve. The values in the parentheses are the range of p-values from all the possible block sizes.} 
\label{Tab:applicationpvals}
\centering
\setlength{\tabcolsep}{10pt} 
\renewcommand{\arraystretch}{1.25} 
\begin{tabular}{cccc}
\toprule
&& \multicolumn{2}{c}{Null Hypothesis} \\
country & biasedness scenario &  (1) straight line & (2) quadratic \\
\midrule
Spain &  biased case & 0.000 & N/A \\ 
 & de-biased case & 1.000 &  N/A \\\midrule
United Kingdom & biased case & 0.000 & (0.000,0.706) \\
 & de-biased case (using 10 blocks) & 0.000 &  (0.000,0.447) \\
 & de-biased case (using 15 blocks) & 0.000 &  (0.500,1.000) \\
 & de-biased case (using 20 blocks) & 0.000 &  1.000 \\
\bottomrule
\end{tabular}
\end{table} 

\section{Discussion and future work} \label{sec:Discussion}  

In this article we have considered the use of tempered or semi-long memory regressor processes within the context of cointegrating regression models for time series analysis.  We have developed limit theory that allows the construction of pointwise confidence intervals for a nonlinear regression function when estimation is accomplished through the use of a kernel smoother.  We have also demonstrated that the limit distribution of a test statistic for the specification of the regression function does not depend on the differencing parameter $d$ in semi-long memory.  Through simulation studies, we have shown that performance of interval estimators of a regression function is more stable and has generally better performance for the semi-long memory case than for long memory. We also developed the limit distribution of the test statistic for their parametric forms and showed that it is free from the unknown parameter of $d$.

Some directions for future work could be related to
multivariate nonparametric cointegrating regression with multiple regressors such that they have a semi-long memory property with error terms belonging to the $\alpha$-stable finite-dimensional distributions with infinite variance and infinite lower-order moments. We assume the regression disturbances are serially and cross-correlated with the regressors and permit the associated bandwidth in the regression model to be random and data dependent. One can investigate the uniform convergence rate for the kernel regression estimator and provide model specification test for multivariate cointegrating regressions where the asymptotic distribution of the test statistic is free of any unknown parameters and is proportional to a local time of linear stable motions. 

\section*{Acknowledgments}
We would like to thank the Editor, Associate Editor, and two anonymous reviewers for their excellent suggestions and comments that have substantially improved the quality of our manuscript. 

\section*{Disclosure statement}
No potential conflict of interest was reported by the authors.

\clearpage

\bibliographystyle{ims}
\bibliography{Bibliography}

\newpage
\clearpage

\setcounter{secnumdepth}{0}

\section{Appendix A: Technical Details and Proofs} \label{App.A}

\subsection{A.1 \quad Preliminaries and Lemmas} \label{App.A.1}
\renewcommand{\theequation}{A.\arabic{equation}}
\setcounter{equation}{0}
\renewcommand{\thelemma}{A.\arabic{lemma}}
\renewcommand{\thedefinition}{A.\arabic{definition}}

$\bullet$ Throughout this section, we set $d_N:= [\mathbb{E}(x_N^2)]^{1/2}$. In the semi-long memory setting, the asymptotic form of $d_N$ is given by $d_N \sim \sqrt{N}/\lambda^d$ as $N \rightarrow \infty$. Then,
\begin{equation}\label{strongsmooth1}
\frac{x_N}{d_N} := \frac{\lambda^d}{\sqrt{N}}\sum_{s=1}^{N} X_{d,\lambda}(s) =
\frac{1}{\sqrt{N}}\sum_{s=1}^{N} \zeta(s)+ o_{P}(1),
\end{equation}
see expression (A.7) in \citet{Sabzikar2020}. For any $t \in [0,1]$, the following weak convergence applies on $D[0,1]$:
\begin{equation*} \label{stdBrownian}
\frac{x_{\floor{Nt}}}{d_N} \Rightarrow B(t), 
\end{equation*}
where $\floor{Nt}$ is the floor function, and $B(t)$ is the standard Brownian motion.

\vspace{0.25cm}

\noindent $\bullet$ We decompose $x_k$ as follows:
\begin{flalign*}
x_k \equiv \sum_{j=1}^{k} X_{d, \lambda}(j) & =\sum_{j=1}^{k}\sum_{i=-\infty}^{j} \phi_N(j-i) \zeta(i) , \quad \phi_N(j):=e^{-\lambda_N j}b_{d}(j) && \\[0.5em]
& = x_s+\sum_{j=s+1}^{k}\sum_{i=-\infty}^{s}\phi_N(j-i) \zeta(i)+\sum_{j=s+1}^{k}\sum_{i=s+1}^{j} \phi_N(j-i) \zeta(i) && \\[0.5em]
& =: x^*_{s,k}+x'_{s,k}, \qquad s<k,
\end{flalign*}
such that $x^*_{s,k}=x_s+\sum_{j=s+1}^{k} \sum_{i=-\infty}^{s} \phi_N(j-i) \zeta(i)$ and it depends on $\{\cdots, \zeta(s-1), \zeta(s)\}$, and $x'_{s,k}=\sum_{j=1}^{k-s} \sum_{i=1}^{j} \phi_N(j-i) \zeta(i+s).$ If we define $a_N:=\sum_{j=0}^{N}\phi_N(j)$, then we can show
$a_N \asymp O(\lambda^{-d})$ as $N\lambda \rightarrow \infty$ and recall that $\lambda_N \equiv \lambda$ is a sample size dependent parameter; see Lemma A.5 part (b) in \citet{Sabzikar2020}.


\begin{definition} \label{def.smootharray}
Let $\Omega_N(\eta) \equiv \{(l,k): \eta N \leq k \leq (1-\eta) N, k+ \eta N \leq l \leq N \}$ for all $0 \leq k < l \leq N$ and $N \geq 1$, where $0<\eta<1$. A random array $\{x_{k,N}\}_{k \geq 1, N \geq 1}$ is called \textit{strong smooth} if there exists a sequence of constants $d_{l,k,N}>0$ and a sequence of $\sigma$-fields $\mathcal{F}_{k,N}$ (define $\mathcal{F}_{0,N}= \sigma\{\phi, \Omega \}$, the trivial $\sigma$-field) such that
\begin{itemize}
\item[(a)] For some $m_0>0$ and $C>0$, $\inf_{(l,k) \in \Omega_N(\eta)} d_{l,k,N} \geq \eta^{m_0}/C$ as $N \rightarrow \infty$,
\begin{itemize}
\item[(i)] $\lim_{\eta \rightarrow 0} \lim_{N \rightarrow \infty} \frac{1}{N} \sum_{l=(1-\eta)N}^{N} (d_{l,0,N})^{-1}=0$,
\item[(ii)] $\lim_{\eta \rightarrow 0} \lim_{N \rightarrow \infty} \frac{1}{N} \max_{0 \leq k \leq (1-\eta)N} \sum_{l=k+1}^{k+\eta N} (d_{l,k,N})^{-1}=0$, and
\item[(iii)] $\limsup_{N \rightarrow \infty} \frac{1}{N} \max_{0 \leq k \leq N-1} \sum_{l=k+1}^{N} (d_{l,k,N})^{-1}< \infty$.
\end{itemize}
\item[(b)] $x_{k,N}$ is adapted to $\mathcal{F}_{k,N}$, and conditional on $\mathcal{F}_{k,N}$, $(x_{l,N}-x_{k,N})/d_{l,k,N}$ has a density $h_{l,k,N}(x)$ which is uniformly bounded by a constant $K$ and
\begin{flalign*}
\lim_{\kappa \rightarrow 0} \lim_{N \rightarrow \infty} \sup_{(l,k) \in \Omega_N(\kappa^{1/(2m_0)})} \sup_{|u| \leq \kappa} |h_{l,k,N}(u)-h_{l,k,N}(0)|=0. &&
\end{flalign*}
\end{itemize} 
\end{definition}

\begin{lemma} \label{Lemma}
Let's define $\eta_i=(\zeta(i),\epsilon(i))'$ where $\mathbb{E}(\eta_0)=0$ and $\mathbb{E}||\eta_0||^2 < \infty$. Let $\mathcal{F}_s=\sigma(\eta_s,\eta_{s-1},...)$, $\Lambda= \Lambda(\eta_1,...,\eta_{m_0})$ for some $m_0>0$ be a real function where $\mathbb{E}(\Lambda(\eta_1))=0$, and $g(x)$ is a bounded function such that $\int_{\mathbb{R}}|g(x)|dx < \infty$.

\begin{itemize}
\item[(i)] For a constant $A_0$, we have
\begin{flalign} \label{App1}
\sup_{x} \mathbb{E}\Big|\sum_{k=1}^{N} g\Big[\frac{x_k-x}{h}\Big] \Big|^m \leq A_0^m (m+1)! \, \Big(\sqrt{N} \lambda^d h\Big)^m, \quad m \geq 1. &&
\end{flalign}
\item[(ii)] As $\{\sqrt{N} \lambda^d h\}^{-1} \rightarrow 0$ and $N\lambda \rightarrow \infty$, we have
\begin{flalign} \label{App2}
\sup\limits_{x} \mathbb{E} \Big|\sum_{k=1}^{N} \Lambda(\eta_{k-j}) g\Big[\frac{x_k-x}{h}\Big] \Big|^2  & \leq C \, \mathbb{E}\Lambda^2(\eta_1) \sqrt{N} \lambda^d h \, \Big\{1+h \sqrt{j} \Big\}. &&
\end{flalign}
\item[(iii)] If we define $u_k:= \sum_{j=0}^{\infty} \psi_j \eta_{k-j}$, then
\begin{flalign} \label{App3}
\mathbb{E}\Big|\sum_{k=1}^{N} u_k \, g\Big(\frac{x_k}{h}\Big) \Big|^2 \leq C \, \mathbb{E}||\eta_1||^2 \sqrt{N} \lambda^d h. &&
\end{flalign}
\item[(iv)] If we define $u_{k,m_0}:=\sum_{j=m_0}^{\infty} \psi_j \Lambda'_{k-j}$ where $\psi_j=(\psi_{1j},\psi_{2j})$ and $\Lambda_j=(\Lambda_1(\eta_j),\Lambda_2(\eta_j))$, then we have
\begin{flalign} \label{App4}
\sup_{x}\mathbb{E}\Big|\sum_{k=1}^{N}u_{k,m_0} \, g\Big[x_k-\frac{x}{h}\Big]\Big|^2  \leq C \, \mathbb{E}||\Lambda_1||^2 \sqrt{N} \lambda^d h \, \Big\{\sum_{j=m_0}^{\infty} j^{1/4} (|\psi_{1j}|+|\psi_{2j}|) \Big\}^2.  &&
\end{flalign}
\end{itemize}

\begin{proof}
For the proof of expression (\ref{App1}), see Lemma 5.1. of \cite{chan2014uniform}.
To prove (\ref{App2}), note that some of the details are very similar to the ones given for the proof of expression (8.11) in Lemma 8.2. of \cite{WP2016}, and therefore we only give an outline for the different parts as follows. Let
\begin{flalign*}
\Delta_N & \equiv \Big|\sum_{k=1}^{N} \Lambda(\eta_{k-j}) g\Big[\frac{x_k-x}{h}\Big] \Big|^2  \leq 2 \Big|\sum_{k=A_0}^{N} \Lambda(\eta_{k-j}) g\Big[\frac{x_k-x}{h}\Big] \Big|^2+C \Big(\sum_{k=1}^{A_0} |\Lambda(\eta_{k-j})| \Big)^2 && \\[0.5em]
& = 2 \Big(\sum_{k=A_0}^{N}\sum_{l=1,|k-l|<A_0}^{N} + 2 \sum_{k=A_0}^{N-1} \sum_{l=k+A_0}^{N} \Big) \Lambda(\eta_{k-j}) \Lambda(\eta_{l-j}) && \\[0.5em]
& \quad \times g\Big[\frac{x_k-x}{h}\Big] g\Big[\frac{x_l-x}{h}\Big]+C \Big(\sum_{k=1}^{A_0}|\Lambda(\eta_{k-j})| \Big)^2 && \\[0.5em]
& = \Delta_{1N}+\Delta_{2N}+\Delta_{3N},
\end{flalign*}
where $A_0$ is a positive constant. Then by letting $d_N \sim \sqrt{N}/\lambda^d$, we have

\begin{flalign*}
\sup\limits_{x} \mathbb{E}|\Delta_{1N}| & \leq C \, \mathbb{E} \Lambda^2(\eta_1) \sqrt{N} \lambda^d h, && \\[0.5em]
\sup\limits_{x} \mathbb{E}|\Delta_{2N}| & \leq C \, \mathbb{E}\Lambda^2(\eta_1) h^2 \sum_{k=A_0}^{N-1}d_k^{-1} \Big(\sum_{l=k+A_0}^{N \land (k+j)} d_{l-k}^{-1} + \sum_{k=0}^{j} |\phi_N(k)| \sum_{l=k+j}^{N}d_{l-k}^{-2} \Big) && 
\\[0.5em]
& \leq C \, \mathbb{E}\Lambda^2(\eta_1) \sqrt{N} \lambda^d h^2
\Big(\frac{j}{d_j}+ \sum_{k=0}^{j} |\phi_j(k)| \sum_{l=k+j}^{N} \frac{\lambda_{l-k}^{2d}}{l-k} \Big)  && \\[0.5em]
& \leq C \, \mathbb{E}\Lambda^2(\eta_1) \sqrt{N} \lambda^d h^2 \Big(\frac{j}{d_j}+\sum_{k=0}^{j} |\phi_j(k)| \Big) && \\[0.5em]
& \leq C \, \mathbb{E}\Lambda^2(\eta_1) \sqrt{N} \lambda^d h^2 
\Big( \frac{\lambda_j^d}{\sqrt{j}} j + \lambda_j^{-d} \Big) && \\[0.8em]
& \leq C \, \mathbb{E}\Lambda^2(\eta_1) \sqrt{N} \lambda^d h^2 \sqrt{j} \quad \text{for} \quad j \geq 1, \quad \text{and}  && \\[1em]
\sup\limits_{x} \mathbb{E}|\Delta_{3N}| & \leq C \, \mathbb{E}\Lambda^2(\eta_1). &&
\end{flalign*}

\noindent Eventually, we can conclude expression (\ref{App2}).
Expression (\ref{App3}) easily follows from expression (\ref{App2}), and expression (\ref{App4}) can be followed by a similar path from the proof of expression (8.13) in Lemma 8.2. of \cite{WP2016}.
\end{proof}

\end{lemma}

\subsection{A.2 \quad Proofs of Theorems} \label{App.A.2}
\renewcommand{\theequation}{A.\arabic{equation}}

\textbf{Proof of Theorem \ref{Theo.f}.}
First, in order to establish the asymptotic distribution of $[\hat{f}(x)-f(x)]$, we decompose it as follows:
\begin{flalign} \label{App5}
\hat{f}(x)-f(x)=\frac{\sum_{k=1}^{N} u_k K[(x_k-x)/h]}{\sum_{k=1}^{N}K[(x_k-x)/h]} + \frac{\sum_{k=1}^{N}[f(x_k)-f(x)]K[(x_k-x)/h]}{\sum_{k=1}^{N}K[(x_k-x)/h]}. &&
\end{flalign}
Then, only the first term in expression (\ref{App5}) determines the asymptotic distribution of $[\hat{f}(x)-f(x)]$ as the second term is negligible under mild conditions.
To prove the self-normalized expression (\ref{normalized}), let $\Theta_{1N}:= \sum_{k=1}^{N} u_k K[\frac{x_k-x}{h}]$, $\Theta_{2N}:=\sum_{k=1}^{N}(f(x_k)-f(x))K[\frac{x_k-x}{h}]$, and $\Theta_{3N}:= 1/\sum_{k=1}^{N} K[\frac{x_k-x}{h}]$. 
Now, we have
\begin{flalign}
& \Big\{h \sum_{k=1}^{N} K_h(x_k-x) \Big\}^{1/2} \Big(\hat{f}(x)-f(x)\Big) \nonumber && \\[1em]
& \qquad = \frac{\sum_{k=1}^{N}u_k K[(x_k-x)/h]}{\{\sum_{k=1}^{N}K[(x_k-x)/h]\}^{1/2}} + \frac{\sum_{k=1}^{N}[f(x_k)-f(x)]K[(x_k-x)/h]}{\{\sum_{k=1}^{N}K[(x_k-x)/h]\}^{1/2}} \nonumber && \\[1em]
& \qquad = \frac{\Theta_{1N}}{\Theta_{3N}^{-1/2}}+\Theta_{2N} \sqrt{\Theta_{3N}} =
\frac{\{\sqrt{N} \lambda^d h\}^{-1/2} \, \Theta_{1N}}{\{\sqrt{N} \lambda^d h\}^{-1/2} \,  \Theta_{3N}^{-1/2}} +\Theta_{2N} \, \, \sqrt{\Theta_{3N}} \nonumber && \\[1em]
&  \qquad \rightarrow_D \frac{d_0 \, N(0,1) \, L_{B}^{1/2}(1,0)}{L_{B}^{1/2}(1,0)}
 = d_0 N(0,1)=:N(0,\sigma^2) \label{th2.1proof},
\end{flalign}
where $\sigma^2=d_0^2$. Note that we have used the fact that $\Theta_{2N} \sqrt{\Theta_{3N}} \rightarrow_P 0$ from Proposition \ref{Prop8}. Also, the last line in \eqref{th2.1proof} comes from (a) Slutsky's theorem, (b) $\{\sqrt{N} \lambda^d h\}^{-1} \sum_{k=1}^{\lfloor Nt \rfloor}K[\frac{x_k-x}{h}]$ \newline 
$\rightarrow_{D} L_{B}(t,0)$, and (c) the following result:
\begin{flalign} \label{App6}
\Big\{\sqrt{N} \lambda^d h\Big\}^{-1/2} \sum_{k=1}^{N} K\Big[\frac{x_k-x}{h}\Big] u_k \rightarrow_{D} d_0 \, N(0,1) \, L_{B}^{1/2}(1,0). &&
\end{flalign}
The convergence in distribution in (\ref{App6}) comes from Proposition \ref{Prop9} part (iv).
Expression (\ref{normalized}) easily implies expression (\ref{maintheo}) and the fact that

\begin{flalign*}
\Big\{\sqrt{N} \lambda^d h\Big\}^{1/2} \Big(\hat{f}(x)-f(x)\Big) &= 
\Big\{\sqrt{N} \lambda^d h\Big\}^{1/2} \frac{ \{h \sum_{k=1}^{N} K_h(x_k-x)\}^{1/2} (\hat{f}(x)-f(x))} {\{h \sum_{k=1}^{N} K_h(x_k-x)\}^{1/2}}&& \\[0.5em]
& \rightarrow_D d_0 \, L_{B}^{-1/2}(1,0) \, N(0,1),
\end{flalign*}
thus the proof is complete.
\QEDA

\vspace{0.25cm}

\noindent \textbf{Proof of Theorem \ref{Theo.test}.} 
First, $T_{N}$ can be decomposed into three terms under $H_0$, i.e. $y_k=g(x_k,\theta_0)+u_k$, as follows
\begin{flalign*}
T_{N} & = \int_{\mathbb{R}} \Big\{\sum_{k=1}^{N} K\Big[\frac{x_k-x}{h} \Big] \Big[g(x_k,\theta_0)+u_k-g(x_k,\hat{\theta}_N)\Big] \Big\}^2 \pi(x) dx && \\[0.5em]
& = T_{N}^{\text{I}}+T_{N}^{\text{II}}+T_{N}^{\text{III}}, \quad \text{such that} && \\[0.5em]
T_{N}^{\text{I}} & := \int_{\mathbb{R}} \Big\{\sum_{k=1}^{N} K\Big[\frac{x_k-x}{h} \Big] u_k \Big\}^2 \pi(x) dx, && \\[0.5em]
T_{N}^{\text{II}} & := \int_{\mathbb{R}} \Big\{\sum_{k=1}^{N} K \Big[\frac{x_k-x}{h} \Big] \Big(g(x_k,\theta_0)-g(x_k,\hat{\theta}_N) \Big) \Big\}^2 \pi(x) dx, \quad \text{and} && \\[0.5em]
T_{N}^{\text{III}} & := 2 \int_{\mathbb{R}} \Big\{\sum_{1 \leq k < j \leq N} K \Big[\frac{x_k-x}{h} \Big] K \Big[\frac{x_j-x}{h} \Big] u_k \Big(g(x_j,\theta_0)-g(x_j,\hat{\theta}_N) \Big) \Big\} \pi(x) dx.
\end{flalign*}

\noindent Based on Assumption \ref{ass:gs}, there exists $g_1(x)$ such that for each $\theta, \theta_0 \in \Theta$, we have $|g(x,\theta)-g(x,\theta_0)| \leq C ||\theta-\theta_0|| g_1(x)$. Therefore,
\begin{flalign} \label{App7}
T_{N}^{\text{II}} \leq C ||\hat{\theta}_N - \theta_0 ||^2 \int_{\mathbb{R}} \Big\{ \sum_{k=1}^{N} K \Big[\frac{x_k-x}{h} \Big] g_1(x_k) \Big\}^2 \pi(x) dx = o_P \Big(\sqrt{N} \lambda^d h \Big). &&
\end{flalign}
Expression (\ref{App7}) comes from (a) Assumption \ref{ass:H0}, $||\hat{\theta}_N-\theta_0 || = o_P(\{ \sqrt{N} \lambda^d h \}^{-1/2})$ and (b) the following result:

\begin{flalign} \label{App8}
\Big\{\sqrt{N} \lambda^d h \Big\}^{-2} \int_{\mathbb{R}} \Big\{\sum_{k=1}^{N} K \Big[\frac{x_k-x}{h} \Big] m(x_k) \Big\}^2 \pi(x) dx \rightarrow_{D} d_{(1)}^2 L_B^2(1,0),  &&
\end{flalign}
where $d^2_{(1)}=\int_{\mathbb{R}} m^2(x) \pi(x) dx \int_{\mathbb{R}} K(s) ds$. The expression (\ref{App8}) can be obtained by a similar path to the expression (7.12) in Proposition 7.3 of \cite{WP2016}.

For a fixed $A>0$, we define the followings: $\hat{\eta_i} = \eta_i \, I(||\eta_i|| \leq A)$, $\tilde{\eta}_i = \hat{\eta}_i - \mathbb{E} (\hat{\eta}_i)$, $\mathring{\eta}_i= \eta_i-\tilde{\eta}_i$, 
$\tilde{u}_k = \sum_{j=0}^{\infty} \psi_j \tilde{\eta}_{k-j}$, and $\mathring{u}_k= \sum_{j=0}^{\infty} \psi_j \mathring{\eta}_{k-j}$. Let $u_k = \tilde{u}_k+ \mathring{u}_k$, $\tilde{u}_{1k}=\sum_{j=0}^{m_0} \psi_j \tilde{\eta}_{k-j}$, and $\tilde{u}_{2k}=\tilde{u}_k-\tilde{u}_{1k}$. 
Also, note that
\begin{flalign} \label{App9}
\sum_{1 \leq k < j \leq N} \tilde{u}_{1k} \tilde{u}_{1j} \Big\{\int_{\mathbb{R}} K \Big[\frac{x_k-x}{h} \Big] K \Big[\frac{x_j-x}{h} \Big] \pi(x) dx \Big\} = o_P \Big(\sqrt{N} \lambda^d h \Big), &&
\end{flalign}
which can be followed by a similar path from the proof of expression (8.18) in Lemma 8.3 of \cite{WP2016}. Therefore, we can conclude $T_{N}^{\text{III}}=o_P(\sqrt{N} \lambda^d h)$.
Now, it suffices to show $\{\sqrt{N} \lambda^d h\}^{-1} T_{N}^{\text{I}} \rightarrow_D d^2_{(0)} L_B(1,0)$. 
From expression (\ref{App4}) of Lemma \ref{Lemma}, we have
\begin{flalign*}
\mathbb{E} \int_{\mathbb{R}} \Big\{\sum_{k=1}^{N} K \Big[\frac{x_k-x}{h}\Big] \mathring{u}_k  \Big\}^2 \pi(x) dx  \leq C \, \mathbb{E} ||\mathring{\eta}_1||^2 \, \sqrt{N} \lambda^d h \leq C \, \mathbb{E} ||\eta_1 ||^2 I(||\eta_1 || > A) \sqrt{N} \lambda^d h, &&
\end{flalign*}
where $\mathbb{E}(\mathring{\eta}_1)=0$. Note that $\mathbb{E} ||\eta_1 ||^2 I(||\eta_1||>A) \rightarrow 0$ as $A \rightarrow \infty$, therefore it suffices to show
\begin{flalign*}
\Big\{ \sqrt{N} \lambda^d h \Big\}^{-1} T_{N}^{\text{I}} := 
\Big\{ \sqrt{N} \lambda^d h \Big\}^{-1} \int_{\mathbb{R}} \Big\{\sum_{k=1}^{N} K \Big[\frac{x_k-x}{h} \Big] \tilde{u}_k \Big\}^2 \pi(x) dx \rightarrow_D d^2_{(0)} L_B(1,0). &&
\end{flalign*}
Then, we have
 $T_{N}^{\text{I}}=T^{\text{I}(a)}_{N}+T^{\text{I}(b)}_{N}+T^{\text{I}(c)}_{N}$ where

\begin{flalign*}
T^{\text{I}(a)}_{N} & = \int_{\mathbb{R}} \Big\{\sum_{k=1}^{N} K \Big[\frac{x_k-x}{h} \Big] \tilde{u}_{1k} \Big\}^2 \pi(x) dx, &&\\[0.5em] 
T^{\text{I}(b)}_{N} & = \int_{\mathbb{R}} \Big\{ \sum_{k=1}^{N} \Big[\frac{x_k-x}{h} \Big] \tilde{u}_{2k} \Big\}^2 \pi(x) dx, \quad \text{and} 
&&\\[0.5em]
T^{\text{I}(c)}_{N} & = 2 \int_{\mathbb{R}} \Big\{ \sum_{1 \leq k < j \leq N} K \Big[\frac{x_k-x}{h} \Big] K \Big[\frac{x_j-x}{h}\Big] \tilde{u}_{1k} \tilde{u}_{2k} \Big\} \, \pi(x) dx. &&
\end{flalign*}

\noindent Now based on expression (\ref{App4}) in Lemma \ref{Lemma}, we have
\begin{flalign*}
\mathbb{E}(T^{\text{I}(b)}_{N}) & \leq C \, \sup_{x} \mathbb{E} \Big\{ \sum_{k=1}^{N} K \Big[\frac{x_k-x}{h} \Big] \tilde{u}_{2k} \Big\}^2 && 
\\[0.5em]
& \leq C \, \mathbb{E} || \eta_1||^2 \sqrt{N} \lambda^d h \, \Big\{\sum_{k=m_0}^{\infty} k^{1/4} (|\psi_{1k}|+|\psi_{2k}|) \Big\}^2, \quad \text{as} \quad N \rightarrow \infty \quad \text{and} \quad m_0 \rightarrow 0.
\end{flalign*}
Under Assumption \ref{ass:coef}, $\sum_{k=m_0}^{\infty}k^{1/4}(|\psi_{1k}|+|\psi_{2k}|)< \infty$. Therefore, we conclude $T^{\text{I}(b)}_{N}=o_P(\sqrt{N} \lambda^d h)$. Based on expression (\ref{App9}), $T^{\text{I}(c)}_{N}=o_P(\sqrt{N} \lambda^d h)$.
We can furtherly write $T^{\text{I}(a)}_{N}$ as
\begin{flalign*}
T^{\text{I}(a)}_{N} &= \int_{\mathbb{R}} \Big\{ \sum_{k=1}^{N} K \Big[\frac{x_k-x}{h} \Big] \tilde{u}_{1k} \Big\}^2 \pi(x) dx
 = \mathbb{E}(\tilde{u}_{10}^2) \sum_{k=1}^{N} \int_{\mathbb{R}} K^2 \Big[\frac{x_k-x}{h} \Big] \pi(x) dx && \\[0.5em]
    & \quad + \sum_{k=1}^{N} (\tilde{u}^2_{1k}- \mathbb{E}(\tilde{u}_{1k}^2)) \int_{\mathbb{R}} K^2 \Big[\frac{x_k-x}{h} \Big] \pi(x) dx && \\[0.5em]
   & \quad + 2 \sum_{1 \leq k < j \leq N} \tilde{u}_{1k} \tilde{u}_{1j} \int_{\mathbb{R}} K \Big[\frac{x_k-x}{h} \Big] K \Big[\frac{x_j-x}{h} \Big] \pi(x) dx && \\[0.5em]
&  = R^{\text{I}(a)}_{N} + R^{\text{II}(a)}_{N}+R^{\text{III}(a)}_{N}.   
\end{flalign*}

\noindent From
\begin{flalign*}
\sum_{k=1}^{N} \Big(\tilde{u}^2_{1k}- \mathbb{E}(\tilde{u}^2_{1k}) \Big) \int_{\mathbb{R}} K^2 \Big[\frac{x_k-x}{h} \Big] \pi(x) dx = O_P \Big(\{\sqrt{N} \lambda^d h\}^{1/2}\Big), &&
\end{flalign*}
which can be followed by a similar path from the expression (8.17) in Lemma 8.3 of \cite{WP2016} and expression (\ref{App9}), we can conclude $|R^{\text{II}(a)}_{N}|+|R^{\text{III}(a)}_{N}|=o_P (\sqrt{N} \lambda^d h)$. Eventually, by virtue of $\mathbb{E}(\tilde{u}_{10}^2) \rightarrow \mathbb{E}(\tilde{u}_0^2)$, we only need to prove
\begin{flalign} \label{App10}
\Big\{ \sqrt{N} \lambda^d h \Big\}^{-1} R^{\text{I}(a)}_{N} \rightarrow_D \mathbb{E}(\tilde{u}_0^2) \int_{\mathbb{R}} K^2(x) dx \int_{\mathbb{R}} \pi(s) ds \, L_B(1,0). &&
\end{flalign}
If we define $\int_{\mathbb{R}} g_h(y)dy=\int_{\mathbb{R}} K^2(z)dz \int_{\mathbb{R}} \pi(x) dx$ such that $|g_h(y)| \leq C \int_{\mathbb{R}} |\pi(x)| dx < \infty$, then (\ref{App10}) could be followed from the proof of Theorem 2.1 in \cite{WP2009a}, and we omit it. Therefore, the proof is complete.
\QEDA

\vspace{0.25cm}

\noindent \textbf{Proof of Theorem \ref{Theo.normalizedtest}.}
The proof is similar to the proof of Theorem 3.2 in \cite{WP2016}. Therefore, we give an outline. We first decompose $T_{N}$ under $H_A$, i.e. $y_k=g(x_k,\theta_0) + \rho_N m(x_k)+u_k$ as follows:
\begin{flalign*}
T_{N} & = \int_{\mathbb{R}} \Big\{\sum_{k=1}^{N} K \Big[\frac{x_k-x}{h} \Big] \Big[g(x_k,\theta_0) + \rho_N m(x_k) + u_k - g(x_k,\hat{\theta}_N) \Big] \Big\}^2 \pi(x) dx && \\[0.5em]
& = T_{N}^{\text{I}}+T_{N}^{\text{II}}+T_{N}^{\text{III}}+2 \rho_N T_{N}^{\text{IV}} + \rho_N^2 T_{N}^{\text{V}},
\end{flalign*}
where
\begin{flalign*}
T_{N}^{\text{IV}} & := \int_{\mathbb{R}} \Big\{\sum_{1 \leq k<j \leq N} K \Big[\frac{x_k-x}{h}\Big] K\Big[\frac{x_j-x}{h} \Big] m(x_k) \Big[u_j-\Big(g(x_j,\hat{\theta}_N)-g(x_j,\theta_0)\Big) \Big] \Big\} \, \pi(x) dx, \, \text{and}
&&\\[0.5em]
T_{N}^{\text{V}} & :=\int_{\mathbb{R}} \Big\{\sum_{k=1}^{N} K \Big[\frac{x_k-x}{h} \Big] m(x_k) \Big\}^2 \pi(x) dx, 
\end{flalign*}
such that $|T_{N}^{\text{IV}}| \leq [T_{N}^{\text{I}}+T_{N}^{\text{II}}+T_{N}^{\text{III}}]^{1/2} \times [T_{N}^{\text{V}}]^{1/2}$ by H\"{o}lder's inequality. The remaining terms have already been defined in the proof of Theorem \ref{Theo.test}.
From the proof of Theorem \ref{Theo.test}, $T_{N}^{\text{I}}+T_{N}^{\text{II}}+T_{N}^{\text{III}}=O_P(\sqrt{N} \lambda^d h)$. Following part (iv) of the Proposition \ref{Prop9} $\{ \sqrt{N} \lambda^d h \}^{-1} \sum_{k=1}^{N} K (\frac{x_k}{h})$ $\rightarrow_{D} L_{B}(1,0)$ under the assumption of $\int_{\mathbb{{R}}} K(x) dx=1$. Therefore, from the continuous mapping theorem and expression (\ref{App1}), we have $\{ \sqrt{N} \lambda^d h \}^{-2} T_{N}^{\text{V}} \rightarrow_{D} d_{(1)}^2 \, L_{B}^2(1,0)$, where $d_{(1)}^2=\int_{\mathbb{R}} m^2(x) \pi(x) dx$, and $P(0 < L^2_{B}(1,0)< \infty)=1$. We conclude that for any $T_0>0$ and as $N \rightarrow \infty$, the following holds
\begin{flalign*}
\lim\limits_{N \rightarrow \infty} P \Big(\{ \sqrt{N} \lambda^d h \}^{-1} T_{N} \geq T_0 \Big) \geq  \lim\limits_{N \rightarrow \infty} P \Big(\{ \sqrt{N} \lambda^d h \}^{-2} T_{N}^{\text{V}} \geq \epsilon_N^{-1/4} \Big) = 1, &&
\end{flalign*}
under the assumption of $\epsilon_N:= \sqrt{N} \lambda^d h \rho_N^2 \rightarrow \infty$.

\noindent Note that $\{ \sqrt{N} \lambda^d h \}^{-2} T_{N}^{V} \rightarrow_{D} d_{(1)}^2 \, L_{B}^2(1,0)>0$ a.s., which confirms the divergence of $\{ \sqrt{N} \lambda^d h \}^{-1} T_{N}$ and the consistency of the test under $H_A$.
\QEDA

\subsection{A.3 \quad Propositions and Their Proofs} \label{App.A.3}
\renewcommand{\theequation}{A.\arabic{equation}}
\renewcommand{\theproposition}{A.\arabic{proposition}}

In this section, we provide several propositions. These propositions are required for the proofs of theorems \ref{Theo.f}, \ref{Theo.test}, and \ref{Theo.normalizedtest} given in the paper.

\begin{proposition} \label{Prop1}
Let $x_{k,N}=x_k/d_N$. A random array $\{ x_{k,N} \}_{k \geq 1, N \geq 1}$ is called strong smooth.

\begin{proof}
We verify this proposition from the Definition \ref{def.smootharray}. Assume $\{x_{k,N}\}_{k \geq 1, N \geq 1}$ is a random triangular array such that $x_{0,N} \equiv 0$. Let $d_{l,k} \equiv \lambda_{l-k}^{-d} \, (l-k)^{1/2}$, therefore
\begin{flalign*}
d_{l,k,N} \equiv \frac{d_{l,k}}{d_N} \propto \Big(\frac{\lambda_{l-k}}{\lambda_N} \Big)^{-d} \,
\sqrt{\frac{l-k}{N}}, \qquad N \lambda_N \rightarrow \infty. &&
\end{flalign*}
Recall that $\lambda_N \equiv \lambda>0$ is a sample size dependent parameter.

\begin{itemize}
\item[(a)] $\inf_{(l,k) \in \Omega_N(\eta)} d_{l,k,N} \equiv
\inf_{(l,k)\in \Omega_N(\eta)}(\lambda_{l-k}/\lambda_N)^{-d} \sqrt{(l-k)/N} \geq \sqrt{\frac{\eta N}{N}} \lambda_{\eta N/N}^d>0$.
\end{itemize}
\begin{flalign*}
\text{(i)} \, & \lim_{\eta \rightarrow 0} \lim_{N \rightarrow \infty} \frac{1}{N} \sum_{l=(1-\eta)N}^{N}\Big\{\Big(\frac{\lambda_l}{\lambda_N}\Big)^{-d} \sqrt{l/N} \Big\}^{-1} && \\[0.5em] 
& \quad \leq \lim_{\eta \rightarrow 0} \lim_{N \rightarrow \infty} \frac{1}{\sqrt{N} \lambda_N^d} \int_{(1-\eta)N}^{N} \frac{\lambda_l^d}{\sqrt{l}} dl 
&& \\[0.5em]
& \quad \leq \lim_{\eta \rightarrow 0} \lim_{N \rightarrow \infty} \frac{1}{\sqrt{N} \lambda_N^d} \Big\{\sqrt{N} \lambda_N^d- \sqrt{(1-\eta)N} \lambda_{(1-\eta)N}^d\Big\}=0. 
\end{flalign*}

\begin{flalign*}
\text{(ii)} \, & \lim_{\eta \rightarrow 0} \lim_{N \rightarrow \infty} \frac{1}{N} \max_{0 \leq k \leq (1-\eta)N} \sum_{l=k+1}^{k+\eta N} \Big\{ \Big( \frac{\lambda_{l-k}}{\lambda_N} \Big)^{-d}\sqrt{(l-k)/N} \Big\}^{-1} 
&& \\[0.5em]
&  \leq \lim_{\eta \rightarrow 0} \lim_{N \rightarrow \infty} \frac{1}{\sqrt{N} \lambda_N^d} \max\limits_{0 \leq k \leq (1-\eta)N} \int_{k+1}^{k+\eta N} \frac{\lambda_{l-k}^d}{\sqrt{l-k}} dl 
&& \\[0.5em]
\qquad & = \lim_{\eta \rightarrow 0} \lim_{N \rightarrow \infty} \frac{1}{\sqrt{N} \lambda_N^d} \Big\{\sqrt{\eta N} \lambda_{\eta N}^d - 1 \Big\}=0.
\end{flalign*}
\begin{flalign*}
\text{(iii)} \, & \limsup\limits_{N \rightarrow \infty} \frac{1}{N} \max\limits_{0 \leq k \leq N-1} \sum\limits_{l=k+1}^{N} 
\Big\{\Big(\frac{\lambda_{l-k}}{\lambda_N} \Big)^{-d} \sqrt{(l-k)/N} \Big\}^{-1} &&\\[0.5em]
& \quad \leq \limsup_{N \rightarrow \infty} \frac{1}{\sqrt{N} \lambda_N^d} \max\limits_{0 \leq k \leq N-1} \int_{k+1}^{N} \frac{\lambda_{l-k}^d}{\sqrt{l-k}} dl && \\[0.5em]
& \quad \leq \limsup\limits_{N \rightarrow \infty} \frac{1}{\sqrt{N} \lambda_N^d} \max\limits_{0 \leq k \leq N-1} \Big\{\sqrt{N-k} \lambda_{N-k}^d -1  \Big\} && \\[0.5em]
& \quad = \limsup\limits_{N \rightarrow \infty} \Big\{1- \frac{1}{\sqrt{N} \lambda_N^d} \Big\} = 1  < \infty.
\end{flalign*}
\begin{itemize}
\item[(b)] If we define $f_{l,k,N}(t)=\mathbb{E}\Big(e^{i t [x_{l,N}-x_{k,N}]/d_{l,k,N}} \Big)$, then
\begin{flalign*}
\sup\limits_{x \in \mathbb{R}} \Big|h_{l,k,N}(x)-n(x) \Big| \leq C \int_{\mathbb{R}} \Big|f_{l,k,N}(t)-e^{-t^2/2} \Big| dt \rightarrow 0, &&
\end{flalign*}
\end{itemize}

\noindent where $n(x)=e^{-x^2/2}\big/\sqrt{2 \pi}$ is the density of standard normal (see proof of Corollary 2.2 in \cite{WP2009a} for a similar argument). Thus, all the conditions given in Definition \ref{def.smootharray} hold, and ${x_{k,N}}$ is a strong smooth array. 
The alternative way to prove this proposition is to use expression (\ref{strongsmooth1}), where $\{\zeta(k)\}_{ k\in\mathbb{Z} }$ is a sequence of i.i.d. random variables with $\mathbb{E}(\zeta(0))=0$  and $\mathbb{E}(\zeta^2(0))=1$; see expression (A.7) in \citet{Sabzikar2020}. Secondly, the triangular array $N^{-1/2} \sum_{k=1}^{N} \zeta(k)$ satisfies the strong smooth conditions; see \cite{WP2009a}. This together with \eqref{strongsmooth1} can complete the proof of Proposition \ref{Prop1}.
\end{proof}

\end{proposition}

\begin{proposition} \label{Prop2}
Let
\begin{flalign*}
S_N(t) =\frac{1}{[\sqrt{N}h]^{1/2}} \sum\limits_{k=1}^{\lfloor Nt \rfloor} u_k K\Big[\frac{x_k-x}{h}\Big], \quad \text{and} \quad
\psi_N(t)  =\frac{1}{\sqrt{N}h} \sum\limits_{k=1}^{\lfloor Nt \rfloor} u^2_k K^2\Big[\frac{x_k-x}{h}\Big]. &&
\end{flalign*}
For any fixed $0 \leq t \leq 1$, $S_N(t)$, $S^2_N(t)$, and $\psi_N(t)$ are uniformly integrable.

\begin{proof}
The proof is similar to Proposition 7.3 in \citet{WP2009b};  
hence, we only give an outline. If we let
\begin{flalign*}
\psi_N''(t)=\frac{1}{\sqrt{N}h} \sum_{k=1}^{\lfloor Nt \rfloor} K^2\Big[\frac{x_k-x}{h}\Big]\mathbb{E}(u_k^2), &&
\end{flalign*}
for $0 \leq t \leq 1$ and define $r_y(t):=K^2(\frac{y}{h}+t)$, then $\sup_{N} \mathbb{E}[\psi''_N(t)]^2 < \infty$ as follows
\begin{flalign*}
\mathbb{E}[\psi''_N(t)]^2 & = \mathbb{E} \Big\{\frac{1}{\sqrt{N}h} \sum_{k=1}^{\lfloor Nt \rfloor} K^2\Big[\frac{x_k-x}{h}\Big]\mathbb{E}(u_k^2) \Big\}^2 &&\\[0.5em]
& \leq \frac{C}{N h^2} \Big\{\sum_{k=1}^{N} \mathbb{E}\Big\{ K^4\Big[\frac{x_k-x}{h}\Big]\Big\}
+ 2 \sum_{1 \leq k< l \leq N} \mathbb{E} \Big\{ K^2 \Big[\frac{x_k-x}{h} \Big] K^2\Big[\frac{x_l-x}{h} \Big] \Big\} \Big\} 
&&\\[0.5em]
& \leq \frac{C}{Nh^2} \Big\{ \sum_{k=1}^{N} \sup_y \mathbb{E} r^2_y\Big(\frac{x'_{0,k}}{h}\Big)+ 2 \sum_{1 \leq k < l \leq N} \sup_y \mathbb{E} r_y \Big(\frac{x'_{0,k}}{h} \Big) \sup_{y} \mathbb{E} r_y \Big(\frac{x'_{k,j}}{h} \Big) \Big\} &&\\[0.5em]
& \leq \frac{C}{Nh^2} \Big\{\sum_{k=1}^{N} \frac{h}{\sqrt{k}} + 2 \sum_{1 \leq k < l \leq N} \frac{h^2}{\sqrt{k} \sqrt{l-k}} \Big\} < \infty, &&
\end{flalign*}
uniformly on $N$. The last line could be followed by a similar path from the proof of Lemma 7.1 part (a) in \cite{WP2009b}. Note that $\sup_{0 \leq t \leq 1} \mathbb{E}|\psi_N(t)-\psi_N''(t)|^2=o(1)$, which can be verified by a similar argument given in Proposition \ref{Prop6}. This implies $\sup_N\mathbb{E}[\psi_N(t)]^2< \infty$, and therefore $\psi_N(t)$ is uniformly integrable. Furthermore, we can write
\begin{flalign*}
|\psi_N(t)-S^2_N(t)|=\frac{2}{\sqrt{N} h}\sum\limits_{1 \leq k < l \leq \lfloor Nt \rfloor } u_k u_l K\Big[\frac{x_k-x}{h}\Big] K\Big[\frac{x_l-x}{h}\Big], &&
\end{flalign*}
which implies $\sup_{0 \leq t \leq 1}\mathbb{E}|\psi_N(t)-S^2_N(t)|=o(1)$ and can be verified by a similar argument given in Proposition \ref{Prop6}. Therefore, $S^2_N(t)$ is uniformly integrable, which implies that $S_N(t)$ is uniformly integrable.
\end{proof}

\end{proposition}

\begin{proposition} \label{Prop3}
Let $\Theta_{1N}:=\sum_{k=1}^{N}u_k K[\frac{x_k-x}{h}]$. Then, $\Theta_{1N}=O_P(\{\sqrt{N}h\}^{1/2})$.

\begin{proof}

Based on Proposition \ref{Prop2}, uniformly in $N$, we have $\mathbb{E}S^2_N(1) \leq C$, where $C$ is a generic constant. Therefore, based on the  Markov's inequality
\begin{flalign*}
P(\Theta_{1N} \geq \{\sqrt{N}h\}^{1/2}) \leq \frac{\mathbb{E}(\Theta^2_{1N})}{\sqrt{N}h} = \mathbb{E}S^2_N(1) \leq C. &&
\end{flalign*}
As a result, $\Theta_{1N}=O_P(\{\sqrt{N}h\}^{1/2})$.
\end{proof}
\end{proposition}

\begin{proposition} \label{Prop4}
Let $\Theta_{2N} :=\sum_{k=1}^{N}[f(x_k)-f(x)]K[(x_k-x)/h]$. Then, 
$\Theta_{2N}=O_P(Nh^{\gamma+1}/d_N)$ where $d_N= \sqrt{N}/\lambda^d$.

\begin{proof}
\begin{flalign*}
\mathbb{E}|\Theta_{2N}| & \leq \sum_{k=1}^{N} \mathbb{E} \Big\{|f(x_{k,N}d_N)-f(x)| K\Big[\frac{x_{k,N}d_N-x}{h}\Big] \Big\} && \\[0.5em]
& = \sum_{k=1}^{N} \int_{\mathbb{R}} \Big\{|f(d_N d_{k,0,N}y)-f(x)| \times K\Big(\frac{d_N d_{k,0,N}y}{h}-\frac{x}{h}\Big)\Big\} h_{k,0,N}(y) dy 
&& \\[0.5em]
& \leq \frac{h}{d_N} \sum_{k=1}^{N} \frac{1}{d_{k,0,N}} \int_{\mathbb{R}} \{|f(hy+x)-f(x)| K(y)\}dy && \\[0.5em]
& \leq \frac{Nh^{\gamma+1}}{d_N} \frac{1}{N} \sum_{k=1}^{N} \frac{1}{d_{k,0,N}} \int_{\mathbb{R}} K(s) f_1(s,x) ds \leq A \frac{N h^{\gamma+1}}{d_N},
\end{flalign*}
where $A$ is a constant and $\gamma \in (0,1]$. The last line of proof follows from Assumption \ref{ass:function} and note that we have considered $x_{k,N}=d_{k,0,N}y$ and $d_N d_{k,0,N} y = hy+x$. Therefore, $\Theta_{2N}=O_P(Nh^{\gamma+1}/d_N)$.
\end{proof}
\end{proposition}

\begin{proposition} \label{Prop5}
Let $\Theta_{3N}:= \{\sum_{k=1}^{N} K[(x_k-x)/h]\}^{-1}$. Then, $\Theta_{3N}=o_P (\{ \sqrt{N} \lambda^d h \}^{-1})$.

\begin{proof} 
From Proposition \ref{Prop4}, we can easily conclude this under the assumption of $\sqrt{N} \lambda^d h \rightarrow \infty$.
\end{proof}
\end{proposition}

\begin{proposition} \label{Prop6}
Assume $h \rightarrow 0$ and $\sqrt{N} \lambda^d h \rightarrow \infty$. Let
\begin{flalign*}
T_{iN}(t) :=\{\sqrt{N} \lambda^d h\}^{-1} \sum_{k=1}^{\lfloor Nt \rfloor} K^i\Big[\frac{x_k-x}{h}\Big], \quad \text{and} \quad
T'_{iN}(t) := \{\sqrt{N} \lambda^d h\}^{-1} \sum_{k=1}^{\lfloor Nt \rfloor} K^2\Big[\frac{x_k-x}{h}\Big] u^2_k. &&
\end{flalign*}
Then, $T_{iN}(t)\rightarrow_{D} d_i L_{B}(t,0)$ and $T'_{iN}(t) \rightarrow_{D} d_0^2 L_{B}(t,0)$ for $i=1,2$,
where $d_i=\int_{\mathbb{R}}K^i(s)ds$ and $d_0^2=\mathbb{E}(u^2_{0}) \int_{\mathbb{R}}K^2(s)ds$.

\begin{proof}
Let's define $x_{k,N}:=\hat{\xi}_N(k/N)$ for $1 \leq k \leq N$ such that
$\hat{\xi}_N(t) \Rightarrow B(t)$ on $D[0,1]$ following Theorem 4.3 in \cite{sabzikar2018invariance}. There exists an equivalent process $\hat{\xi}_N(k/N)$ of $\xi_N(k/N)$ i.e. $\hat{\xi}_N(k/N)=_{D}\xi_N(k/N)$
such that $\sup_{0 \leq t \leq 1} |\hat{\xi}_N(t)-B(t)|=o_P(1)$.
Now, apply part (i) of Proposition \ref{Prop9} by setting $c_N=\frac{d_N}{h}$ and $g(t)=K^i[t-\frac{x}{h}]$ for $i=1,2$; then we have
\begin{flalign*} 
\sup\limits_{0 \leq t \leq 1} \Big|\{\sqrt{N} \lambda^d h\}^{-1} \sum_{k=1}^{\lfloor Nt \rfloor} K^i\Big[\frac{d_N \hat{\xi}_N(k/N)-x}{h} \Big] - L_{B}(t,0) \int_{\mathbb{R}}K^i(s)ds \Big| \rightarrow_P 0, &&
\end{flalign*}
as $c_N \rightarrow \infty$, $\frac{c_N}{N} \rightarrow 0$, and $N \rightarrow \infty$. 
This together with the fact that $\hat{\xi}_N(\frac{k}{N})=_D \xi_N(\frac{k}{N})=\frac{x_k}{d_N}$ for $1 \leq k \leq N$ imply for $N \geq 1$, $T_{iN}(t)$ is f.d.d. for $d_i L_B(t,0)$.
Additionally, $T_{iN}(t)$ is tight for $N \geq 1$ following Proposition \ref{Prop7}, i.e.
\begin{flalign*}
\max\limits_{1 \leq k \leq N} \Big|K^i\Big[\frac{x_k-x}{h}\Big] \Big|=o_P\Big(\sqrt{N} \lambda^d h\Big), &&
\end{flalign*}
under the assumption of $\sup_{0 \leq t \leq 1}|x_{\lfloor Nt \rfloor,N}- B(t)|=o_P(1)$.
Therefore,
\begin{flalign*}
T_{iN}(t) \rightarrow_{D} d_i \, L_{B}(t,0) \quad \text{for} \quad i=1,2. &&
\end{flalign*}
Now, in order to prove
$T'_{iN}(t):=\{\sqrt{N} \lambda^d h\}^{-1} \sum_{k=1}^{\lfloor Nt \rfloor}K^2[\frac{x_k-x}{h}]u^2_k \rightarrow_{D} d_0^2 \, L_{B}(t,0)$, we define
\begin{flalign*}
T_{iN}''(t) := \{\sqrt{N} \lambda^d h\}^{-1} \sum_{k=1}^{\lfloor Nt \rfloor} K^2\Big[\frac{x_k-x}{h}\Big] \mathbb{E}u^2_k, &&
\end{flalign*}
and need to show $\sup_{0 \leq t \leq 1}\mathbb{E}|T'_{iN}(t)-T''_{iN}(t)|^2=o(1)$.
Based on the Preliminary given in Section A.1, $x_k=x^*_{0,k}+x'_{0,k}$ for $s=0$, we have:
\begin{flalign*}
 & \mathbb{E}\Big[\Big|T'_{iN}(t)-T''_{iN}(t) \Big|^2 \Big| \zeta(0), \zeta(-1), ... \Big] && \\[0.5em]
 & \quad \leq \Big\{\frac{1}{\sqrt{N} \lambda^d h}\Big\}^2 \sup\limits_{y, 1 \leq M \leq N} \mathbb{E}\Big[\sum_{k=1}^{M}K^2\Big[\frac{y+x'_{0,k}}{h}\Big] (u^2_k-\mathbb{E}(u^2_k))^2 \Big]^2 \quad \text{a.s.} && \\[0.5em]
& \quad \leq \Big\{\frac{1}{\sqrt{N} \lambda^d h}\Big\}^2 \sup\limits_{y} \Big[ \sum_{k=1}^{N} \mathbb{E}r^2\Big(\frac{x'_{0,k}}{h}\Big) g^2(u_k)
 + 2 \sum\limits_{1 \leq k < l \leq N} \Big|\mathbb{E}r\Big(\frac{x'_{0,k}}{h}\Big) r\Big(\frac{x'_{0,l}}{h}\Big) g(u_k) g_1(u_l) \Big|\Big]
\end{flalign*}
as $g(t)=t^2-\mathbb{E}u^2_k$, $g_1(t)=t^2-\mathbb{E}u^2_l$, and $r(t)=K^2(\frac{y}{h}+t)$. Now based on the last part of proof of Proposition 7.2 in \cite{WP2009b} and as $\{ \sqrt{N} \lambda^d h \}^{-1} \rightarrow 0$, we have
\begin{flalign*}
\mathbb{E}\Big[\Big|T'_{iN}(t)-T''_{iN}(t) \Big|^2 | \zeta(0), \zeta(-1), \cdots \Big] \rightarrow 0 \qquad \text{a.s.} &&
\end{flalign*}
Therefore, $T'_{iN}(t)$ and $T''_{iN}(t)$ have the same f.d.d. Since $\mathbb{E}u^2_k=\mathbb{E}u^2_{m_0}$ for $k \geq m_0$ and $T''_{iN}(t) \rightarrow_{D} d_0^2 L_{B}(t,0)$ from the modification of expression (\ref{App10}), we can easily show that  $T'_{iN}(t)$ is tight from Proposition \ref{Prop7}.
Therefore, $T'_{iN}(t) \rightarrow_{D} d_0^2 L_{B}(t,0)$ on $D[0,1]$.
\end{proof}
\end{proposition}

\begin{proposition} \label{Prop7}
$S_N(t) =\{\sqrt{N}h\}^{-1/2} \sum_{k=1}^{\lfloor Nt \rfloor} u_k K[\frac{x_k-x}{h}]$ is tight on $D[0,1]$.

\begin{proof}
According to Theorem 4 of \cite{billingsley1974conditional}, we need to show
\begin{flalign} \label{App12}
\max\limits_{1 \leq k \leq N} \Big|u_k K\Big[\frac{x_k-x}{h}\Big] \Big| = o_P\Big(\{\sqrt{N}h\}^{1/2}\Big). &&
\end{flalign}
The proof follows from Proposition 7.4 in \cite{WP2009b}, and therefore we give an outline. There exists a sequence of $\alpha_N(\nu, \kappa)$ such that $\lim_{\kappa \rightarrow 0} \limsup_{N \rightarrow \infty} \alpha_{N}(\nu, \kappa)=0$ for each $\nu > 0$ and $0 \leq t_1 \leq t_2 \leq \cdots \leq t_m \leq t \leq 1$ where $t-t_m \leq \kappa$. We have
\begin{flalign*}
P \Big[\Big| S_N(t)-S_N(t_m) \Big| \geq \nu \Big|S_N(t_1) ... S_N(t_m) \Big] \leq \alpha_N(\nu, \kappa) \quad \text{a.s.} &&
\end{flalign*}
Therefore, we need to show
\begin{flalign*}
\sup\limits_{|t-t_m| \leq \kappa} P\Big( \Big| \sum_{k=\lfloor Nt_m \rfloor+1}^{\lfloor Nt \rfloor} u_k K\Big[\frac{x_k-x}{h}\Big] \Big| \geq \nu \, N^{1/2} \lambda^{-d} \Big| \zeta(\lfloor Nt_m \rfloor),\zeta({\lfloor Nt_m \rfloor}-1), \cdots, \zeta(1) \Big) \leq \alpha_N(\nu, \kappa). &&
\end{flalign*}
We may choose $\alpha_N(\nu,\kappa)$ such that
\begin{flalign*}
\alpha_N(\nu, \kappa) & := \frac{1}{\nu^2} \frac{1}{\sqrt{Nh^2}} \sup\limits_{y, 0 \leq t \leq \kappa} \mathbb{E} \Big\{\sum_{k=1}^{\lfloor Nt \rfloor} u_k K\Big[\frac{y+x'_{0,k}}{h}\Big] \Big\}^2 && \\[0.5em]
& \leq \frac{1}{\nu^2} \frac{1}{\sqrt{N h^2}} \sup\limits_{y} \sum_{k=1}^{\lfloor N \kappa \rfloor} \mathbb{E} \Big\{u^2_k K^2\Big[\frac{y+x'_{0,k}}{h}\Big] \Big\} && \\[0.5em]
& \quad + \frac{1}{\nu^2} \frac{1}{\sqrt{Nh^2}} \sup\limits_{y} 2 \sum\limits_{1 \leq k < l \leq \lfloor N\kappa \rfloor} \Big|\mathbb{E} \Big\{ u_k u_l K\Big[\frac{y+x'_{0,k}}{h}\Big] K\Big[\frac{y+x'_{0,l}}{h}\Big] \Big\} \Big|.
\end{flalign*}
Following expression (\ref{App2}) in Lemma \ref{Lemma}, we can show
\begin{flalign*}
\alpha_N(\nu, \kappa) \leq  \frac{C}{\nu^2 \sqrt{N h^2}} \times \frac{\lfloor N\kappa \rfloor h \lambda^d}{(\lfloor N\kappa \rfloor)^{1/2}} \Big[1+h \sqrt{j}\Big] = \frac{C}{\nu^2 \sqrt{N}} \times (\lfloor N\kappa \rfloor)^{1/2} \lambda^d \Big[1+h \sqrt{j}\Big], &&
\end{flalign*}
which yields $\lim_{\kappa \rightarrow 0} \limsup_{N \rightarrow \infty} \alpha_N(\nu,\kappa)=0$.
Note that 
\begin{flalign*}
\max_{1 \leq k \leq N} \Big| u_k K\Big[\frac{x_k-x}{h}\Big] \Big| \leq 
\Big\{\sum_{j=1}^{N}u_j^4 K^4 \Big[\frac{x_j-x}{h}\Big] \Big\}^{1/4}. &&
\end{flalign*}
Following Lemma \ref{Lemma}, one can show that $\mathbb{E}u_j^4 K^4[\frac{x_j-x}{h}] \leq C h/\sqrt{j}$, which concludes (\ref{App12}). As a result, $S_N(t)$ is tight.
\end{proof}
\end{proposition}

\begin{proposition} \label{Prop8}
Let $\Theta_{2N}:=\sum_{k=1}^{N}(f(x_k)-f(x))K[\frac{x_k-x}{h}]$ and $\Theta_{3N}:= 1/\sum_{k=1}^{N} K[\frac{x_k-x}{h}]$. Then, $\Theta_{2N} \sqrt{\Theta_{3N}} \rightarrow_P 0$.

\begin{proof}
First, note that
\begin{flalign*}
\Theta_{2N} \sqrt{\Theta_{3N}}=O_P \Big(\frac{Nh^{\gamma+1}}{d_N} \Big) o_P \Big(\{ \sqrt{N} \lambda^d h \}^{-1/2} \Big)=o_P \Big((N/d_N)^{1/2} h^{\gamma+1-1/2}\Big), &&
\end{flalign*}
 which is equal to $o_P ((N/d_N)^{1/2}h^{\gamma+1/2})$. Under assumptions $N h^{2 \gamma+1}/d_N \rightarrow 0$ and $h \rightarrow 0$ as $N \rightarrow \infty$, we conclude $\Theta_{2N} \sqrt{\Theta_{3N}} \rightarrow_P 0$.
\end{proof}
\end{proposition}

\begin{proposition} \label{Prop9}
Assume $g(x)$ is a bounded function such that $\int_{\mathbb{R}}|g(x)|dx < \infty$. Let $c_N:=\frac{d_N}{h}$ such that $c_N \rightarrow \infty$ and $\frac{c_N}{N} \rightarrow 0$. Then, we have

\begin{itemize}
\item[(i)] $\frac{c_N}{N} \sum_{k=1}^{N} g(c_N x_{k,N}) \rightarrow_D \int_{\mathbb{R}} g(x) dx \, L_{B}(1,0)$,
\item[(ii)] $\frac{d_N}{N} \sum_{k=1}^{N} |g(x_k)| (1+|u_k|) = O_P(1)$,
\item[(iii)] $(\frac{d_N}{N})^{1/2} \sum_{k=1}^{N}g(x_k)u_k=
O_P(1)$, \quad \text{and}
\item[(iv)] For $\sqrt{N} \lambda^d h \rightarrow \infty$, we have
\begin{flalign*}
 & \Big\{( \sqrt{N} \lambda^d h )^{-1/2} \sum_{k=1}^{N} K[(x_k-x)/h]u_k, 
 ( \sqrt{N} \lambda^d h )^{-1} \sum_{k=1}^{N} g\Big[\frac{x_k-x}{h}\Big] \Big\} && \\
 & \quad \rightarrow_D \Big\{d_0 N(0,1) L_{B}^{1/2}(1,0), \int_{\mathbb{R}}g(s) ds L_{B}(1,0) \Big\} \quad \text{on} \quad D^2[0,1].
\end{flalign*}
\end{itemize}

\begin{proof}
The proof of (i) is similar to Theorem 2.1 in \citet{WP2009a} and Lemma 7 of \cite{jeganathan2004convergence}. The proof of (ii) follows from the expression (7.2) in Proposition 7.1 of \cite{WP2016}, and the proof of (iii) follows from expression (\ref{App3}) in Lemma \ref{Lemma}. The proof of (iv) is similar to the proof of expression (3.8) in \cite{WP2009b} with some minor modifications.
\end{proof}
\end{proposition}

\clearpage

\section{Appendix B: Additional Simulation Results} \label{App.B}
\setcounter{table}{0}
\renewcommand{\thetable}{B.\arabic{table}}
\setcounter{figure}{0}
\renewcommand{\thefigure}{B.\arabic{figure}}

In this section, we provide additional numerical results for the regression function and the test statistic when the sample size is small. Also, we illustrate the values of size and power for the test statistic in the semi-long memory case with $d \geq 0.5$.

\begin{figure}[ht]
\centering
\begin{tabular}{ cc }
 \includegraphics[width=0.45\textwidth]{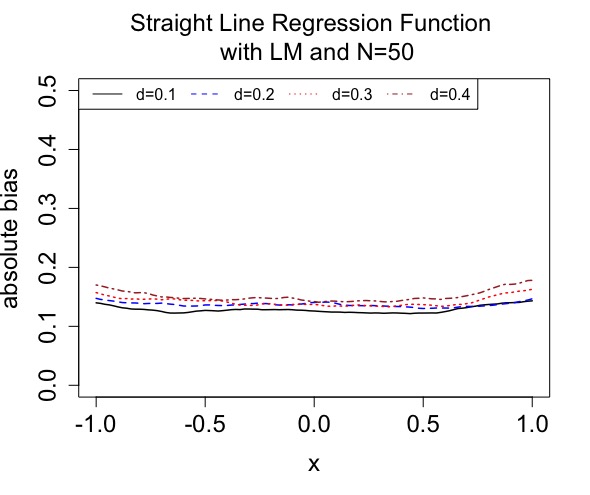} &
 \includegraphics[width=0.45\textwidth]{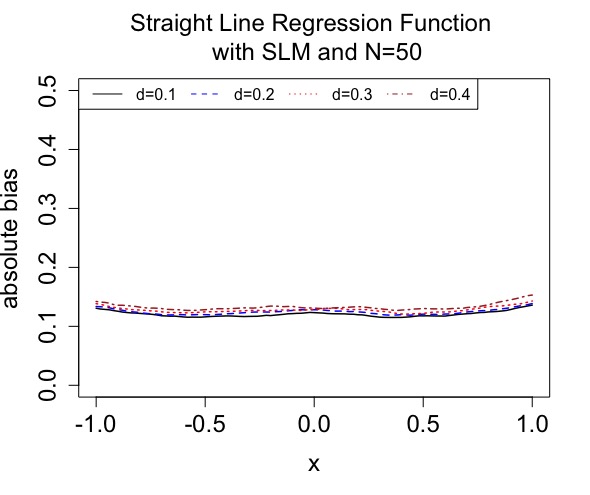}\\
  \includegraphics[width=0.45\textwidth]{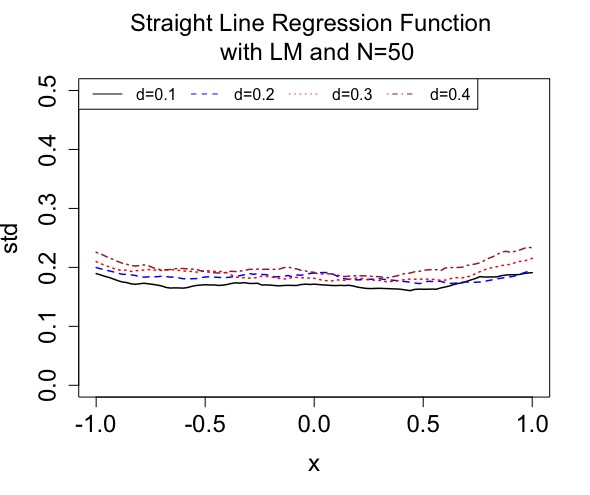} &
 \includegraphics[width=0.45\textwidth]{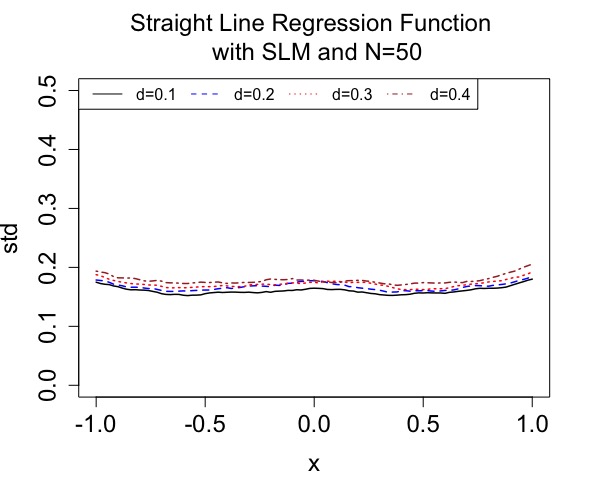} \\
  \includegraphics[width=0.45\textwidth]{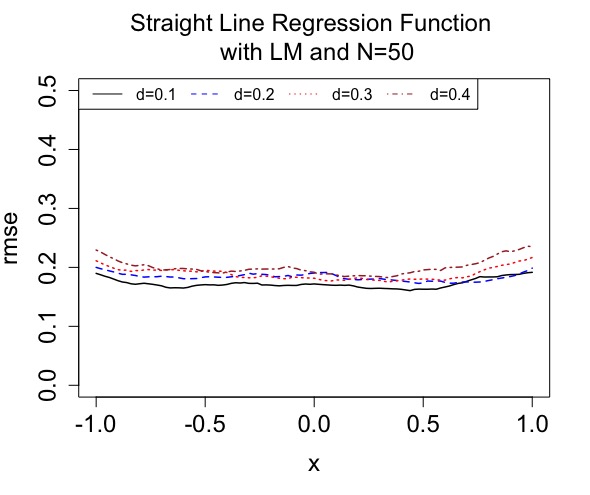} &
 \includegraphics[width=0.45\textwidth]{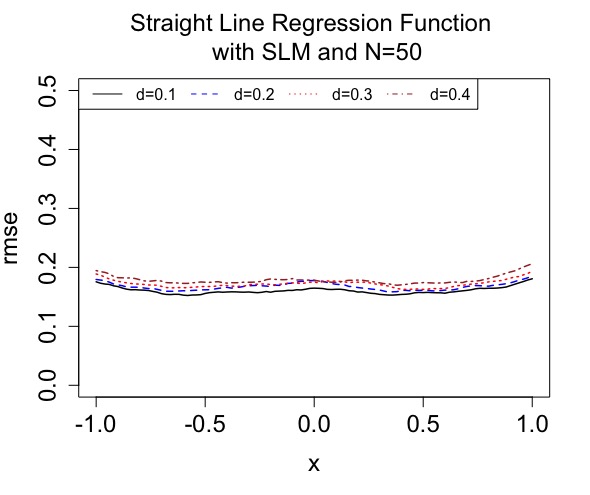}
 \end{tabular}
\caption{Comparison of empirical absolute bias, std, and rmse for the straight line regression function with $h=N^{-1/7}$ and $\lambda=N^{-1/5}$ under SLM.}
\label{Fig.line_N50_1}
\end{figure}

\begin{figure}[ht]
\centering
\begin{tabular}{ cc }
 \includegraphics[width=0.45\textwidth]{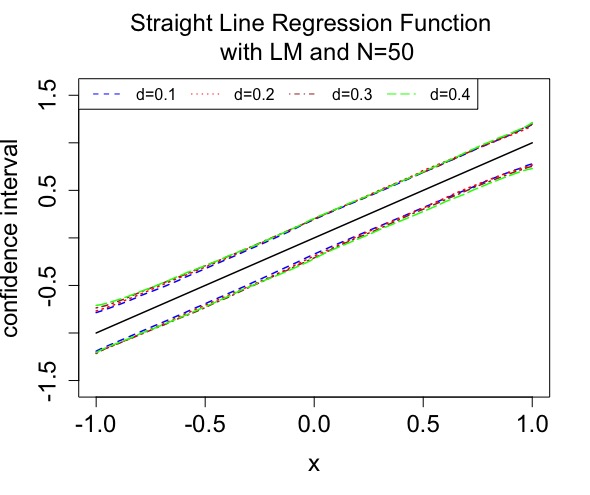} &
 \includegraphics[width=0.45\textwidth]{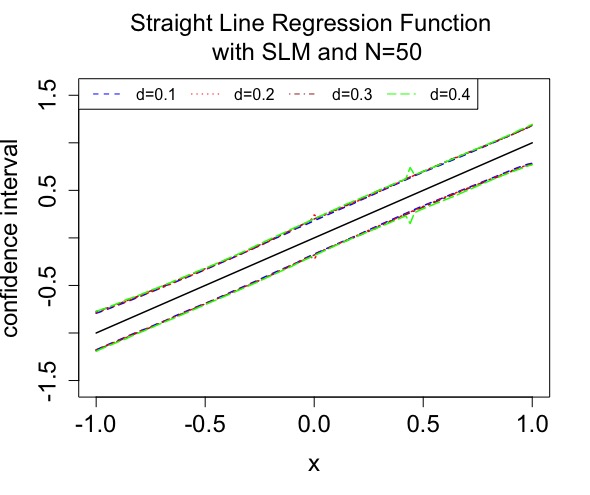}\\
  \includegraphics[width=0.45\textwidth]{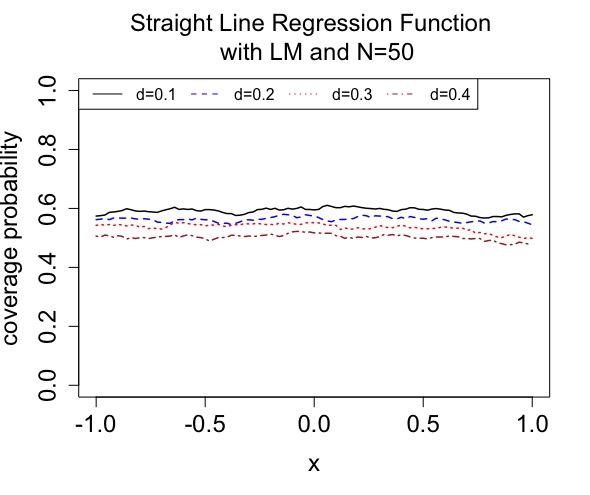} &
 \includegraphics[width=0.45\textwidth]{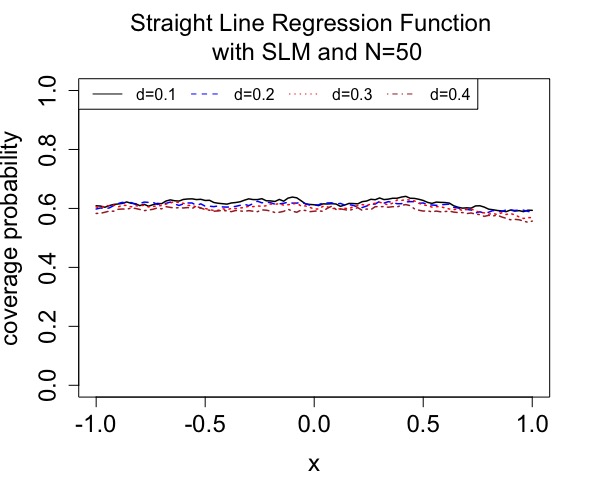} \\
  \includegraphics[width=0.45\textwidth]{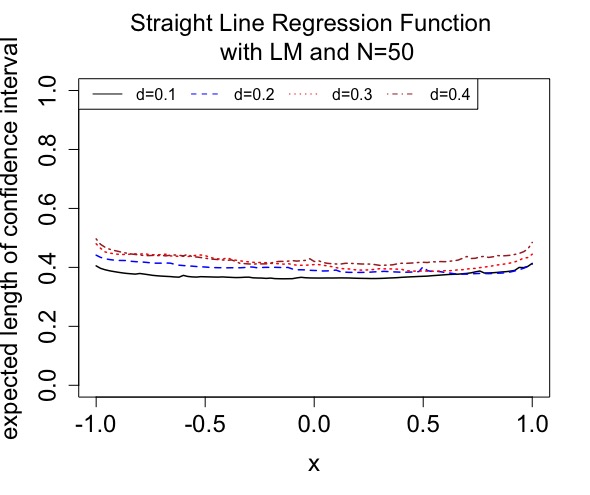} &
 \includegraphics[width=0.45\textwidth]{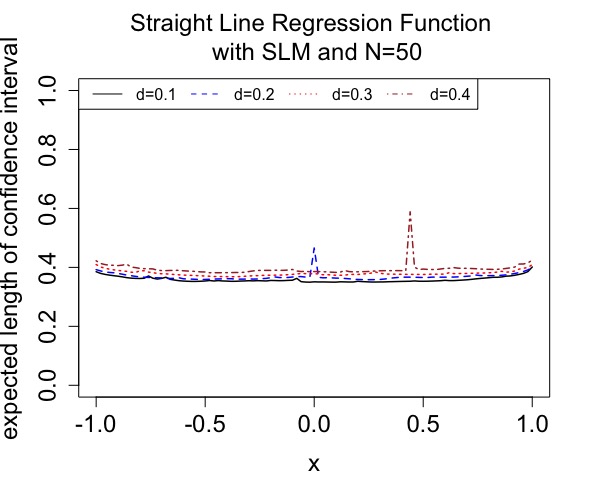}
 \end{tabular}
\caption{Comparison of empirical confidence intervals, coverage probabilities and expected lengths of confidence intervals for the straight line regression function with $h=N^{-1/7}$ and $\lambda=N^{-1/5}$ under SLM.}
\label{Fig.line_N50_2}
\end{figure}

\begin{figure}[ht]
\centering
\begin{tabular}{ cc }
 \includegraphics[width=0.45\textwidth]{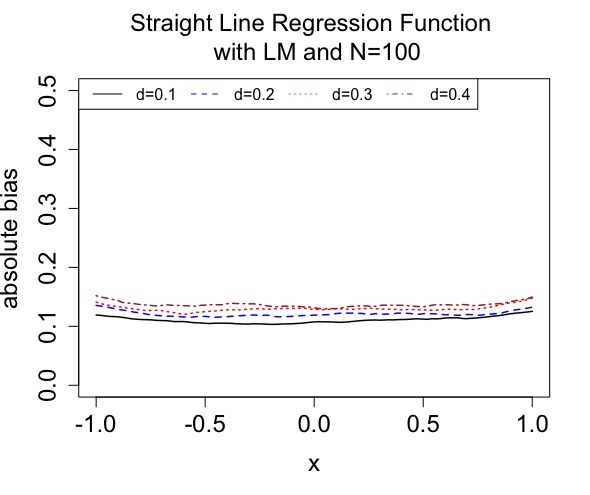} &
 \includegraphics[width=0.45\textwidth]{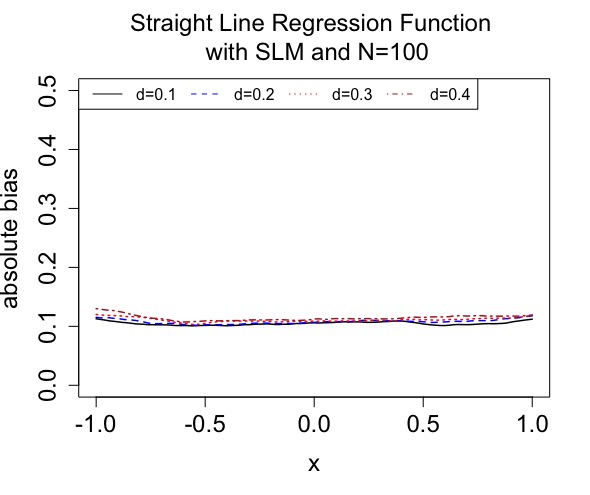}\\
  \includegraphics[width=0.45\textwidth]{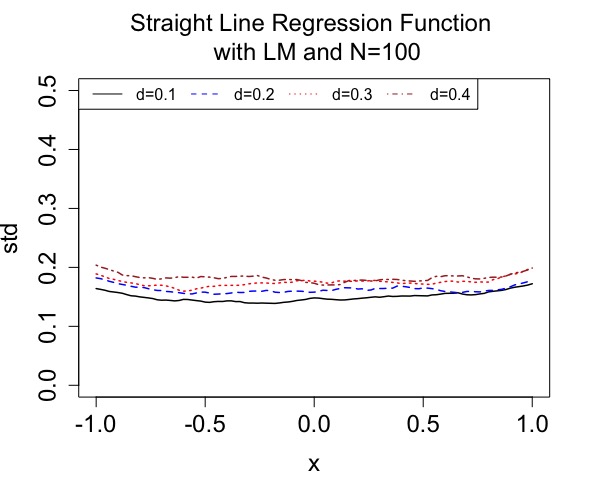} &
 \includegraphics[width=0.45\textwidth]{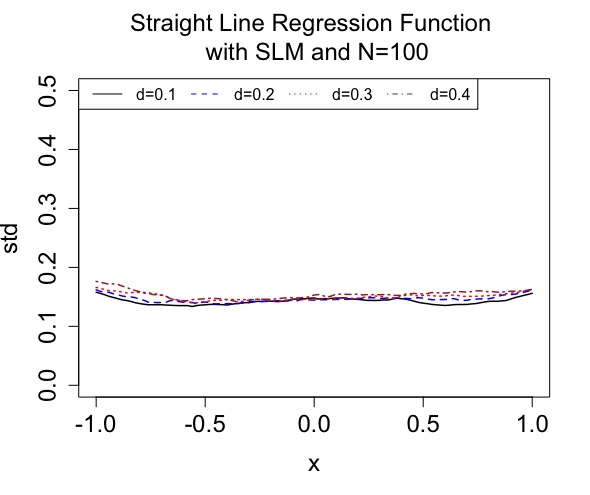} \\
  \includegraphics[width=0.45\textwidth]{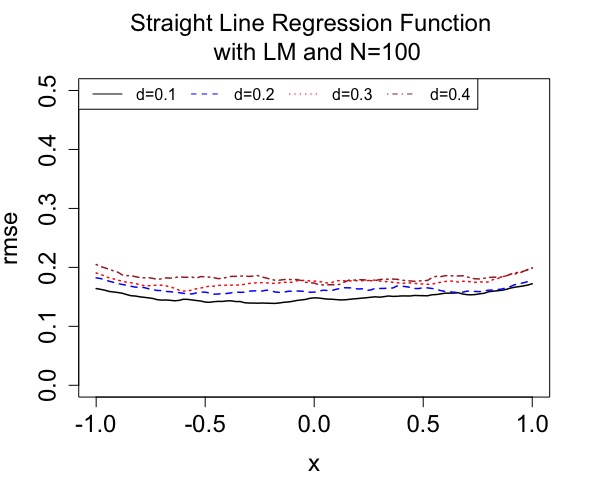} &
 \includegraphics[width=0.45\textwidth]{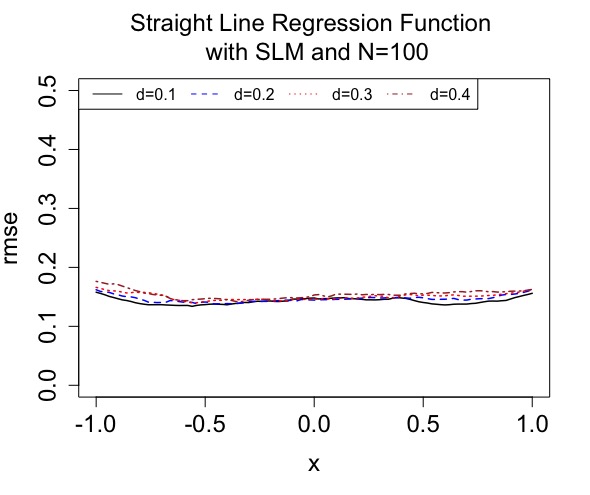}
 \end{tabular}
\caption{Comparison of empirical absolute bias, std, and rmse for the straight line regression function with $h=N^{-1/7}$ and $\lambda=N^{-1/5}$ under SLM.}
\label{Fig.line_N100_1}
\end{figure}

\begin{figure}[ht]
\centering
\begin{tabular}{ cc }
 \includegraphics[width=0.45\textwidth]{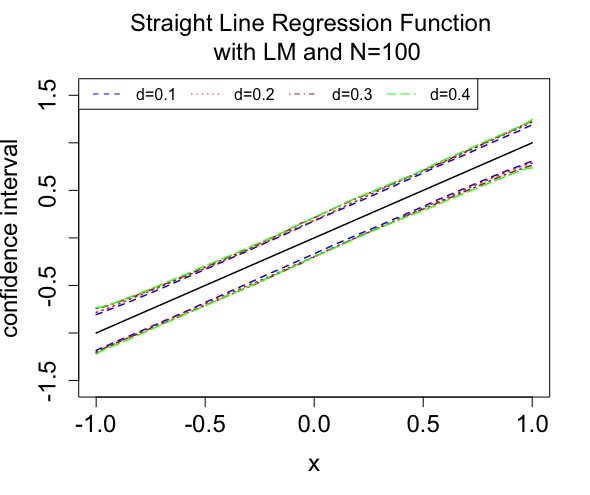} &
 \includegraphics[width=0.45\textwidth]{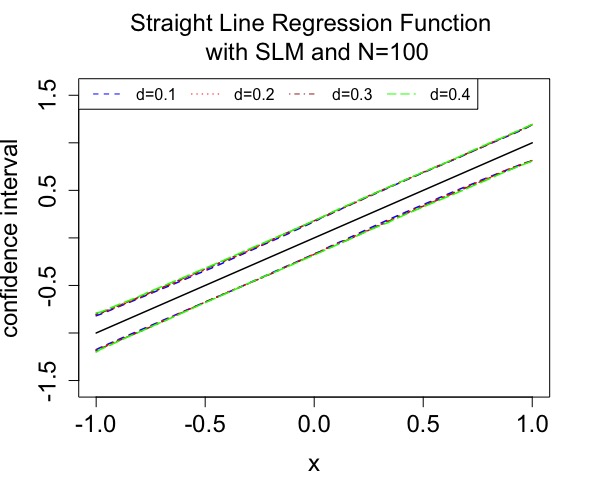}\\
  \includegraphics[width=0.45\textwidth]{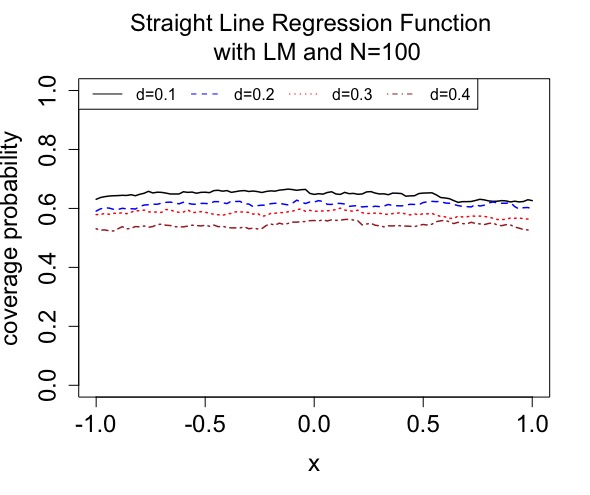} &
 \includegraphics[width=0.45\textwidth]{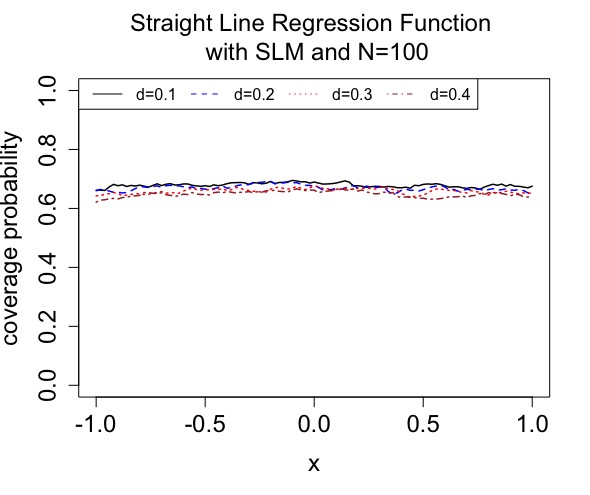} \\
  \includegraphics[width=0.45\textwidth]{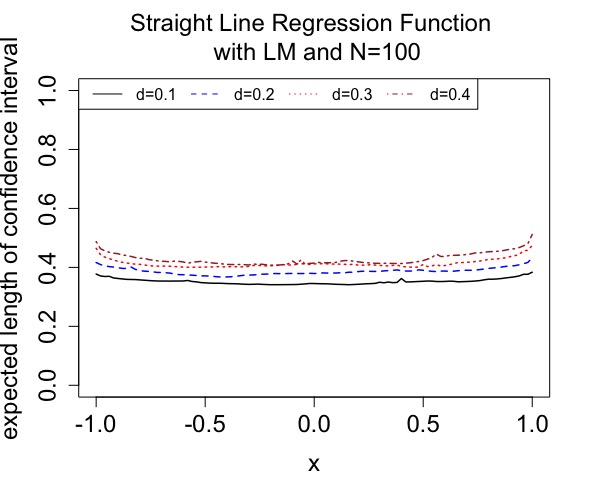} &
 \includegraphics[width=0.45\textwidth]{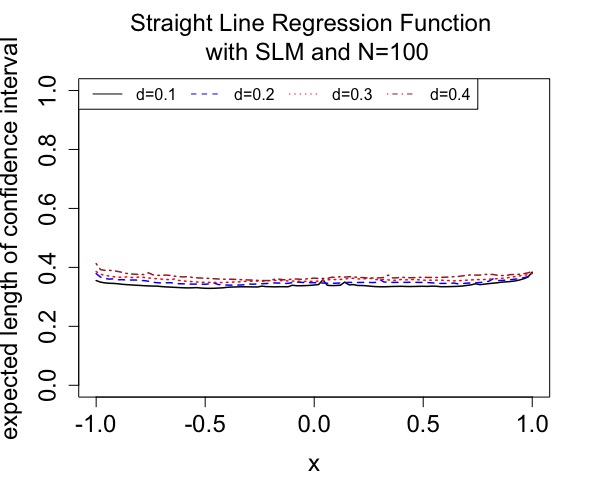}
 \end{tabular}
\caption{Comparison of empirical confidence intervals, coverage probabilities and expected lengths of confidence intervals for the straight line regression function with $h=N^{-1/7}$ and $\lambda=N^{-1/5}$ under SLM.}
\label{Fig.line_N100_2}
\end{figure}

\begin{figure}[ht]
\centering
\begin{tabular}{ cc }
 \includegraphics[width=0.45\textwidth]{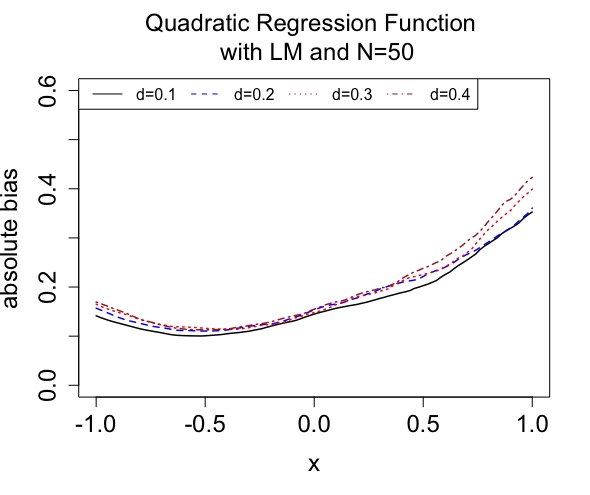} &
 \includegraphics[width=0.45\textwidth]{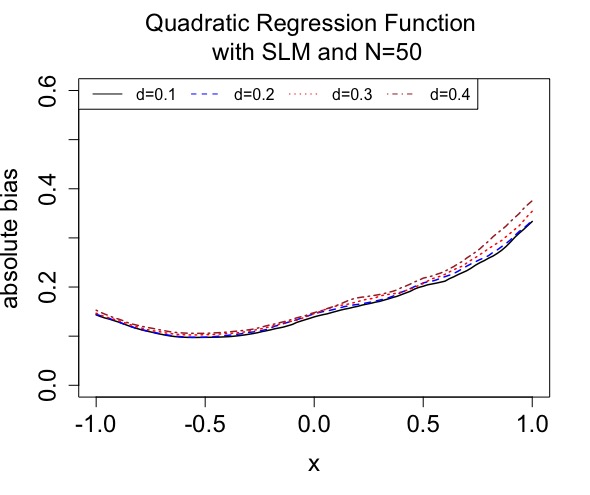}\\
  \includegraphics[width=0.45\textwidth]{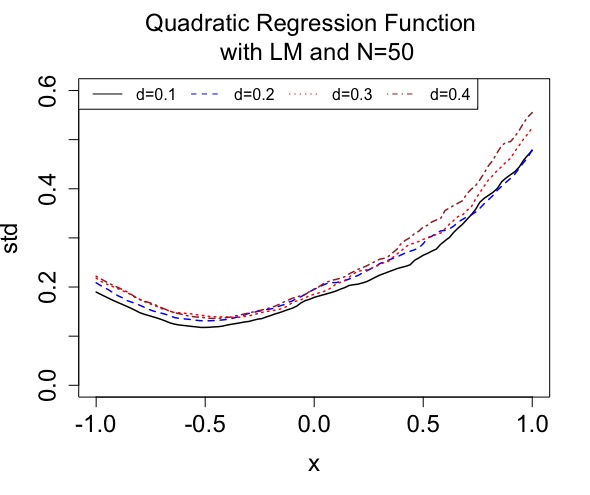} &
 \includegraphics[width=0.45\textwidth]{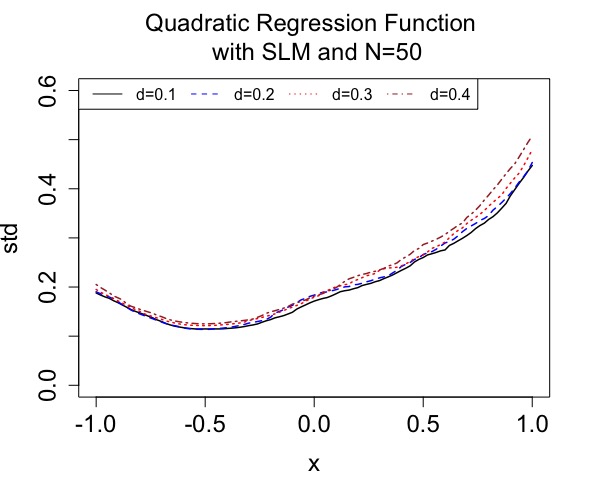} \\
  \includegraphics[width=0.45\textwidth]{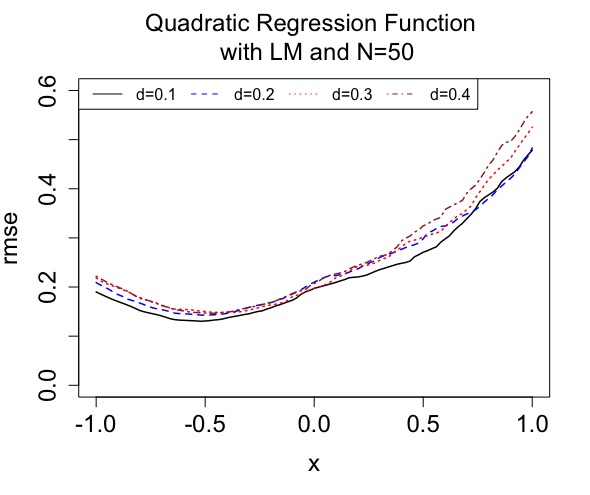} &
 \includegraphics[width=0.45\textwidth]{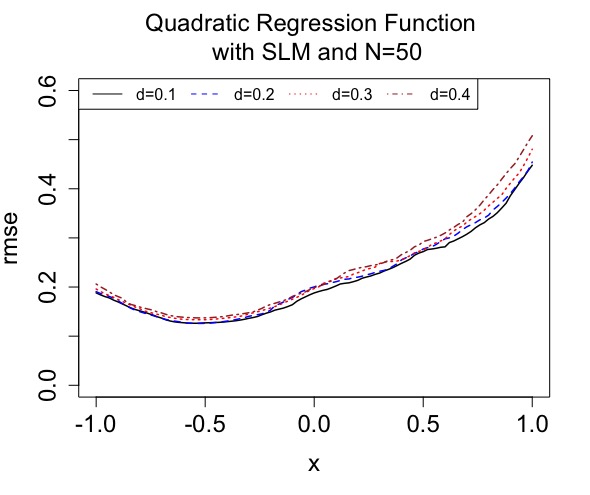}
 \end{tabular}
\caption{Comparison of empirical absolute bias, std, and rmse for the quadratic regression function with $h=N^{-1/7}$ and $\lambda=N^{-1/5}$ under SLM.}
\label{Fig.quad_N50_1}
\end{figure}

\begin{figure}[ht]
\centering
\begin{tabular}{ cc }
 \includegraphics[width=0.45\textwidth]{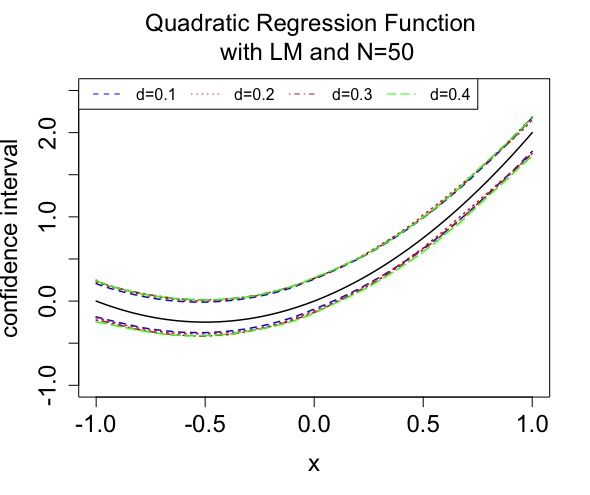} &
 \includegraphics[width=0.45\textwidth]{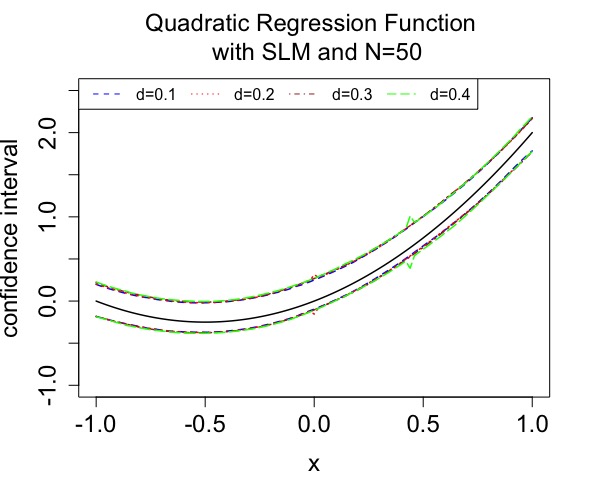}\\
  \includegraphics[width=0.45\textwidth]{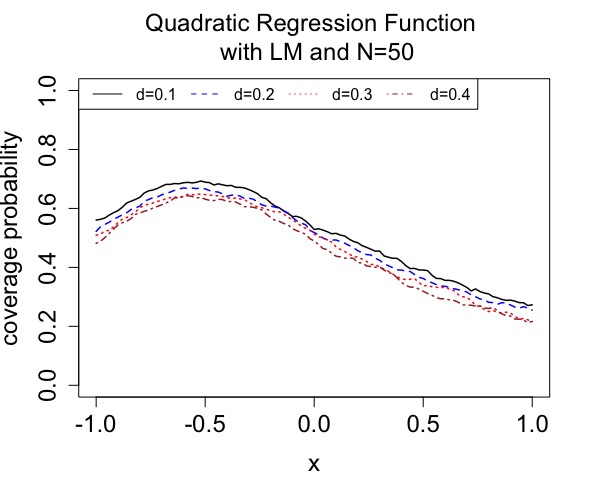} &
 \includegraphics[width=0.45\textwidth]{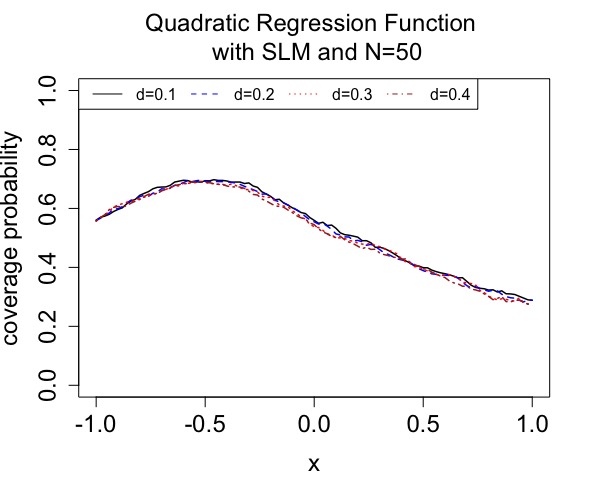} \\
  \includegraphics[width=0.45\textwidth]{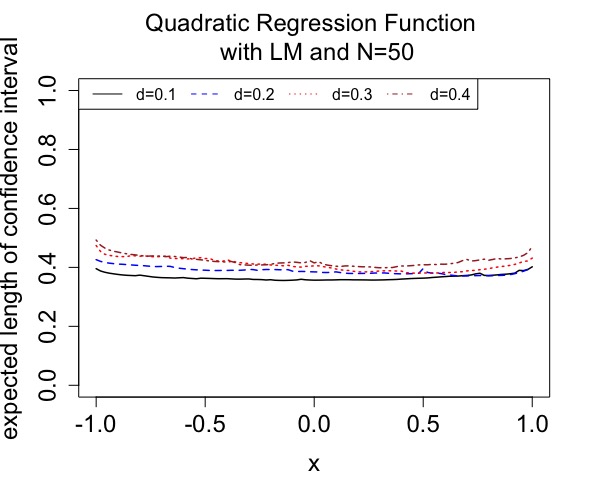} &
 \includegraphics[width=0.45\textwidth]{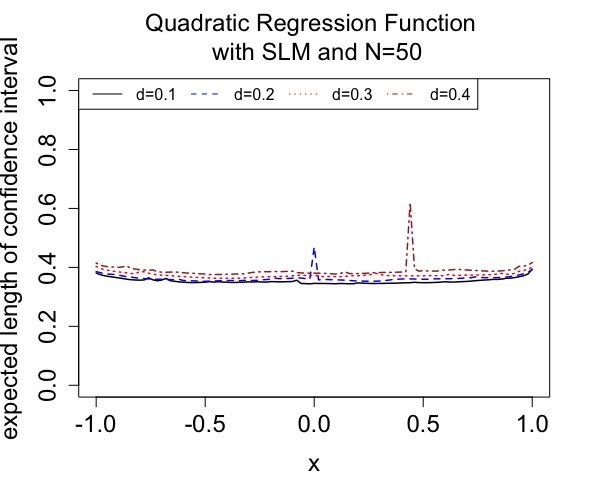}
 \end{tabular}
\caption{Comparison of empirical confidence intervals, coverage probabilities and expected lengths of confidence intervals for the quadratic regression function with $h=N^{-1/7}$ and $\lambda=N^{-1/5}$ under SLM.}
\label{Fig.quad_N50_2}
\end{figure}

\begin{figure}[ht]
\centering
\begin{tabular}{ cc }
 \includegraphics[width=0.45\textwidth]{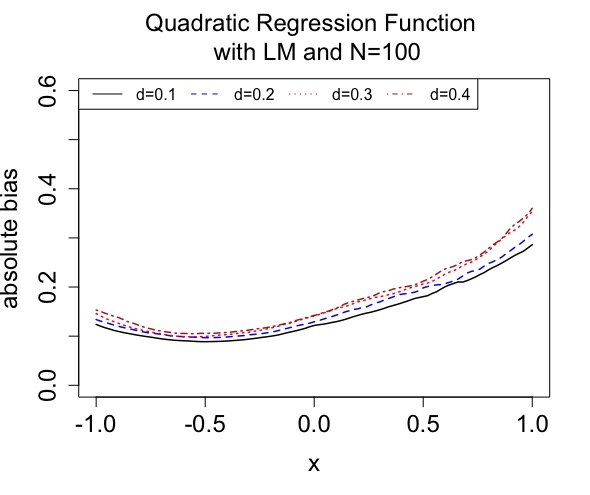} &
 \includegraphics[width=0.45\textwidth]{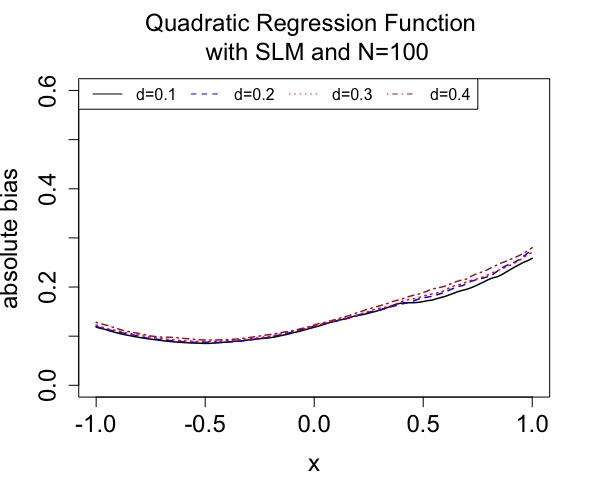}\\
  \includegraphics[width=0.45\textwidth]{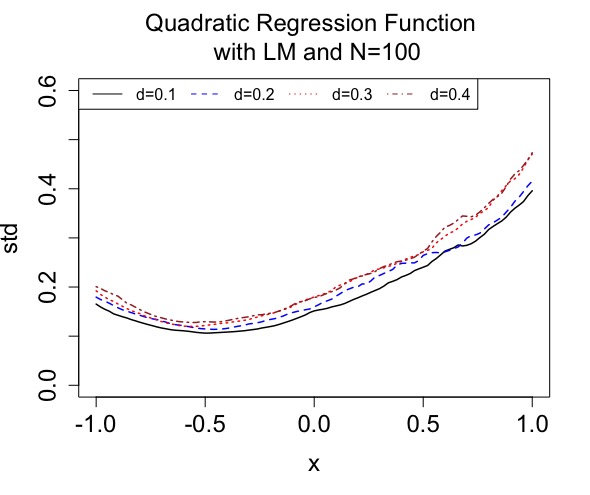} &
 \includegraphics[width=0.45\textwidth]{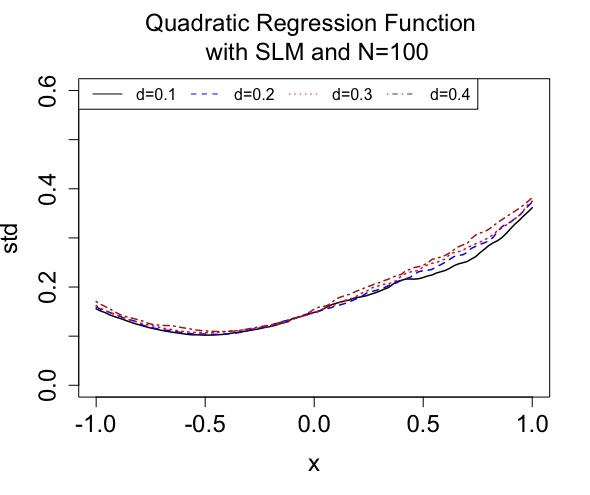} \\
  \includegraphics[width=0.45\textwidth]{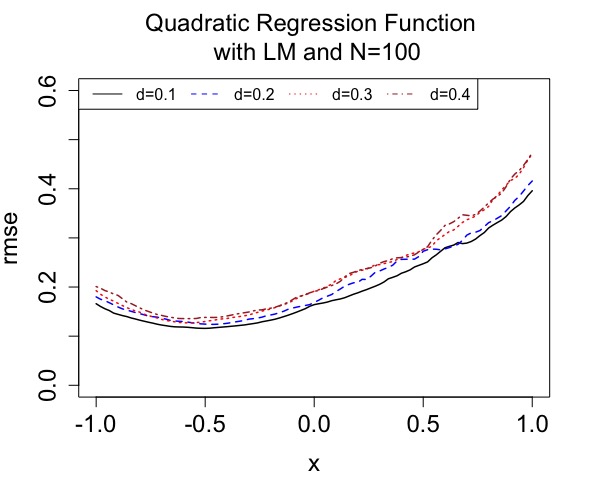} &
 \includegraphics[width=0.45\textwidth]{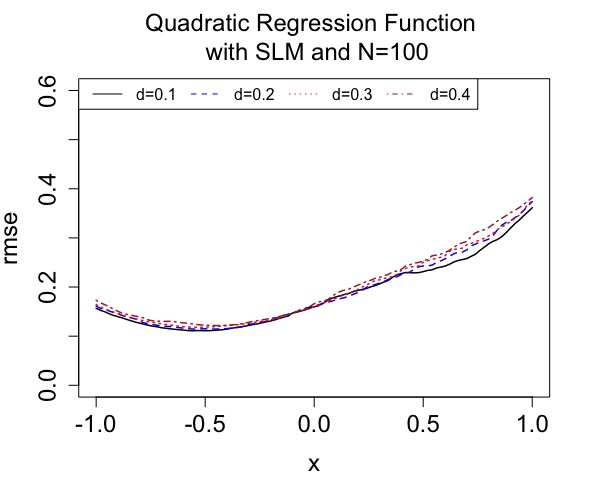}
 \end{tabular}
\caption{Comparison of empirical absolute bias, std, and rmse for the quadratic regression function with $h=N^{-1/7}$ and $\lambda=N^{-1/5}$ under SLM.}
\label{Fig.quad_N100_1}
\end{figure}

\begin{figure}[ht]
\centering
\begin{tabular}{ cc }
 \includegraphics[width=0.45\textwidth]{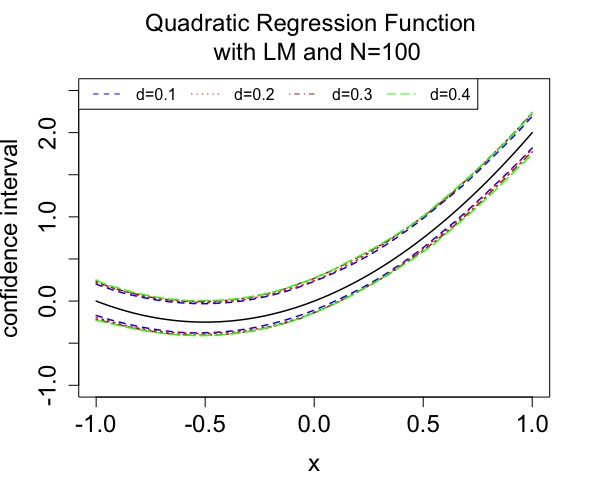} &
 \includegraphics[width=0.45\textwidth]{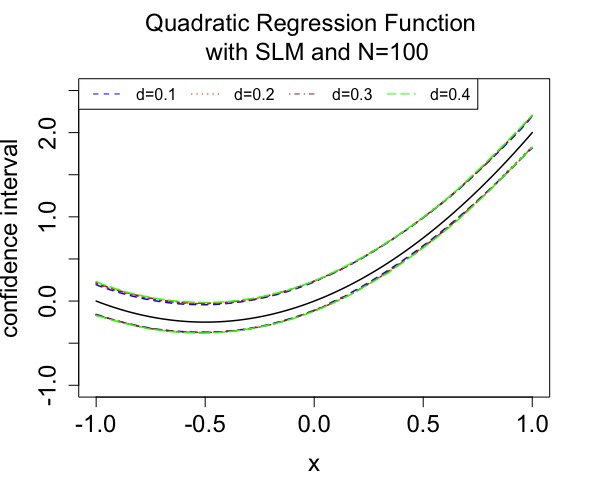}\\
  \includegraphics[width=0.45\textwidth]{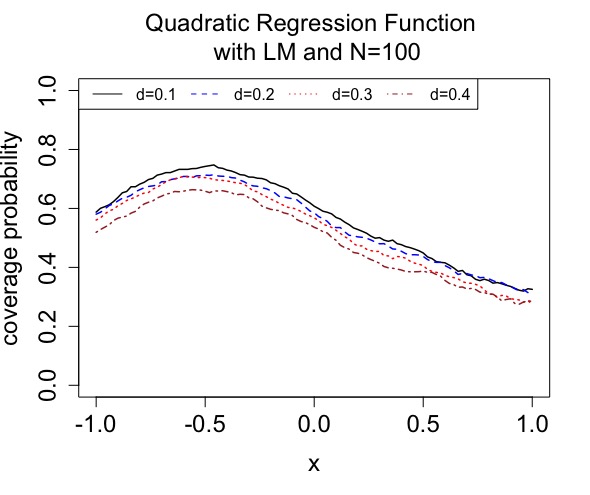} &
 \includegraphics[width=0.45\textwidth]{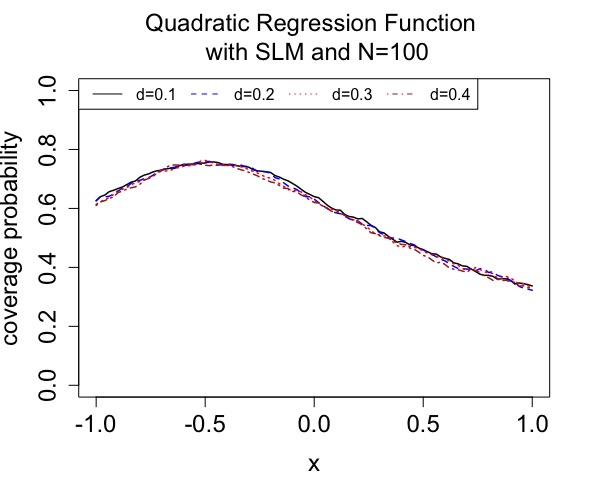} \\
  \includegraphics[width=0.45\textwidth]{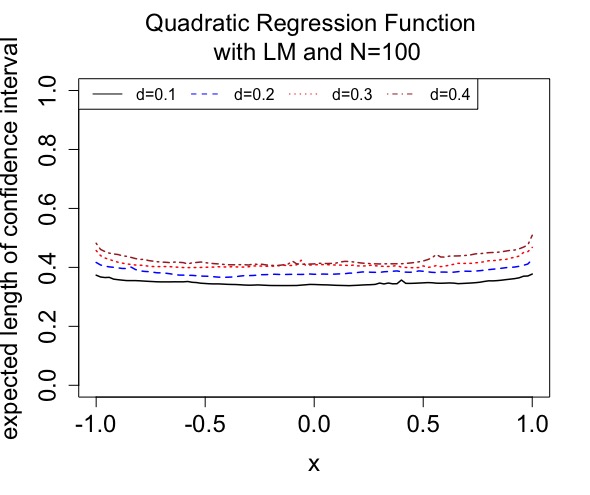} &
 \includegraphics[width=0.45\textwidth]{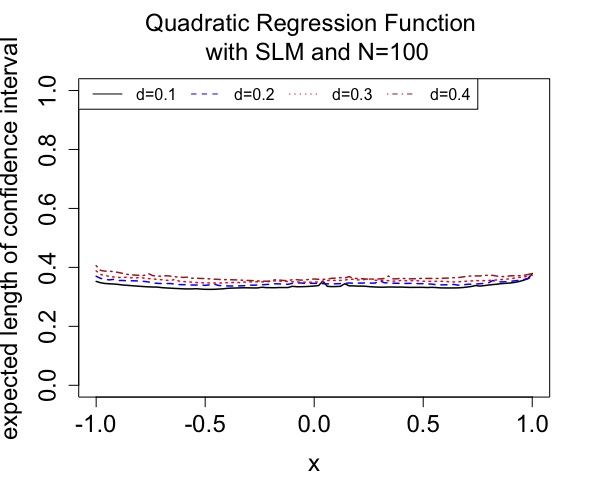}
 \end{tabular}
\caption{Comparison of empirical confidence intervals, coverage probabilities and expected lengths of confidence intervals for the quadratic regression function with $h=N^{-1/7}$ and $\lambda=N^{-1/5}$ under SLM.}
\label{Fig.quad_N100_2}
\end{figure}

\begin{figure}[ht]
\centering
\begin{tabular}{ cc }
 \includegraphics[width=0.45\textwidth]{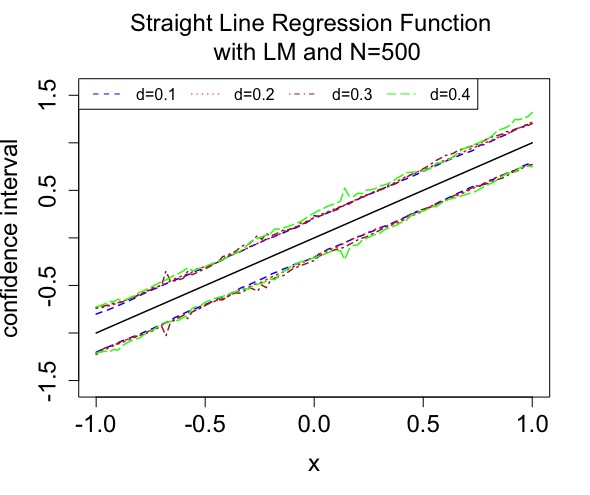} &
 \includegraphics[width=0.45\textwidth]{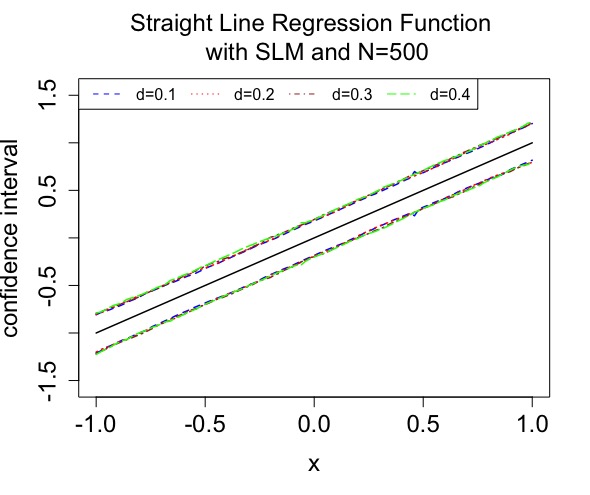}\\
  \includegraphics[width=0.45\textwidth]{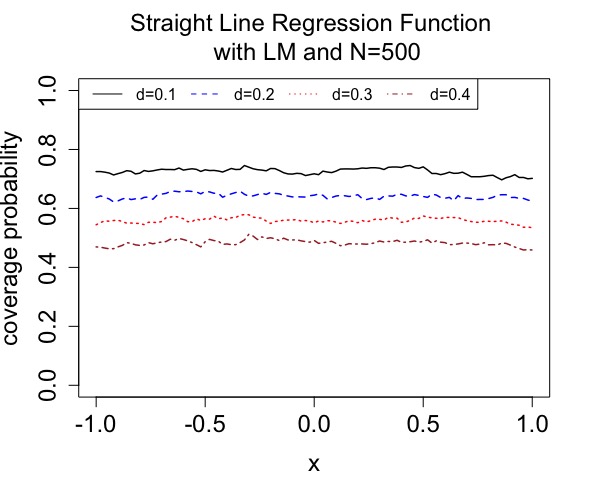} &
 \includegraphics[width=0.45\textwidth]{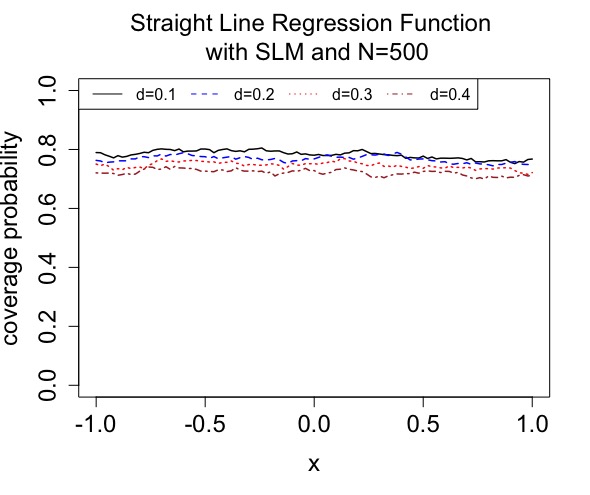} \\
  \includegraphics[width=0.45\textwidth]{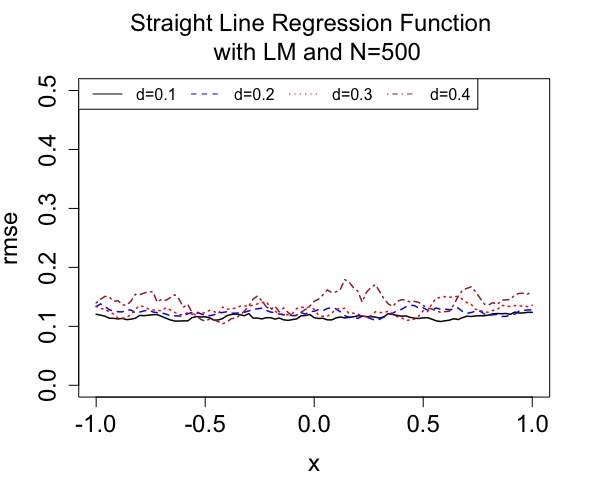} &
 \includegraphics[width=0.45\textwidth]{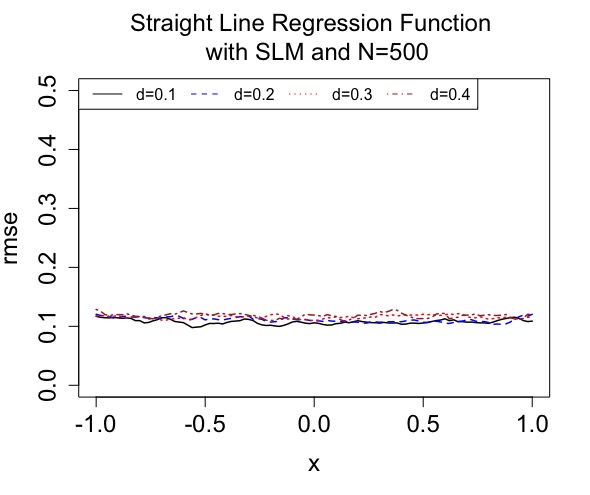}
 \end{tabular}
\caption{Comparison of empirical confidence intervals, coverage probabilities and expected lengths of confidence intervals for the straight line regression function with $h=N^{-1/3}$ and $\lambda=N^{-1/5}$ under SLM.}
\label{Fig.line_N500_smallh}
\end{figure}

\begin{figure}[ht]
\centering
\begin{tabular}{ cc }
 \includegraphics[width=0.45\textwidth]{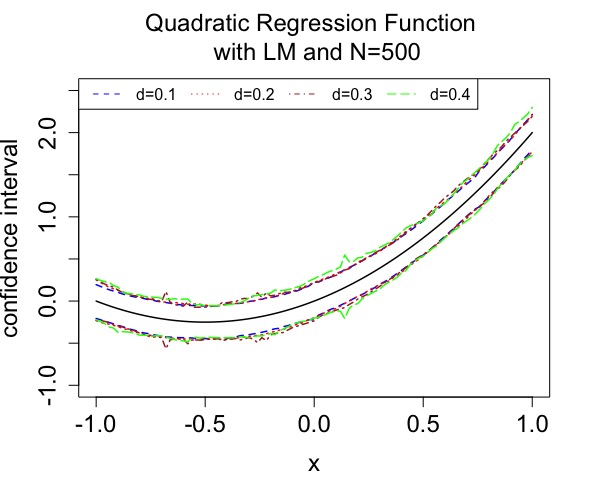} &
 \includegraphics[width=0.45\textwidth]{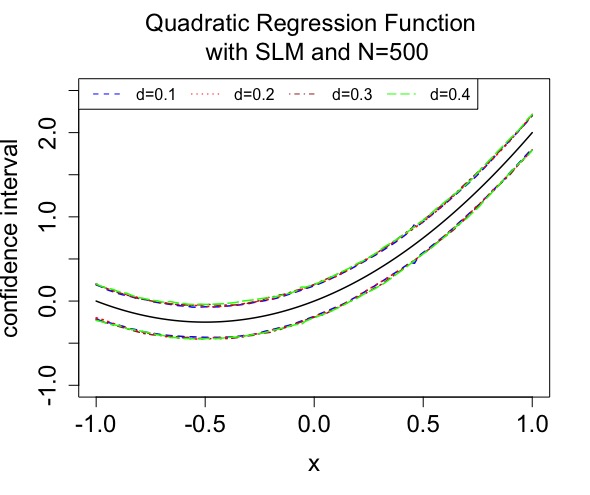}\\
  \includegraphics[width=0.45\textwidth]{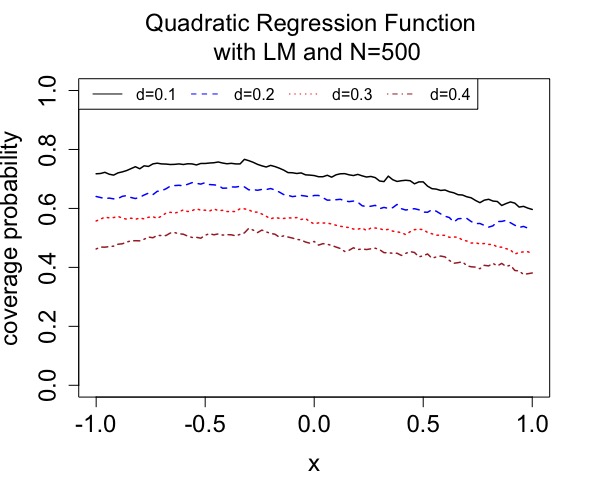} &
 \includegraphics[width=0.45\textwidth]{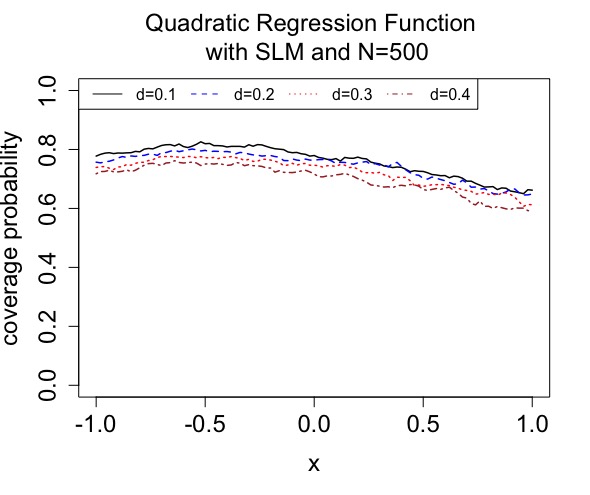} \\
  \includegraphics[width=0.45\textwidth]{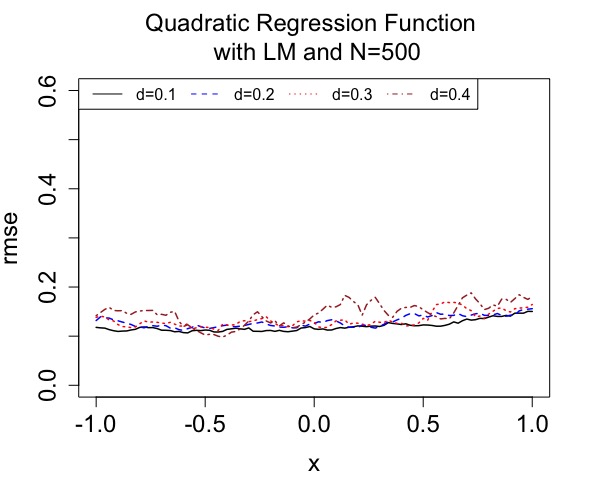} &
 \includegraphics[width=0.45\textwidth]{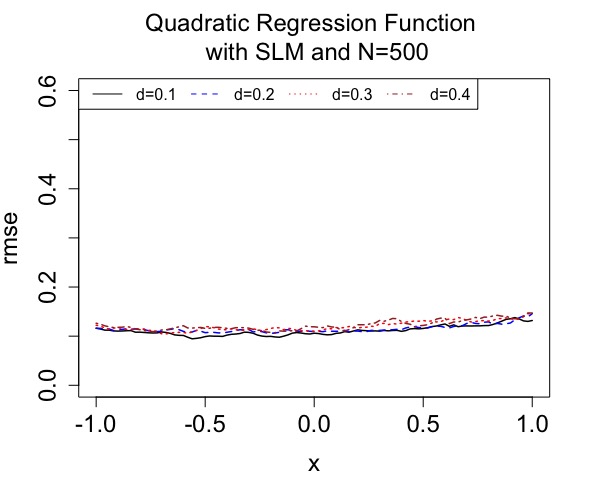}
 \end{tabular}
\caption{Comparison of empirical confidence intervals, coverage probabilities and expected lengths of confidence intervals for the quadratic regression function with $h=N^{-1/3}$ and $\lambda=N^{-1/5}$ under SLM.}
\label{Fig.quad_N500_smallh}
\end{figure}

\begin{table}[!ht]
\footnotesize
\caption{The values of size and power for de-biased test statistic at $5\%$ level for the straight line regression function $f(x)$.} 
\label{Tab:sizepower_allNs_line}
\centering
\setlength{\tabcolsep}{10pt} 
\renewcommand{\arraystretch}{1.25} 
\begin{tabular}{ccccc}
\toprule
N & size/power & d & LM & SLM \\
\midrule
50 & size & 0.1 & $\{0.338, 0.196, 0.108, 0.136 \}$ & $\{ 0.342,       0.201, 0.105, 0.144 \}$ \\
 && 0.2 & $\{0.322, 0.200, 0.113, 0.143 \}$ & $\{ 0.349, 0.196,       0.102, 0.145 \}$ \\
 && 0.3 & $\{0.322, 0.206, 0.114, 0.139 \}$ & $\{0.350, 0.186,       0.104, 0.143 \}$ \\
 && 0.4 & $\{0.324, 0.212, 0.119, 0.133 \}$ & $\{ 0.335, 0.185,      0.098, 0.142 \}$ \\ \midrule
50 & power & 0.1 & $\{ 0.350, 0.216, 0.121, 0.149 \}$ & $\{0.351, 0.215, 0.118, 0.150 \}$ \\
  && 0.2 & $\{ 0.334, 0.217, 0.125, 0.153 \}$ & $\{ 0.362, 0.207, 0.118, 0.150 \}$ \\
 && 0.3 & $\{0.347, 0.221, 0.139, 0.153 \}$ & $\{ 0.357, 0.206,       0.122, 0.159 \}$ \\
 && 0.4 & $\{0.348, 0.240, 0.156, 0.158 \}$ & $\{0.351, 0.204,       0.115, 0.158 \}$ \\ \midrule
 100 & size & 0.1 & $\{ 0.203, 0.118, 0.076, 0.075 \}$ & $\{0.201, 0.122, 0.074, 0.084 \}$ \\
 && 0.2 & $\{ 0.207, 0.125, 0.082, 0.088 \}$ & $\{ 0.206, 0.116,      0.069, 0.083 \}$ \\
 && 0.3 & $\{0.205, 0.130, 0.087, 0.075 \}$ & $\{0.200, 0.105,      0.066, 0.080 \}$ \\
 && 0.4 & $\{0.208, 0.143, 0.104, 0.091 \}$ & $\{0.192, 0.109,      0.073, 0.077 \}$ \\ \midrule
 100 & power & 0.1 & $\{0.213, 0.128, 0.081, 0.086 \}$ & $\{0.211,     0.120, 0.079, 0.085 \}$ \\
  && 0.2 & $\{0.224, 0.141, 0.099, 0.102 \}$ & $\{0.212, 0.123,      0.076, 0.089 \}$ \\
 && 0.3 & $\{0.231, 0.153, 0.108, 0.096 \}$ & $\{0.210, 0.116,      0.075, 0.089 \}$ \\
 && 0.4 & $\{0.247, 0.186, 0.138, 0.128 \}$ & $\{0.212, 0.127,      0.084, 0.087 \}$ \\
\bottomrule
\end{tabular}
\end{table} 

\begin{table}[!ht]
\footnotesize
\caption{The values of size and power for de-biased test statistic at $5\%$ level for the straight line regression function $f(x)$ with SLM and large values of $d$.} 
\label{Tab:sizepower_bigd_line}
\centering
\setlength{\tabcolsep}{10pt} 
\renewcommand{\arraystretch}{1.25} 
\begin{tabular}{ccccc}
\toprule
size/power & d & N=50 & N=100 & N=500 \\
\midrule
size & 0.5 & $\{0.322, 0.181, 0.100, 0.144 \}$ & $\{0.202, 0.109, 0.076, 0.086\}$ & $\{0.057, 0.034, 0.023, 0.027 \}$ \\
 & 1 & $\{0.307, 0.170, 0.105, 0.138 \}$ & $\{0.186, 0.117, 0.082, 0.094\}$ & $\{0.064, 0.040, 0.029, 0.031 \}$ \\
 & 1.5 & $\{0.303, 0.185, 0.108, 0.145 \}$ & $\{0.209, 0.139,       0.102, 0.106 \}$ & $\{0.061, 0.042, 0.029, 0.025 \}$ \\
 \midrule
power & 0.5 & $\{ 0.351, 0.208, 0.115, 0.156 \}$ & $\{0.221,       0.133, 0.090, 0.103 \}$ & $\{0.063, 0.041, 0.026, 0.032 \}$ \\
  & 1 & $\{ 0.345, 0.211, 0.135, 0.171 \}$ & $\{0.234, 0.156, 0.116, 0.126 \}$ & $\{ 0.093, 0.069, 0.053, 0.052 \}$ \\
 & 1.5 & $\{0.359, 0.252, 0.177, 0.201 \}$ & $\{0.298, 0.230,       0.191, 0.181 \}$ & $\{0.180, 0.152, 0.127, 0.116 \}$ \\
\bottomrule
\end{tabular}
\end{table} 

\begin{table}[!ht]
\footnotesize
\caption{The values of size and power for de-biased test statistic at $5\%$ level for the quadratic regression function $f(x)$.} 
\label{Tab:sizepower_allNs_quad}
\centering
\setlength{\tabcolsep}{10pt} 
\renewcommand{\arraystretch}{1.25} 
\begin{tabular}{ccccc}
\toprule
N & size/power & d & LM & SLM \\
\midrule
50 & size & 0.1 & $\{ 0.287, 0.210, 0.109, 0.121 \}$ & $\{0.307,       0.212, 0.114, 0.128 \}$ \\
 && 0.2 & $\{0.292, 0.211, 0.117, 0.118 \}$ & $\{ 0.308,0.214,       0.108, 0.122 \}$ \\
 && 0.3 & $\{0.295, 0.207, 0.113, 0.117 \}$ & $\{0.302, 0.207,       0.104, 0.116 \}$ \\
 && 0.4 & $\{0.309, 0.207, 0.121, 0.124 \}$ & $\{0.298, 0.201,      0.097, 0.110 \}$ \\ \midrule
50 & power & 0.1 & $\{0.294, 0.214, 0.109, 0.124\}$ & $\{ 0.304,       0.216, 0.116, 0.129 \}$ \\
  && 0.2 & $\{ 0.287, 0.215, 0.116, 0.118 \}$ & $\{0.309, 0.215,      0.109, 0.120 \}$ \\
 && 0.3 & $\{ 0.299, 0.209, 0.121, 0.119 \}$ & $\{0.307, 0.208,       0.106, 0.121 \}$ \\
 && 0.4 & $\{ 0.312, 0.213, 0.132, 0.133 \}$ & $\{0.301, 0.200,       0.105, 0.117 \}$ \\ \midrule
 100 & size & 0.1 & $\{ 0.164, 0.105, 0.064, 0.056 \}$ & $\{0.174,     0.110, 0.066, 0.057\}$ \\
 && 0.2 & $\{ 0.175, 0.106, 0.069, 0.071 \}$ & $\{ 0.165, 0.105,      0.061, 0.058 \}$ \\
 && 0.3 & $\{ 0.164, 0.119, 0.084, 0.072 \}$ & $\{ 0.177, 0.102,      0.060, 0.057 \}$ \\
 && 0.4 & $\{ 0.182, 0.131, 0.090, 0.077\}$ & $\{0.172, 0.108,      0.063, 0.061 \}$ \\ \midrule
 100 & power & 0.1 & $\{ 0.168, 0.109, 0.065, 0.058 \}$ & $\{ 0.174,   0.116, 0.066, 0.062 \}$ \\
  && 0.2 & $\{ 0.180, 0.114, 0.070, 0.071 \}$ & $\{ 0.170, 0.111,     0.063, 0.061 \}$ \\
 && 0.3 & $\{ 0.174, 0.122, 0.086, 0.073 \}$ & $\{ 0.184, 0.109,      0.062, 0.060 \}$ \\
 && 0.4 & $\{ 0.188, 0.136, 0.098, 0.085 \}$ & $\{ 0.177, 0.113,      0.069, 0.064 \}$ \\
\bottomrule
\end{tabular}
\end{table}

\begin{table}[!ht]
\footnotesize
\caption{The values of size and power for de-biased test statistic at $5\%$ level for the quadratic regression function $f(x)$ with SLM and large values of $d$.} 
\label{Tab:sizepower_bigd_quad}
\centering
\setlength{\tabcolsep}{10pt} 
\renewcommand{\arraystretch}{1.25} 
\begin{tabular}{ccccc}
\toprule
size/power & d & N=50 & N=100 & N=500 \\
\midrule
size & 0.5 & $\{0.293, 0.195, 0.098, 0.118\}$ & $\{0.184, 0.110,      0.064, 0.074\}$ & $\{0.024, 0.011, 0.009, 0.009 \}$ \\
 & 1 & $\{0.301, 0.175, 0.088, 0.114\}$ & $\{0.171, 0.107, 0.068,      0.063 \}$ & $\{0.031, 0.020, 0.009, 0.010 \}$ \\
 & 1.5 & $\{ 0.334, 0.179, 0.105, 0.123 \}$ & $\{0.185, 0.121,      0.084, 0.083 \}$ & $\{0.076, 0.017, 0.011, 0.008\}$ \\
 \midrule
power & 0.5 & $\{0.305, 0.202, 0.100, 0.122 \}$ & $\{0.186, 0.116,      0.070, 0.076 \}$ & $\{ 0.025, 0.012, 0.010, 0.011 \}$ \\
  & 1 & $\{0.311, 0.180, 0.096, 0.125 \}$ & $\{ 0.180, 0.112, 0.076, 0.076 \}$ & $\{ 0.032, 0.019, 0.010, 0.011 \}$ \\
 & 1.5 & $\{ 0.356, 0.202, 0.125, 0.141 \}$ & $\{0.202, 0.141,       0.101, 0.097 \}$ & $\{ 0.088, 0.027, 0.018, 0.013 \}$ \\
\bottomrule
\end{tabular}
\end{table}

\end{document}